\documentclass[fleqn,usenatbib]{mnras}

\usepackage{newtxtext,newtxmath}

\usepackage[T1]{fontenc}
\usepackage{ae,aecompl}

\DeclareRobustCommand{\VAN}[3]{#2}
\let\VANthebibliography\thebibliography
\def\thebibliography{\DeclareRobustCommand{\VAN}[3]{##3}\VANthebibliography}

\usepackage{graphicx}	
\usepackage{subfig}
\usepackage{amsmath}	

\usepackage{amssymb}	
\usepackage{amsfonts}
\usepackage{multicol}        
\usepackage{bm}		
\usepackage{pdflscape}	
\usepackage{natbib}
\usepackage{epstopdf}
\usepackage{bm}
\usepackage{dcolumn}
\usepackage{hyperref}
\usepackage{cancel}
\usepackage{color}
\usepackage{latexsym}
\usepackage{nicefrac}
\usepackage{latexsym}
\usepackage{booktabs}

\title[Forecasts on Interacting Dark Energy with BINGO and SKA]{Forecasts on Interacting Dark Energy from 21-cm Angular Power Spectrum with BINGO and SKA observations
}

\author[L. Xiao, A. A. Costa, B. Wang]{
Linfeng Xiao,$^{1}$\thanks{E-mail: hartley@sjtu.edu.cn}
Andre A. Costa,$^{2}$\thanks{E-mail: andrecosta@yzu.edu.cn (corresponding author)}
Bin Wang$^{3,2}$\thanks{E-mail: wang\_b@sjtu.edu.cn}
\\
$^{1}$Department of Astronomy, School of Physics and Astronomy, Shanghai Jiao Tong University, Shanghai, 200240, China\\
$^{2}$Center for Gravitation and Cosmology, College of Physical Science and Technology, Yangzhou University, Yangzhou 225009, China\\
$^{3}$School of Aeronautics and Astronautics, Shanghai Jiao Tong University, Shanghai 200240, China.
}

\date{Accepted XXX. Received YYY; in original form ZZZ}

\pubyear{2020}

\begin{document}
\label{firstpage}
\pagerange{\pageref{firstpage}--\pageref{lastpage}}
\maketitle

\begin{abstract}
Neutral hydrogen (HI) intensity mapping is a promising technique to probe the large-scale structure of the Universe, improving our understanding on the late-time accelerated expansion. In this work, we first scrutinize how an alternative cosmology, interacting dark energy (IDE), can affect the 21-cm angular power spectrum relative to the concordance $\Lambda$CDM model. We re-derive the 21-cm brightness temperature fluctuation in the context of such interaction and uncover an extra new contribution. Then we estimate the noise level of three upcoming HI intensity mapping surveys, BINGO, SKA1-MID Band\,1 and Band\,2, respectively, and employ a Fisher matrix approach to forecast their constraints on the IDE model. We find that while {\it Planck} 2018 maintains its dominion over early-Universe parameter constraints, BINGO and SKA1-MID Band\,2 put complementary bounding to the latest CMB measurements on dark energy equation of state $w$, the interacting strength $\lambda_i$ and the reduced Hubble constant $h$, and SKA1-MID Band\,1 even outperforms {\it Planck} 2018 in these late-Universe parameter constraints. The expected minimum uncertainties are given by SKA1-MID Band\,1+{\it Planck}: $\sim 0.34\%$ on $w$, $\sim 0.22\%$ on $h$, $\sim 0.64\%$ on HI bias $b_{\rm HI}$, and an absolute uncertainty of about $3\times10^{-4}$ ($7\times10^{-4}$) on $\lambda_{1}$ ($\lambda_{2}$). Moreover, we quantify the effects from systematics of the redshift bin number, redshift-space distortions, foreground residuals and uncertainties on the measured HI fraction, $\Omega_{\mathrm{HI}}(z)$. Our results indicate a bright prospect for HI intensity mapping surveys in constraining IDE, whether on their own or further by synergies with other measurements.
\end{abstract}




\section{Introduction}\label{sec.intro}

Understanding the late-time accelerated expansion is one of the major challenges in modern cosmology. Within the framework of General Relativity (GR), such expansion is driven by an exotic form of energy with negative pressure, called Dark Energy (DE). Supported by observational evidences, a cosmological constant $\Lambda$ is still the prevailing DE candidate, albeit two decades of research. Another cosmological component with unknown physical nature giving rise to galaxy clusters and large-scale structures is cold Dark Matter (DM).  DE and DM dominate the energy budget  occupying $\sim 95\%$ of the total energy of our Universe nowadays. The common $\Lambda$CDM model, composed of these dark components and a small amount of ordinary matter, has  succeeded in accounting for numerous astronomical  observations, such as the temperature and polarization anisotropies in Cosmic Microwave Background (CMB) and the properties of large-scale structures.

In the present era of precision cosmology, the CMB measurement from {\it Planck} satellite can constrain the parameters of standard $\Lambda$CDM model to an accuracy level $\leq 1\%$~\citep{Aghanim:2018eyx}. Nevertheless, some inconsistencies between the CMB measurement and other low-redshift observations have been revealed in the $\Lambda$CDM model, such as the $H_0$ tension~\citep{Riess:2011,Riess:2016jrr}, the $\sigma_8$ tension~\citep{Ade:2015xua,Hamann:2013iba,Battye:2013xqa,Petri:2015ura}, discrepancies in measuring distances $D_A$ and $D_H$~\citep{Delubac:2014aqe}, discordance found in Kilo Degree Survey in weak lensing~\citep{Joudaki:2016kym}, 21-cm signal observed by EDGES~\citep{Bowman:2018yin} and the missing satellite~\citep{Klypin:1999uc,Simon:2007dq}. In spite of these observational challenges, the $\Lambda$CDM model also suffers two serious  theoretical problems: 1) The cosmological constant problem, namely why the value of $\Lambda$ is much smaller than that estimated in  quantum field theory~\citep{RevModPhys.61.1}. 2) The coincidence problem, which states why DM and DE can evolve to very similar energy density levels at the current moment~\citep{Chimento:2003iea}. $\Lambda$ is not the end story to account for the cosmic acceleration, there are many attempts to devise exotic fields to explain DE, but until now there is no clear winner at sight (for a review, see for example~\cite{2010deto.book.....A}).

Considering that DM and DE are the two main components of the Universe, a natural understanding from the field theory point of view is that there may have certain interactions between them. Since the physical nature of both DM and DE is not clear, it is very difficult to describe the interaction between dark sectors from first principles. A simple way is to start from a phenomenological description, assuming the coupling as a function of the energy densities of DM or DE. Inevitably such interaction significantly affects the evolution of our Universe in both the expansion history and growth of large-scale structures. In terms of the background evolution, the interacting DE (IDE) model can reproduce the result of a model with  varying effective DE equation of state (EoS)~\citep{Wang:2005jx,Wang:2005ph}. On the other hand, the influence of IDE will lead to the change in the gravitational potential evolution which can leave imprints on the CMB angular power spectrum ~\citep{He:2009pd,He:2010im,Baldi:2010vv,Baldi:2010pq,Xu:2011nr,Xu:2011tsa,Costa:2013sva,Pu:2014goa} and structure formation~\citep{He:2009mz,He:2010ta,Zhang:2018mlj,An:2018vzw}. Furthermore, through the gravitational potential, IDE is able to modulate the in-fall velocity of matter particles and results in modifications to redshift-space distortions (RSD)~\citep{Costa:2016tpb} and kinetic Sunyaev-Zel'dovich (kSZ) effect~\citep{Xu:2013jma}. Also, the change of the gravitational potential will deflect the trajectories of photons emitted from distant objects, which gives rise to a weak gravitational lensing effect~\citep{An:2017kqu,An:2017crg}. For a review on  theoretical challenges, cosmological implications and observational signatures on the IDE can be found in~\cite{Wang:2016lxa} and references therein.

In current observations, CMB measurements are undoubtedly the most powerful. However,  CMB map is a projected snapshot of the last scattering surface at $z \sim 1090$, encoding 2-dimensional information primarily imprinting the early epoch but comparatively limited for the late Universe. Operating low-redshift observations, for instance, BOSS (SDSS III)~\citep{Dawson2013}, eBOSS~\citep{Zhao:2015gua}, DES~\citep{Abbott:2015swa}, DESI~\citep{Levi:2013gra}, J-PAS~\citep{Benitez:2014ibt}, LSST~\citep{LSST:2008ijt}, {\it Euclid}~\citep{Amendola:2016saw}, etc., can supply worthy diverse information of the Universe at small redshifts. Furthermore, measuring the large-scale structures through galaxy number counts and cosmic shear can detect more signatures of the Universe evolution. Combining all different complementary probes driven by different physics, we can understand better on the properties of DM, DE and grasp the signature on the interaction between them.

Besides of conventional observations which have been widely performed, a new technique named neutral hydrogen (HI) intensity mapping (IM) is leading the trend of redshift surveys in the radio wave band. HI IM aims to map the integrated intensity of 21-cm radiation from multiple unresolved galaxies inside some redshift range~\citep{Madau:1996cs,Battye:2004re,Peterson:2006bx,Loeb:2008hg}. Since there is no need to resolve individual galaxies, the IM technique can conduct an extremely large-volume survey in a relatively short observational time, which is a great advantage over traditional optical surveys. On the other hand, HI is expected to be a good tracer of matter distribution in the post-reionization epoch with minimal bias~\citep{Padmanabhan:2014zma}, thanks to the absence of complicated reionization astrophysics. In addition to mapping a 3-dimensional Universe, the one-to-one match between the observed frequency and the source redshift provides the possibility to do a tomographic analysis in HI IM surveys.

HI IM also faces some challenges from astrophysical contamination and systematic effects. At the frequency window $\sim 1$\,GHz, HI observations are predominantly contaminated by foreground emissions, such as the Galactic synchrotron radiation and extra-galactic point sources~\citep{Battye:2012tg}. The signal level of HI IM ($T \sim 1$\,mK) is about 4 orders of magnitude lower than the foregrounds emission ($T \sim 10$\,K), thus preparing an efficient foreground removal technique to separate the signal from contamination is crucial to the HI IM probes~\citep[e.g.,][]{Wolz:2013wna,Alonso:2014dhk,Olivari:2015tka}. On the other hand, systematic effects, primarily related to the instrument, may behave similarly to the signal or even cover it at small scales (e.g. the $1/f$ noise). Besides, they will also make the foreground removal procedure harder. A careful treatment of systematic effects is hence necessary in HI IM experiments.

The Green Bank Telescope (GBT) was the first project that succeeded in HI emission detection at $z \simeq 0.8$ by cross-correlating the HI signal with the DEEP2 optical galaxy redshift survey data. This proved the feasibility of HI IM and built confidence in other experiments on the post-reionization epoch based on such technique. HI IM experiments fall into two major categories: one proposal is to survey the sky with a single dish, taken by experiments as GBT~\citep{Chang:2010jp}, BINGO~\citep{Battye:2012tg,Battye:2016qhf,2019JPhCS1269a2002W} and FAST~\citep{Nan:2011um,Bigot-Sazy:2015tot}; the other is to use an interferometer, adopted by PAPER~\citep{Parsons:2009in}, TIANLAI~\citep{Chen:2012xu}, MWA~\citep{Bowman:2012ef}, LOFAR~\citep{vanHaarlem:2013dsa}, CHIME~\citep{Bandura:2014gwa}, HIRAX~\citep{Newburgh:2016mwi}, HERA~\citep{DeBoer:2016tnn} and LWA~\citep{Eastwood:2017lex}. Another ambitious HI IM project that must be mentioned is the upcoming Square Kilometre Array (SKA), an international project for world's largest radio telescope array with an unprecedented scale, using thousands of dishes and up to a million low-frequency antennas. Its first phase SKA1, designed to cover a wide redshift range of $0 < z < 6$, is under construction and is comprised of two radio telescope arrays: SKA1-LOW working in an interferometric mode and SKA1-MID operating in a single dish mode, with a total survey area over 25000\,deg$^2$~\citep{Bacon:2018dui}.

As a novel approach to mapping large-scale structures, HI IM is of significant importance to revealing the nature of dark sectors and the late acceleration. Numerous tentative tests have been made to verify/estimate the detecability of HI IM on these aspects. For example, \citet{Carucci:2017cnn} is dedicated to distinguishing two quintessences models with dynamical EoS from the concordant $\Lambda$CDM cosmology via HI IM power spectrum, \citet{Dinda:2018uwm,Wu:2021vfz} have showcased the potential of HI IM in constraining $w$CDM, CPL model as well as other variant DE scenarios, and~\citet{CosmicVisions21cm:2018rfq} envisages probing inflation and early DE with a stage II HI IM survey. The hot topic of Axion as a promising DM candidate has also attracted HI IM to be a judge~\citep{Bauer:2020zsj}, and announced by~\citet{Carucci:2015bra}, SKA1-LOW is able to rule out a 4 keV warm DM model with 5000 observing hours at a statistical significance of $> 3\sigma$. On the other hand, how to study modified gravity with HI IM is another interesting subject~\citep{2010PhRvD..81f2001M,Heneka:2018ins,Wang:2021coq}.

Likewise, IDE is expected to leave footprints in the HI IM signal during the post-reionization epoch as a result of modifications in the expansion history and the growth of large-scale structures. Some preliminary studies in this direction was done in~\citep{Costa:2018aoy,Xiao:2018jyl}. In this work, we will carefully discuss how IDE changes the HI IM signal, i.e., the 21-cm angular power spectrum, and then we will examine the ability of two experimental setups, BINGO and SKA1-MID, in constraining the IDE models. The parameter constraints are forecasted through a Fisher matrix analysis together with the covariance matrix from {\it Planck} 2018. We also investigate the effects in the projected constraints from several choices for the frequency channel width and the contribution from RSD.

The paper is organized as follows. In Sect.~\ref{sec.IDE} we give a rapid review of the IDE model. Sect.~\ref{sec.power21cm} presents the formulae for the 21-cm angular power spectrum in the IDE scenario, the experimental parameters of BINGO and SKA1-MID, and a physical analysis of IDE's influence to each component of the 21-cm signal. In Sect.~\ref{sec.forecast} we set up the Fisher matrix and forecast the parameter constraints from BINGO and SKA1-MID, alone or together with {\it Planck} 2018. After that, an extensive analysis of how several kinds of systematics, involving the frequency channel width, the RSD contribution as well as foreground residuals and the measurement uncertainty of $\Omega_{\mathrm{HI}}(z)$, can impact the constraints is appended in the same section. Finally, we draw our conclusions in Sect.~\ref{sec.conclusions}.

Throughout this paper, unless stated otherwise, we assume as our fiducial cosmology the best-fit values from {\it Planck} 2018 TT, TE, EE + lowE + lensing: \{$\Omega_{\mathrm{b}}h^2=0.02237$, $\Omega_{\mathrm{c}}h^{2}=0.1200$, $\tau = 0.0544$, $\ln({10^{10}A_{s}})=3.044$, $n_{s}=0.9649$ and $h=0.6736$\}, besides the interacting parameters $\lambda_1 = \lambda_2 = 0$.


\section{The interacting dark energy model}\label{sec.IDE}

An interaction between DM and DE can serve as a solution to the coincidence problem. In this scenario, the energy momentum tensor of DM and DE do not evolve separately but satisfies
\begin{equation}
\nabla_{\mu} T_{\epsilon}^{\mu\nu} = Q_{\epsilon}^{\nu},
\label{eq.EM_conservation}
\end{equation}
where the subscript $\epsilon$ represents either DM 
($c$) or DE ($d$). The term $Q_{\epsilon}^{\nu}$ is the energy-momentum flux between these two components. Assuming that the dark sector cannot interact with normal matter beyond gravity, the total energy-momentum tensor of the dark sector is conserved, i.e., $Q_{c}^{\nu} + Q_{d}^{\nu} = 0$.

We will consider a Friedmann-Lemaitre-Robertson-Walker (FLRW) Universe with small perturbations on its homogeneuos and isotropic background, therefore the line element for the scalar modes is expressed as
\begin{align}\label{eq.line_element}
\mathrm{d}s^2 &= a^2 [(1+2\psi)\mathrm{d}\eta^2 - 2\partial_iB\mathrm{d}\eta\mathrm{d}x^i \nonumber \\ 
&- (1-2\phi)\delta_{ij}\mathrm{d}x^i\mathrm{d}x^j
- (\partial_i\partial_j - \frac{1}{3}\delta_{ij}\nabla^2)E\mathrm{d}x^i\mathrm{d}x^j ], 
\end{align}
where $a$ is the scalar factor and $\eta$ refers to the conformal time. $\psi$, $B$, $\phi$ and $E$ are functions of space and time describing small perturbations to the metric. In this general expression for the metric, we are not assuming any specific gauge, but in practice it should be restricted to some of them~\citep{He:2010im}. Of course, this choice will not influence the predictions of observables~\citep{Kodama:1985bj}.

If the matter component of the Universe is considered as a perfect fluid, the energy-momentum tensor can be written as
\begin{equation}\label{T}
T^{\mu\nu}(\eta,x,y,z) = (\rho + P)U^\mu U^\nu + Pg^{\mu\nu},
\end{equation}
where, for every species, the energy density reads $\rho(\eta,x,y,z) = \rho(\eta)[1 + \delta(\eta,x,y,z)]$, the pressure is $P(\eta,x,y,z) = P(\eta) + \delta P(\eta,x,y,z)$ and the four-velocity vector is $U^\mu = a^{-1}(1 - \psi, \vec{v}_{\epsilon})$, and we have separated the contributions from the background and small perturbations about it. Substituting the energy-momentum tensor Eq.~(\ref{T}) into the conservation equation Eq.~(\ref{eq.EM_conservation}), together with the line element Eq.~(\ref{eq.line_element}), we have the background  continuity equations
\begin{alignat}{2}\label{eq:intera_fen}
\dot{\rho}_{c} &+ 3\mathcal{H}\rho_{c}= a^2Q^{0}_{c}= +aQ\,,\nonumber \\
\dot{\rho}_{d} &+ 3\mathcal{H}\left(1+w\right)\rho_{d}= a^2Q^{0}_{d}= -aQ\,.
\end{alignat}
Here, $\mathcal{H}$ is the Hubble parameter with respect to the conformal time, $\mathcal{H} \equiv \dot{a}/a = a H$, and the dot denotes a derivative with respect to the conformal time. $w = P_d/\rho_d$ is the equation of state of DE and $Q$ refers to the energy transfer between the dark sectors in cosmic time coordinates. Generally there is no restriction on the formalism of $Q$, and phenomenologically we adopt a widely discussed energy transfer term dependent on the background energy densities of DM and DE, i.e., $Q=3H(\lambda_1\rho_c + \lambda_2\rho_d)$. Given  constant DE EoS, the allowed regions for the interaction and DE EoS have been well discussed in~\citep{He:2008si,Gavela:2009cy}. In Table~\ref{tab.models}, we summarize the phenomenological scenarios under investigation in this study, and the constraints listed in the last column are the stable conditions discussed in ~\citep{He:2008si,Gavela:2009cy}. For IDE Model IV, we have $\lambda \equiv \lambda_1 = \lambda_2$.
\begin{table}
	\centering
	\caption{The four different scenarios of interacting dark energy models used in our analysis together with their stable conditions.}
	\begin{tabular}{|c|c|c|c|}
		\hline
		Model & $Q$ & DE EoS & Constraints \\
		\hline
		I & $3\lambda_{2}  H\rho _{d}$ & $-1 <  w < 0$ & $\lambda_{2} < 0$ \\
		\hline
		II & $3\lambda_{2} H\rho_{d}$  & $w < -1$ & $0 < \lambda_{2} < -2w \Omega_{\mathrm{c}}$ \\
		\hline
		III & $3\lambda_{1} H\rho_{c}$  & $w < -1$ & $0 < \lambda_{1} < -w /4$ \\
		\hline
		IV & $3\lambda H \left(\rho_{d} + \rho_{c} \right) $ & $w < -1$ & $0 < \lambda < -w /4$ \\
		\hline
	\end{tabular}
	\label{tab.models}
\end{table}

Additionally the energy-momentum conservation leads the first-order perturbations in the synchronous gauge  to the system equations~\citep{Costa:2013sva}
\begin{align}
\dot{\delta}_{c} & =  -(kv_{c} + \frac{\dot{h}}{2}) + 3\mathcal{H}\lambda_{2} \frac{1}{r} \left( \delta_{d}-\delta_{c} \right)\,, \label{linear_pert_1} \\
\dot{\delta}_{d} & = -\left(1+w \right) (k v_{d} + \frac{\dot{h}}{2})+3\mathcal{H} (w - c_{e}^{2}) \delta_{d}  \nonumber \\
& +3\mathcal{H} \lambda_{1} r \left( \delta _{d}-\delta _{c} \right)  \nonumber \\
& -3\mathcal{H} \left( c_{e}^{2}-c_{a}^{2} \right) \left[ 3 \mathcal{H} \left( 1+w \right) + 3\mathcal{H} \left( \lambda_{1} r+\lambda_{2} \right) \right]\frac{v_{d}}{k}      \,, \label{linear_pert_2} \\
\dot{v}_{c} & = -\mathcal{H}v_{c} -3\mathcal{H}(\lambda_{1} + \frac{1}{r}\lambda_{2})v_{c} \,, \label{linear_pert_3} \\
\dot{v}_{d} & = -\mathcal{H}\left(1-3c_{e}^{2} \right) v_{d}+\frac{3\mathcal{H}}{1+w} \left( 1+c_{e}^{2} \right) \left(\lambda_{1} r+\lambda_{2} \right) v_{d}  \nonumber \\
& + \frac{kc_{e}^{2}\delta _{d}}{1+w}\,,
\label{linear_pert_4}
\end{align}
where $v_{\rm c}$ ($v_{\rm d}$) is the peculiar velocity of DM (DE) and $h = 6\phi$ refers to the synchronous gauge metric perturbation. Also, we have defined $r \equiv \rho_c/\rho_d$, $c_e$ is the effective sound speed and $c_a$ represents the adiabatic sound speed for the DE fluid in its rest frame. We will solve this set of differential equations together with  the IDE background evolution via a modified version of the {\bf \textsc{CAMB}} code~\citep{Lewis:1999bs}.


\section{The angular power spectra of 21-cm radiation}\label{sec.power21cm}
In this section, we first present the formula for the 21-cm brightness temperature fluctuation and its angular power spectrum in IDE scenarios. Then we give a brief introduction to intensity mapping surveys and the expected noise to be considered in our work. Finally, by comparing the total signal to noise and investigating each contribution to the brightness temperature fluctuation individually, we carefully analyze how 21-cm angular power spectra can be affected by the EoS $w$ and interacting parameters between dark sectors.

\subsection{HI Power Spectra}\label{sec.HI_theory}
The 21-cm line originates from the transition between the hyperfine levels in the ground state of neutral hydrogen atoms, whose frequency in the rest frame is $\nu = 1420$\,MHz. The brightness temperature fluctuations of the redshifted 21-cm signal is of great interest to cosmology since the distribution of HI constitutes a good tracer of the large-scale structures in our Universe. Following ~\cite{Hall:2012wd}, the observed brightness temperature at redshift $z$ reads
\begin{equation}
T_{\mathrm{b}}(z,\hat{\mathbf{n}}) = \frac{3}{32\pi} \frac{(h_{\mathrm{p}}c)^3n_{\mathrm{HI}}A_{10}}{k_{\mathrm{B}}E_{21}} \bigg\vert\frac{\mathrm{d}\zeta}{\mathrm{d}z}\bigg\vert,
\end{equation}
where $\hat{\mathbf{n}}$ is the unit vector along the line of sight, $h_{\mathrm{p}}$ is the Planck's constant, $c$ is the speed of light, $n_{\mathrm{HI}}$ is the number density of neutral hydrogen atoms at a given redshift, $A_{10}=2.869\times 10^{15}\,{\rm s}^{-1}$ is the spontaneous emission coefficient, $k_{\rm B}$ is the Boltzman's constant, $E_{21}=5.88\,\mu{\rm eV}$ is the rest frame energy of the 21-cm transition and $\zeta$ is an affine parameter of the propagation of photons. If we first exclude the contribution from perturbations, the background brightness temperature is given by
\begin{align}
\bar{T}_{\mathrm{b}}(z) &= \frac{3}{32\pi} \frac{(h_{\mathrm{p}}c)^3\bar{n}_{\mathrm{HI}}A_{10}}{k_{\mathrm{B}}E_{21}^2(1+z)H(z)} \\
&= 0.188h \Omega_{\mathrm{HI}}(z) \frac{(1+z)^2}{E(z)} \mathrm{K},
\label{eq:Tb}
\end{align}
where $\Omega_{\mathrm{HI}}$ is the fractional density of neutral hydrogen in our Universe, and $E(z) \equiv H(z) / H_0$. Here $H_0 = 100 h\,\mathrm{km~s}^{-1}\mathrm{Mpc}^{-1}$ is the Hubble parameter at present. In this work, considering the focused low redshift range ($z \lesssim 1$), we take a typical value of $\Omega_{\mathrm{HI}} = 6.2 \times 10^{-4}$~\citep{Prochaska:2008fp,Switzer:2013ewa}. However, generally $\Omega_{\mathrm{HI}}$ is a function of redshift $z$~\citep{Padmanabhan:2014zma,Bull:2014rha,Crighton:2015pza}, and also its uncertainty from the observational side will hamper precise parameter measurements by 21-cm observations~\citep{Olivari:2017bfv,Castorina:2019zho}. We thus have deliberately set aside Sect.~\ref{sec.Omhi} for a discussion on such issue, by incorporating $\Omega_{\mathrm{HI}}$ together with its redshift dependence into our Fisher analysis.

Now we focus on the perturbation of $T_{\mathrm{b}}$ to linear order. Taking~\cite{Hall:2012wd} as a guidance, we repeat the derivation therein in the conformal Newtonian gauge
\begin{equation}
\mathrm{d}s^2 = a^2(\eta)\left[(1+2\Psi)\mathrm{d}\eta^2 - (1-2\Phi)\delta_{ij}\mathrm{d}x^i\mathrm{d}x^j \right],
\label{eq:metric}
\end{equation}
by recasting Eq.~(\ref{eq.line_element}) with $\psi = \Psi$, $\phi = \Phi$, and $B=E=0$, in which $\Psi$ and $\Phi$ are the spacetime-dependent gravitational potentials. In our IDE cases, assuming $\mathbf{v}$, the bulk velocity of HI, still closely traces the total matter velocity $\mathbf{v}_\mathrm{m} \equiv \frac{\rho_{c}\mathbf{v}_{c} + \rho_{b}\mathbf{v}_{b}}{\rho_{c} + \rho_{b}}$ (the subscript $b$ here refers to baryon), the corresponding Euler equation will be written as
\begin{equation}
\dot{\mathbf{v}} + \mathcal{H} \mathbf{v} + \mathbf{\nabla} \Psi = - \mathbf{v} \frac{ aQ}{\rho_{\mathrm{m}}} \,,
\label{eq:EulerEq}
\end{equation}
where $\rho_{\mathrm{m}}$ is the energy density for the total matter and the DM-DE interaction manifests in the new term, $- \mathbf{v} \frac{ aQ}{\rho_{\mathrm{m}}}$, on the right-hand-side here. 
Then the perturbed brightness temperature $\Delta_{T_{\rm b}}$ including interaction between dark sectors  is given by
\begin{align}
\Delta_{T_{\rm b}}(z,\hat{\mathbf{n}}) & = \delta_n - \frac{1}{\mathcal{H}}\hat{\mathbf{n}} \cdot (\hat{\mathbf{n}} \cdot \mathbf{\nabla} \mathbf{v}) + \left(\frac{\rm d \ln (a^3 \bar{n}_{\text{HI}})}{\rm d \eta}-
\frac{\dot{\mathcal{H}}}{\mathcal{H}} - 2\mathcal{H} \right)\delta\eta \nonumber \\
&+ \frac{1}{\mathcal{H}}\dot{\Phi} + \Psi - \frac{1}{\mathcal{H}} \hat{\mathbf{n}} \cdot \mathbf{v} \frac{aQ}{\rho_{\mathrm{m}}},
\label{eq:perturb}
\end{align}
where $\delta_n$ is defined by $n_{\text{HI}} = \bar{n}_{\text{HI}}(1+\delta_n)$ and $\delta\eta$ is the perturbation of the conformal time $\eta$ at redshift $z$. We assume the large-scale clustering of HI gas follows the matter distribution, through some bias, and keep the conventional assumption that the bias is scale-independent. During the period of matter domination, where the comoving gauge coincides with the synchronous gauge, we can write $\delta_n$ in the Fourier space as~\citep{Hall:2012wd}
\begin{equation}
\delta_n = b_{\rm HI}\delta_\mathrm{m}^{\text{syn}} +  \left( \frac{\rm d \ln (a^3
	\bar{n}_{\text{HI}})}{\rm d \eta} - 3 \mathcal{H}\right)\frac{v_\mathrm{m}}{k} ,
\label{eq.bias}
\end{equation}
where $k$ is the Fourier space wavevector, $v_{\mathrm{m}}$ is the Newtonian-gauge total matter velocity with $\mathbf{v} = - k^{-1} \mathbf{\nabla} v_{\mathrm{m}}$, $\delta_\mathrm{m}^{\text{syn}}$ is the total matter overdensity in the synchronous gauge and $b_{\rm HI}$ is the scale-independent bias.

In order to obtain the angular power spectrum of 21-cm line at a fixed redshift, we expand $\Delta_{T_{\rm b}}$ in spherical harmonics
\begin{equation}
\Delta_{T_{\rm b}}(z,\hat{\mathbf{n}})=\sum_{\ell m}\Delta_{T_{\rm b},\ell m}(z)Y_{\ell m}(\hat{\mathbf{n}}),
\end{equation} 
and express these perturbation coefficients $\Delta_{T_{\rm b},\ell m}(z)$ with the Fourier transform of temperature fluctuations, such that
\begin{equation}
\Delta_{T_{\rm b},\ell m}(z) = 4\pi i^l \int \frac{\rm d^3 \mathbf{k}}{(2\pi)^{3/2}}
\Delta_{T_{\rm b},\ell}(\mathbf{k},z) Y_{\ell m}^*(\hat{\mathbf{k}}).
\end{equation}
Following Eq.~(\ref{eq:perturb}), the $\ell$th multipole moment of $\Delta_{T_{\rm b}}$ reads
\begin{align}
\Delta_{T_{\rm b},\ell}(\mathbf{k},z) & = \delta_n\, j_{\ell}(k \chi)+ \frac{kv}{\mathcal{H}}j_{\ell}{}''(k\chi)+\left(\frac{1}{\mathcal{H}}\dot{\Phi} + \Psi\right)j_{\ell}(k \chi) \nonumber \\
& -\left(\frac{1}{\mathcal{H}}\frac{\rm d \ln (a^3 \bar{n}_{\text{HI}})}{\rm d \eta}-
\frac{\dot{\mathcal{H}}}{\mathcal{H}^2} - 2 \right) \left[\Psi\, j_{\ell}(k \chi) \right. \nonumber \\ 
& \left. +v\, j_{\ell}{}'(k \chi) +\int_0^\chi (\dot{\Psi}+\dot{\Phi})j_{\ell}(k \chi')d\chi' \right] \nonumber \\
& + \frac{1}{\mathcal{H}} v\, j_{\ell}{}'(k \chi) \frac{aQ}{\rho_{\mathrm{m}}}, 
\label{eq:perturb2}
\end{align}
where $\chi$ is the comoving distance to redshift $z$ and $j_{\ell}(k \chi)$ is the spherical Bessel Function. A prime on $j_{\ell}(k \chi)$ refers to a derivative with respect to the argument $k \chi$. Each term in Eq.~(\ref{eq:perturb2}) has its own physical meaning: $\delta_n$, in the first term, is the density fluctuation; the second term represents the effect of RSD; within the third term, $\dot{\Phi}/\mathcal{H}$ originates from the part of the ISW effect that is not cancelled by the Euler equation, whereas $\Psi$ arises from increments in redshift from radial distances in the gas frame. The physical meaning of those in the square brackets are very similar to the CMB contributions. The first, second and third terms correspond to the contributions from the usual SW effect, Doppler shift and ISW effect, respectively, from the perturbed time of the observed redshift. They are multiplied by a factor basically characterizing the time derivative of $\bar{T}_{\mathrm{b}}$ (i.e., $d\bar{T}_{\mathrm{b}}/d\eta$). The final term $\propto aQ$, that we have uncovered in this work, is introduced by the interaction between the dark sectors.

We then integrate $\Delta_{T_{\rm b},\ell}(\mathbf{k},z)$ over a redshift (or frequency) normalized window function $W(z)$ as
\begin{equation}
\Delta_{T_{\rm b},\ell}^W(\mathbf{k}) = \int_0^\infty {\rm d} z W(z)
\Delta_{T_{\rm b},\ell}(\mathbf{k},z).
\label{eq:window}
\end{equation}
We assume a rectangular window function centered at redshift $z$ with a redshift bin width $\Delta z$ given by
\begin{equation}
W(z) = \begin{cases}
\frac{1}{\Delta z}, & z-\frac{\Delta z}{2}\leq z \leq z+\frac{\Delta z}{2}\,,\\
0, & \text{otherwise}\,.
\end{cases}
\end{equation}
Then the angular-cross spectrum of $\Delta_{T_{\rm b},\ell}$ between redshift windows can be calculated via
\begin{equation}
C_{\ell}^{WW'} = 4\pi\int {\rm d} \ln k \, {\cal P}_{\cal R}(k) \Delta_{T_{\rm b},\ell}^W(k) \Delta_{T_{\rm b},\ell}^{W'}(k).
\label{eq:Cl}
\end{equation}
${\cal P}_{\cal R}(k)$ is the dimensionless power spectrum of the
primordial curvature perturbation $\cal R$ and we define $\Delta_{T_{\rm b},\ell}^W(k) \equiv \Delta_{T_{\rm b},\ell}^W(\mathbf{k})/\cal R(\mathbf{k})$.

\subsection{Surveys and Noises}\label{sec.surveys}
Assuming a specific cosmological model and parameters, we can predict the corresponding 21-cm angular power spectrum using the formulae presented in the previous subsection. Then, observations from HI IM experiments will lay constraints in our cosmological models or even rule it out. In this work, we will consider two IM facilities: BINGO and SKA.

BINGO will be a single-dish IM telescope located in Brazil, working in the frequency range from 980 to 1260\,MHz ($z = 0.13 - 0.45$). The frequency channel width, also called channel bandwidth, is obtained by equally dividing the frequency range into $N_{\rm bin}$ pieces. Since our model for the HI power spectra is only valid in the linear region, we assume a fiducial channel bandwidth of 8.75\,MHz, which is wide enough to avoid appreciable nonlinear influences. Nevertheless, we refer to Sect.~\ref{sec.analyses} \& \ref{sec.extension} for a discussion on the effect of different channel bandwidth values. BINGO will cover a sky area of about 3000 deg$^2$ excluding the Galactic plane in one year operation. It will have an illuminated aperture $D_{\rm dish} = 34\,\textrm{m}$ with full-width half-maximum (FWHM) beam resolution given by
\begin{equation}
\theta_{\rm FWHM} = 1.2\frac{\lambda_{\rm med}}{D_{\rm dish}}\,,
\label{eq.FWHM}
\end{equation}
where $\lambda_{\rm med}=c/\nu_{\rm med}$ is the wavelength at the medium frequency $\nu_{\rm med}$ of the entire range\footnote{The FWHM beam resolution, $\theta_{\rm FWHM}$, is a constant contingent on survey configurations. The beam effect, however, is indeed redshift dependent and will significantly reduce the signal-to-noise ratio at high $\ell$ range. We clarify our beam correction later in Eq.~(\ref{bll0}) - (\ref{eq.B_l}).}. In this work we fix $\theta_{\rm FWHM}$ to be 40\,arcmin for BINGO, which corresponds to the angular resolution of such instrument at 1\,GHz~\citep{Battye:2012tg}. We assume the telescope is equipped with 50 feed horns and receivers with dual polarization. See Table~\ref{tab.survey_parameter} for BINGO configurations in detail.
\begin{table*}
	\caption{\label{tab.survey_parameter} Survey parameters for BINGO and SKA1-MID.}
	\begin{tabular}{p{140pt}ccc}
		\toprule
		& BINGO & SKA1-MID Band\,1 & SKA1-MID Band\,2\\
		\hline
		Frequency range (MHz)                        & [980, 1260] & [350, 1050] & [950, 1405] \\
		Redshift range                               & [0.13, 0.45]& [0.35, 3.06] & [0.01, 0.49] \\
		System temperature $T_{\mathrm{sys}}$ (K)    & 70          & Eq.~(\ref{eq.T_sys}) & 15 \\
		Number of dishes $n_{\mathrm{d}}$            & 1           & 197          & 197 \\
		Number of beams $n_{\mathrm{beam}}$ (dual pol.) & 50$\times$2 & 1$\times$2   & 1$\times$2 \\
		Illuminated aperture $D_{\mathrm{dish}}$ (m) & 34          & 15           & 15 \\
		Beam resolution $\theta_{\rm FWHM}$ (arcmin) & 40          & 117.9        & 70 \\
		Sky coverage $\Omega_{\rm sur}$ (deg$^2$)    & 3000        & 20000        & 5000 \\
		Observation time $t_{\mathrm{obs}}$ (yr)     & 1           & 1.14         & 1.14 \\
		Channel bandwidth $\delta \nu$ (MHz)                 & 8.75        & 8.75         & 8.75 \\
		Number of channels $N_{\rm bin}$             & 32          & 80           & 52 \\
		\lasthline
	\end{tabular}
\end{table*}

SKA will be the largest radio telescope in the world with a collecting area over a square kilometre. The project is delivered in two phases, with SKA1 under construction now and SKA2 to be configured. SKA1 is made up of two telescope arrays, SKA1-MID and SKA1-LOW. SKA1-MID, sited in South Africa, will work in the frequency range from 350-1750\,MHz, and SKA1-LOW, located in western Australia, will observe between 50-350\,MHz. For a direct comparison with BINGO, we focus on SKA1-MID due to its target redshift range of $z \lesssim 3$.

SKA1-MID is a dish array comprised of 64$\times$13.5\,m MeerKAT dishes and 133$\times$15\,m SKA1 dishes \citep{Bacon:2018dui}. Following~\cite{Chen:2019jms}, we assume each of those movable 197 dishes is of 15\,m in diameter with a dual polarization receiver. The operation of SKA1-MID will be divided into two bands, Band\,1 from 350-1050\,MHz ($0.35 < z < 3.06$) and Band\,2 from 950-1750\,MHz ($0 < z < 0.49$). We consider both bands operating in the single-dish (auto-correlation) mode due to its superiority over the interferometric (cross-correlating the output from the dishes) mode in measuring HI signals at BAO scales as well as a higher sensitivity to HI surface brightness temperature~\citep{Bull:2014rha,Santos:2015gra}. In order to make a comparative analysis with BINGO, we assume the same fiducial channel bandwidth of 8.75\,MHz for both SKA1-MID bands and, as a compromise, we cut off the up-limit frequency of Band\,2 at 1405\,Mhz. The FWHM beam resolution calculated by Eq.~(\ref{eq.FWHM}) gives $\theta_{\rm FWHM} = 1.96^{\circ}$ for Band\,1 at $\nu_{\rm med} = 700$\,MHz and $\theta_{\rm FWHM} = 1.17^{\circ}$ for Band\,2 at  $\nu_{\rm med} = 1177.5$\,MHz, respectively \footnote{Note that our $\theta_{\rm FWHM}$ values here are not the same as those in \cite{Chen:2019jms}.}.

The system temperature of SKA1-MID is calculated via \citep{Bacon:2018dui}
\begin{equation}
T_{\rm sys} = T_{\rm rx} + T_{\rm spl} + T_{\rm CMB} + T_{\rm gal},
\label{eq.T_sys}
\end{equation}
where $T_{\rm CMB}\approx2.73$\,K is the CMB temperature and $T_{\rm spl}\approx3$\,K designates the ``spill-over'' contribution. $T_{\rm gal}$ represents the part from our Galaxy itself as a function of frequency given by
\begin{equation}
T_{\rm gal} = 25\mathrm{\,K}(408\,\mathrm{MHz}/\nu)^{2.75}\,,
\label{tgal}
\end{equation}
and $T_{\rm rx}$ is the receiver noise temperature, which can be described by
\begin{equation}
T_{\rm rx} = 15\,\mathrm{K}+ 30\,\mathrm{K}\left(\frac{\nu}{\mathrm{GHz}}-0.75\right)^2\,
\label{trxb1}
\end{equation}
for Band\,1, but fixed at 7.5\,K for Band\,2. Given that Band\,2 will operate within a high frequency range where the contribution from the galactic part is subdominant, we assume a frequency-independent value of $T_{\rm gal} \approx 1.3$\,K and, then, the system temperature of Band\,2 can be further simplified as a constant value of $T_{\rm sys} = 15$\,K.
The survey parameters for the two bands of SKA1-MID are summarized in Table~\ref{tab.survey_parameter}, together with the total observational time and sky coverage according to~\cite{Bacon:2018dui}.

In practice, together with the cosmological signal, there will be several contaminants. They mainly come from foregrounds, such as galactic synchrotron emission and extragalactic point sources. The amplitudes of those contaminants are much higher than the 21-cm signal and, thus, some foreground removal technique to subtract them is necessary~\citep{Bigot-Sazy:2015jaa,Olivari:2015tka,Zhang:2015mga}. In terms of current techniques, however, it is still very challenging to reach an optimal scenario that foreground contamination could be fully deducted, and hence the foreground residual will degrade and even bias the cosmological constraints from 21-cm intensity mapping. Therefore, in this work, we estimate the degradation by foreground residual after doing the forecast in an optimistic case where all foreground contamination have been removed. In addition, we will consider noises from two aspects: the shot noise in the auto-spectra and an instrumental noise (i.e, the thermal noise), and assume they are uncorrelated at different redshift bins.

The shot noise arises in the measured auto-spectra due to the fact that the HI sources are discrete. Given an angular density of sources $\bar{N}(z)$, the shot noise can be calculated by $C_{\ell}^{\mathrm{shot}} = \bar{T}_{\rm b}^2(z) / \bar{N}(z)$~\citep{Hall:2012wd}, where
\begin{equation}
\bar{N}(z) = \frac{n_0 c}{H_0} \int \frac{\chi^2(z)}{E(z)} \mathrm{d} z.
\end{equation}
Following~\cite{Masui:2009cj}, we will assume a comoving number density of sources $n_0 = 0.03 h^3 \, \mathrm{Mpc}^{-3}$.

The thermal noise originates from the voltages generated by thermal agitations in the resistive components of the receiver. It defines the fundamental sensitivity of the instrument, which can be calculated via the radiometer equation~\citep{Wilson2009}
\begin{equation}
\sigma_{\rm T} = \frac{T_{\rm sys}}{\sqrt{t_{\rm pix}\delta\nu}}\,,
\label{eq:radiometer}
\end{equation}
where $T_{\rm sys}$ is the total system temperature and $\delta\nu$ is the frequency channel width (i.e., the channel bandwidth). $t_{\rm pix}$ is the integration time per pixel given by
\begin{equation}
t_{\rm pix}=t_{\rm obs}\frac{n_{\rm beam}n_{\rm d}\Omega_{\rm pix}}{\Omega_{\rm sur}},
\end{equation}
where $n_{\rm beam}$ is the number of beams, $n_{\rm d}$ denotes the number of dishes, $\Omega_{\rm sur}$ corresponds to the survey coverage and $\Omega_{\rm pix}$ is the pixel area which is proportional to the square of the beam resolution $\theta_{\rm FWHM}$ (i.e., $\Omega_{\rm pix} \propto \theta^2_{\rm FWHM}$). Then, the angular power spectrum of thermal noise reads
\begin{equation}
N_{\ell}(z_i,z_j)= \left(\frac{4 \pi}{N_{\rm pix}} \right)\sigma_{\rm T,i}\sigma_{\rm T,j}\,,
\label{noise}
\end{equation}
with $N_{\rm pix}$ representing the number of pixels in the map and $\sigma_{\rm T,i}$, given by Eq.~(\ref{eq:radiometer}), is the thermal noise for the frequency channel centered at redshift $z_i$. Here we will only consider the auto correlations of thermal noise (i.e., $N_{\ell}(z_i,z_j) = 0$ if $z_i \neq z_j$).

We also need to take into account the resolution of our experiment. Therefore, at each frequency channel $\nu_i$, we apply a beam correction~\citep{Chen:2019jms}
\begin{equation} 
b_{\ell}(z_i) = \exp\left[-\frac{1}{2}\ell^{2} \sigma^2_{b, i}\right]\,,
\label{bll0}
\end{equation}
where $\sigma_{b, i}=\theta_{\rm B}(z_i)/\sqrt{8\ln 2}$~\citep[e.g.,][]{Bull:2014rha} and
\begin{equation}
\theta_{\rm B}(z_i)=\theta_{\rm FWHM}(\nu_{\rm med})\frac{\nu_{\rm med}}{\nu_i}\,.
\end{equation}
This beam correction reduces the signal by a factor of $b_{\ell}^2$, but equivalently we can regard it as an increase in the noise by a factor of
\begin{equation} 
B_{\ell}(z_i, z_j) = \exp\left[\ell^{2} \sigma_{b,i}\sigma_{b,j}\right]\,.
\label{eq.B_l}
\end{equation}
We employ $B_{\ell}(z_i, z_j)$ only to the thermal noise, since the shot noise in reality is a part of the signal itself.

\subsection{Foreground Residuals}\label{sec.foreground}
Analogue to CMB analyses, the 21-cm observations are primarily challenged by foreground contamination, which can be broadly divided into two categories: galactic and extra-galactic. The galactic foregrounds, regarding to its prominent emission sources resided in the galactic plane, are expected to be anisotropic, encompassing synchrotron\footnote{Here we assume galactic synchrotron emission to be unpolarized and thus no polarization leakage into intensity signals. In spite of its frequency dependence, the polarised foregrounds also vary with instrumental designs and survey strategies. Now the prevailing approach to alleviate such trouble is from both sophisticated modelling and hardware-level control. One may refer to~\citet{Alonso:2014dhk} for a detailed discussion.} and free-free emissions, whereas the extra-galactic foregrounds, incorporating free-free and point sources emissions, are reasonably assumed to be isotropic according to the cosmological principle. In fact, the 21-cm signal is at least $\sim 10^4$ of magnitude covered by foregrounds, and precise foreground modelling as well as effective removal algorithm are thus crucial to fully extract cosmological information from 21-cm observations. Fortunately, the feature of approximately coherent frequency dependence renders these foregrounds to be removed through component separation methods, although some problems still remain: 1) foreground residuals, which still produce uncertainties on parameter measurements, and 2) part of the 21-cm power is subtracted together with foregrounds on very large radial modes, causing loss of cosmological information, which is attributed to the difficulty to separate cosmological modes from foregrounds on those scales~\citep{Bull:2014rha}.

A rigorous approach to foreground modelling is intimately related to HI signal reconstruction. A standard procedure is first generating simulated multi-frequency maps, a superposition of mock foreground maps to the theoretical HI maps, and then to apply a foreground subtraction algorithm on the synthetic map for component separation (see BINGO series papers~\citet{Liccardo:2021kbu,Fornazier:2021ini} as a concrete example). However, this study is not dedicated to HI map reconstruction and, instead, we are concerned on how foreground residuals will adversely impact our parameter determinations in IDE scenarios. Therefore, 
following~\citet{Bull:2014rha} and~\citet{Alonso:2014dhk}, we assume Gaussian-distributed foregrounds and characterize the efficiency of foreground removal by a coefficient, $\epsilon_{\rm FG}$, such that the total angular power spectra of foreground residuals are given by
\begin{equation}\label{eq:cl_fg}
C_l^{\rm FG}(\nu_1,\nu_2)= \epsilon^2_{\rm FG} \sum_{X}A_{X}\,\left(\frac{l_{\rm ref}}{l}\right)^{\beta_{X}}\,
\left(\frac{\nu_{\rm ref}^2}{\nu_1\,\nu_2}\right)^{\alpha_{X}}
\exp\left(-\frac{\log^2(\nu_1/\nu_2)}{2\,\xi_{X}^2}\right).
\end{equation}
We consider a summation over four species of foregrounds and the corresponding model parameters are listed in Table~\ref{tab.FG}. The exponential term accounts for the cross-correlation between two frequencies, where $\xi_{X}$ refers to the correlation length for each foreground species, which manifests how smooth a foreground emission is in frequency and, consequently, the difficulty in subtracting it. 
A successful HI intensity mapping survey will require $\epsilon_{\rm FG} \lesssim 10^{-5}$ for clear 21-cm signals.
\begin{table}
\begin{center}
\begin{tabular}{|c|c|c|c|c|}
\hline
Foreground              & A (mK$^2$) & $\beta$ & $\alpha$ & $\xi$ \\
\hline
Galactic synchrotron    & 700        & 2.4 & 2.80 & 4.0 \\
Point sources           &  57        & 1.1 & 2.07 & 1.0 \\
Galactic free-free      & 0.088      & 3.0 & 2.15 & 35  \\
Extragalactic free-free & 0.014      & 1.0 & 2.10 & 35  \\
\hline
\end{tabular}
\end{center}
\caption{Parameters of foreground models taken from \citet{Santos:2004ju} with the condition $l_{\rm ref}=1000$ and $\nu_{\rm ref}=130\,{\rm MHz}$.}
\label{tab.FG}
\end{table}

\subsection{Physical Analyses} \label{sec.analyses}
\begin{figure*}
	\subfloat[]{\label{fig.cl_21}
		\includegraphics[width=0.47\textwidth]{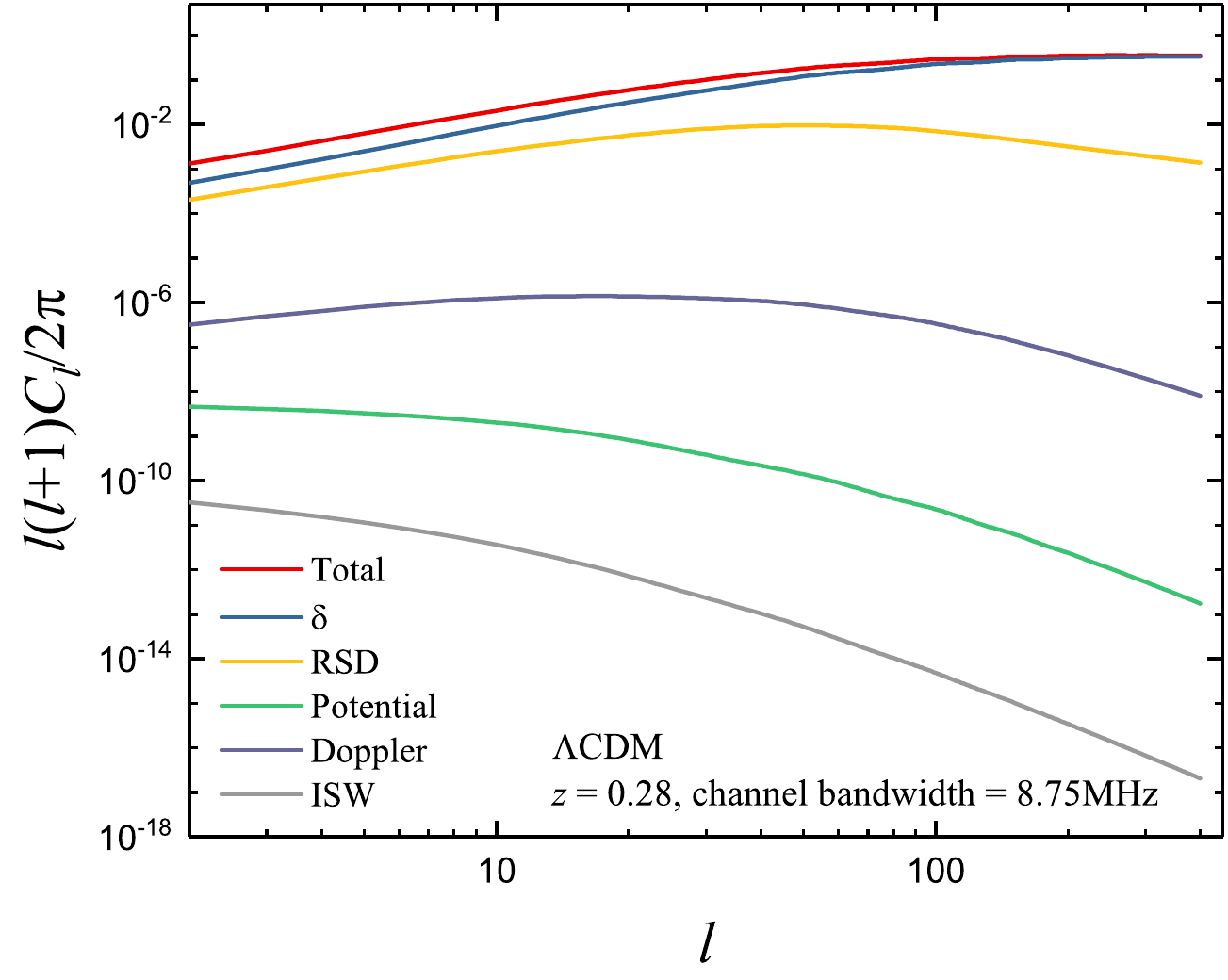}
	}
	\subfloat[]{\label{fig.BINGO_noise}
		\includegraphics[width=0.47\textwidth]{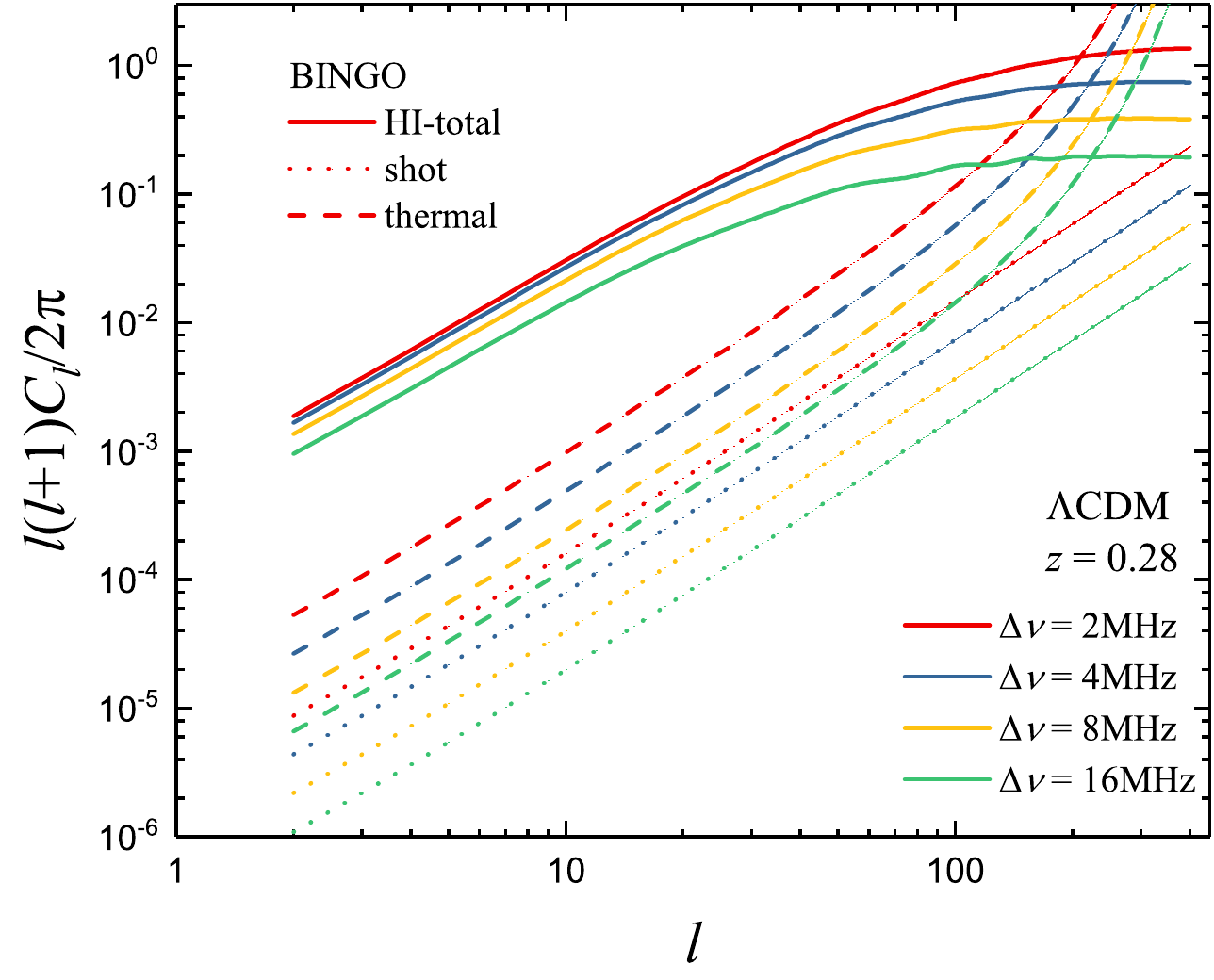}
	}
	
	\subfloat[]{\label{fig.SB1_noise}
		\includegraphics[width=0.47\textwidth]{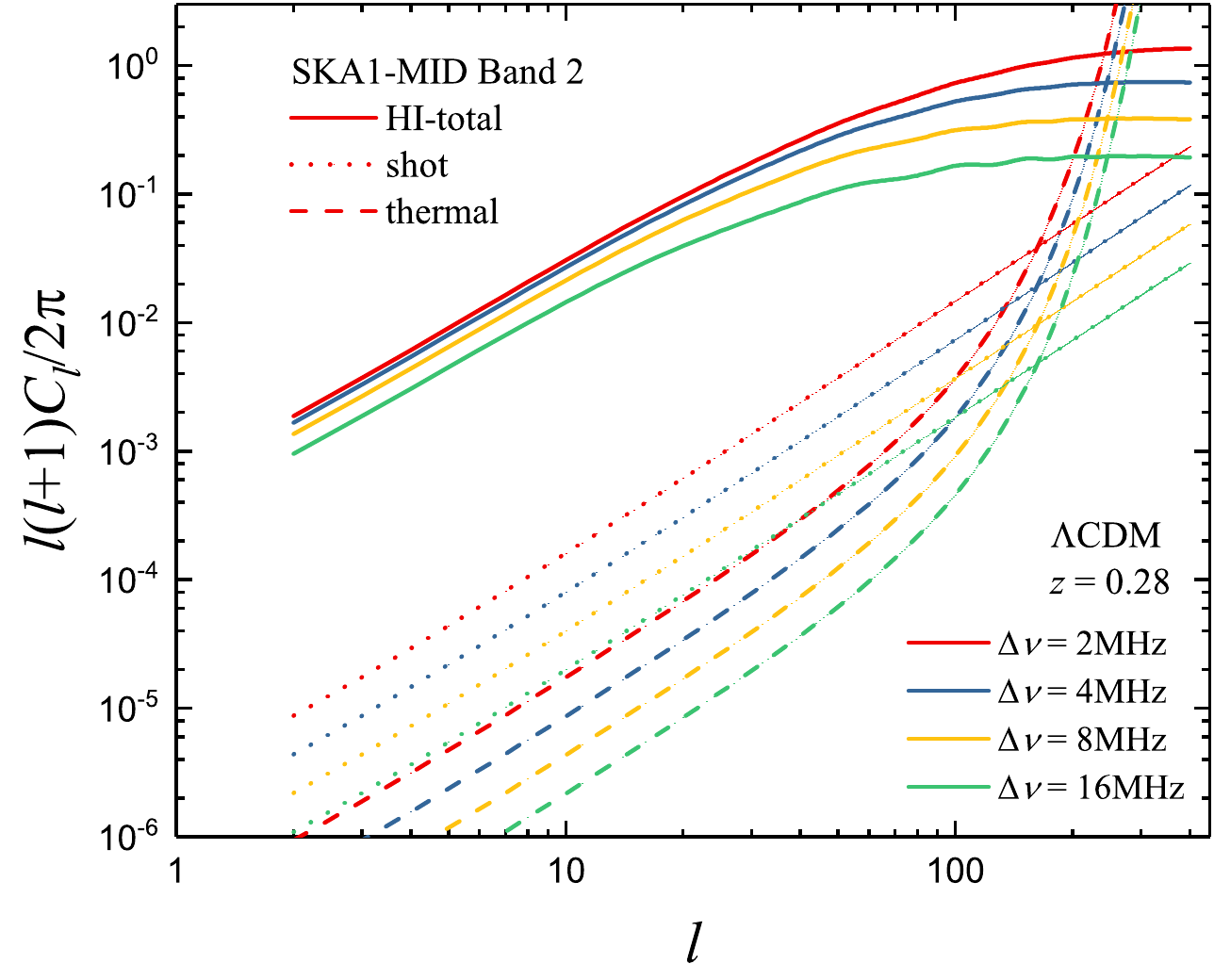}
	}
	\subfloat[]{\label{fig.SB2_noise}
		\includegraphics[width=0.47\textwidth]{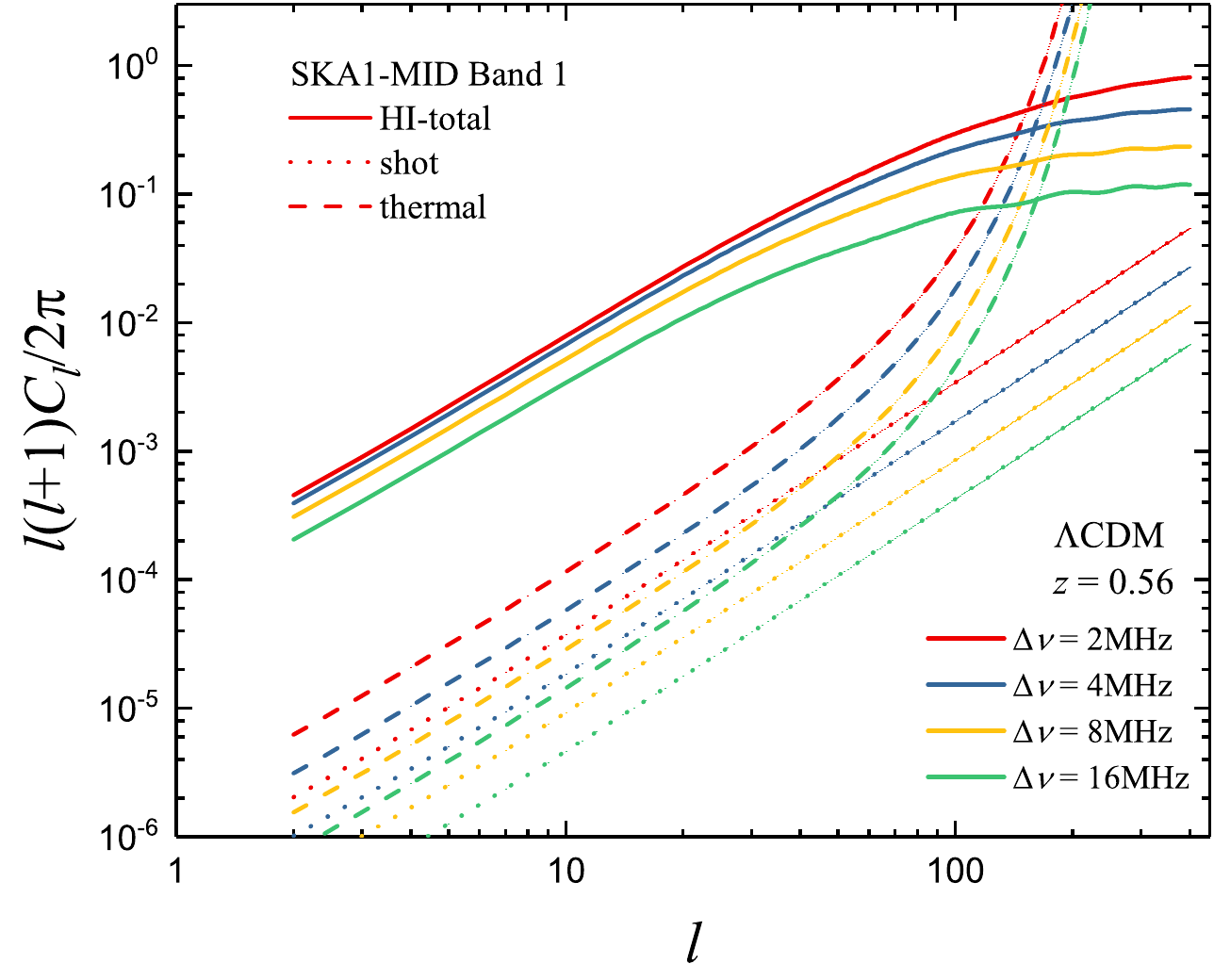}
	}
	\caption{(a) The auto power-spectra of each term in Eq.~\ref{eq:perturb2} with a channel bandwidth $\Delta\nu = 8.75\,\mathrm{MHz}$ centred at $z=0.28$ for the $\Lambda$CDM model, juxtaposed with the total signal in red that has additionally comprised the cross-correlations of these terms. Also to notice that we have designated 'Potential' to the combination of the term, $\dot{\Phi}/\mathcal{H} + \Psi$, and the SW effect. (b)$\sim$(d) Total HI signal, shot noise and thermal noise with respect to different channel bandwidths for BINGO and the two SKA1-MID bands, respectively. The fiducial cosmological parameters are set to be the same as panel (a).}
\end{figure*}

Before analysing the influence to 21-cm signals from DM-DE interactions, we first turn to the $\Lambda$CDM model for some hints. Fig.~\ref{fig.cl_21} shows the auto-spectra for each term in Eq.~(\ref{eq:perturb2}) with a channel bandwidth of $8.75\,\mathrm{MHz}$ at $z=0.28$, parameterized by the {\it Planck} 2018 best-fit values listed in Sect.~\ref{sec.intro}. The density fluctuation and RSD term are the two leading contributions across the whole multipole range we consider here. Especially at $\ell \sim 400$ the total signal is greatly dominated by the $\delta_n$ term. Moreover, besides of the auto-spectra, the total 21-cm signal also encompasses the contributions from cross-correlations of each two terms in Eq.~(\ref{eq:perturb2}), which is in line with the case of real observation. Regarding to their magnitudes, for example, the cross-spectrum of HI overdensity $\times$ RSD should be somewhere between the auto-spectra of $\delta_n$ and RSD terms, although no clear illustration in Fig.~\ref{fig.cl_21}.

In Fig.~\ref{fig.BINGO_noise} we display the total signal, shot noise and thermal noise\footnote{Here we consider the dimensionless shot and thermal noises dividing them by the average brightness temperature in order to be consistent with the 21-cm signal.} with respect to different channel bandwidths with the experimental parameters for BINGO taken from Table~\ref{tab.survey_parameter}. Basically the shot noise is about one order smaller than the thermal noise when $\ell \lesssim 100$, beyond which the thermal noise obtains an additional quick increase attributed to the beam correction Eq.~(\ref{eq.B_l}). The cross-over point of the total signal and the thermal noise at $\ell \sim 200$ indicates that we can ignore nonlinear effects at high $\ell$s. Nevertheless, very narrow channel bandwidths should be avoided to not introduce nonlinear effects along the radial direction, otherwise one should generalize the calculation of perturbations to higher orders, especially for the RSD part. By widening the channel bandwidth, both signal and noise levels decline simultaneously, while the cross-over point does not have a considerable shift. Furthermore, we see that the signal level decreases more on small scales where the signature of BAO wiggles is more prominent. Likewise, the signal and noise levels for the two SKA1 bands are illustrated in Fig.~\ref{fig.SB1_noise} and~\ref{fig.SB2_noise}, respectively, exhibiting very similar features as for BINGO. It is worthy noting, however, that SKA1-MID Band\,2 has a superior thermal noise configuration with the same channel bandwidth as BINGO.

\begin{figure*}
	\subfloat[]{\label{fig.cl_wCDM}
		\includegraphics[width=0.47\textwidth]{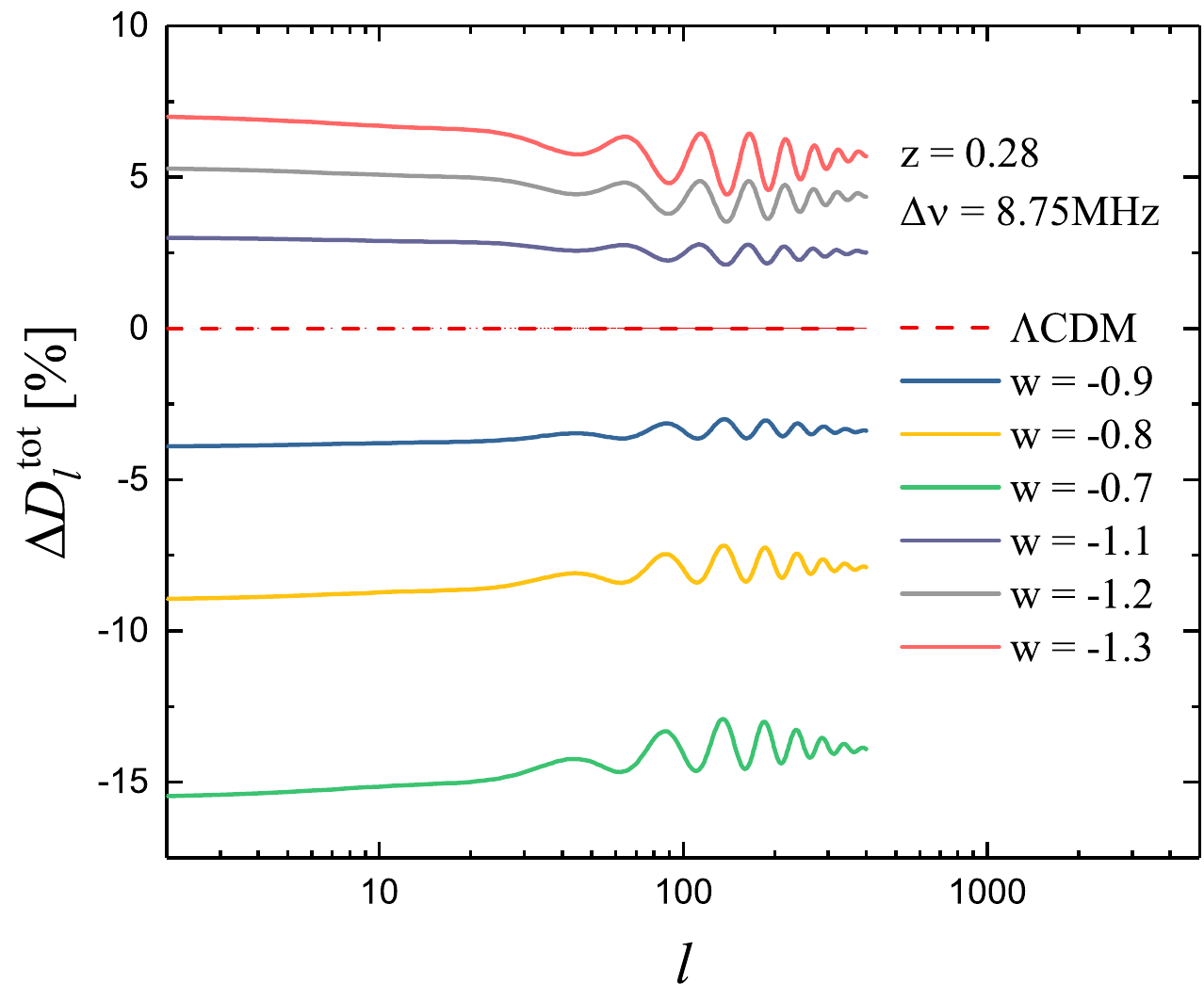}
	}
	\subfloat[]{\label{fig.w_rhode}
		\includegraphics[width=0.47\textwidth]{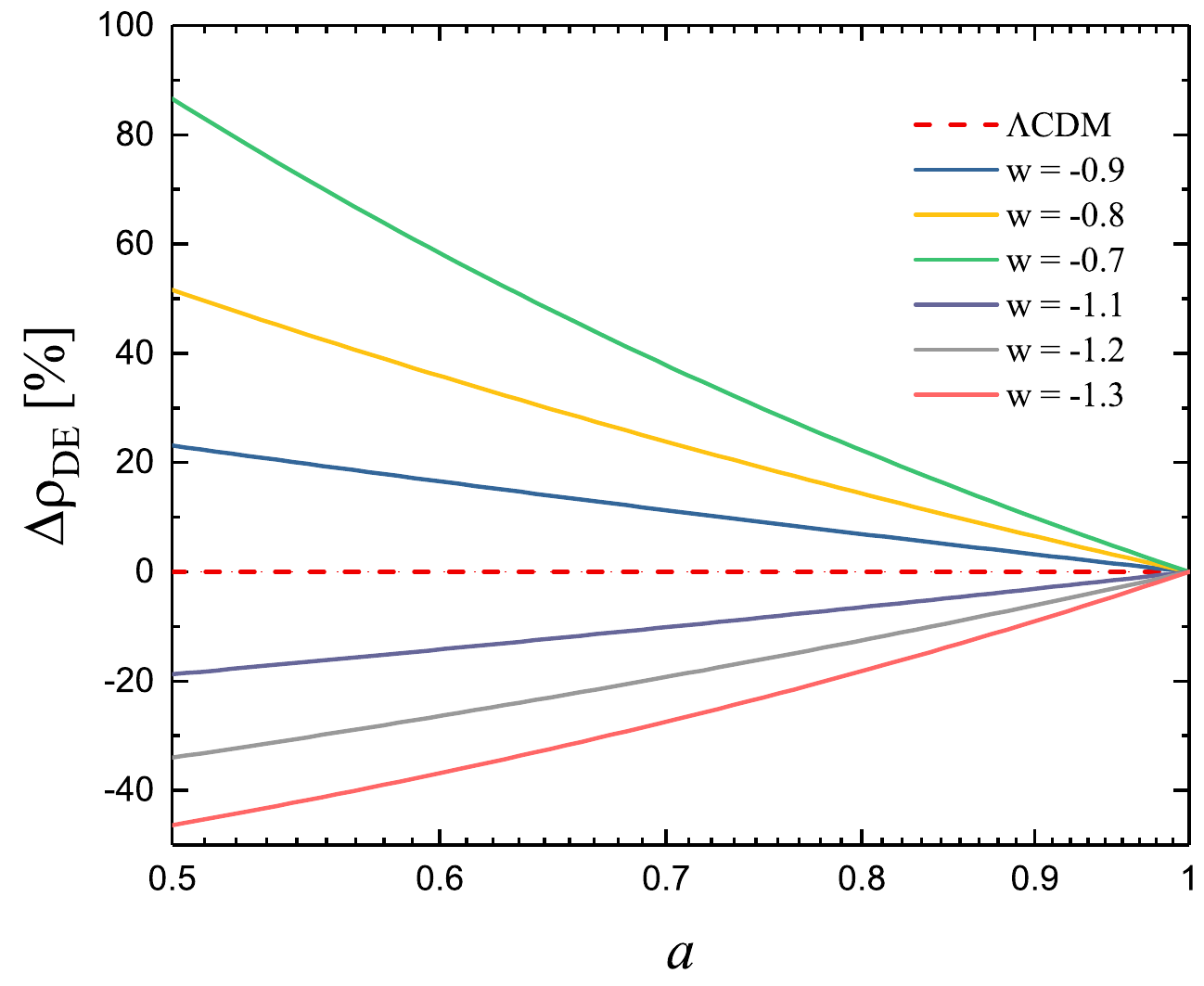}
	}
	\caption{(a) Total 21-cm signal deviation in the $w$CDM model from the $\Lambda$CDM model. A smaller $w$ enhances the signal level. (b) Time evolution of the fractional DE density with respect to the $\Lambda$CDM model. A smaller $w$ corresponds to less $\rho_d$ in the past, which contributes matter inhomogeneities to grow.}
\end{figure*}

Hereafter, we define $D_{\ell} \equiv \ell (\ell +1) C_{\ell}/2\pi$ and $\Delta D_{\ell}^{i} \equiv (D_{\ell}^{i} - D_{\ell,\Lambda\mathrm{CDM}}^{i})/D_{\ell,\Lambda\mathrm{CDM}}^{i}$ to be the fractional angular power spectrum of the $i$th contribution ($i$ corresponds to each term in Fig.~\ref{fig.cl_21} as well as for the IDE extra term in Eq.~(\ref{eq:perturb2})) with respect to the $\Lambda$CDM prediction. If we go further to the scenario of $w$CDM model, Fig.~\ref{fig.cl_wCDM} shows that a smaller value of $w$ leads to a larger 21-cm signal, keeping all other parameters and settings as in Fig.~\ref{fig.cl_21}. These discrepancies, however, are not symmetrical about $w = -1$ due to the time evolution of $\rho_d \propto a^{-3(1+w)}$. In addition, a larger deviation from $w = -1$ leads to prominent BAO wiggles but subtle phase shift in the multipole space. In light of $\rho_{d}$ deviations from the $\Lambda$CDM model shown in Fig.~\ref{fig.w_rhode}, we infer that more DE in the past is not conducive to matter clustering and thus suppress the 21-cm signal. This is an intuitive explanation, yet to some extent, it can shed light on how IDE affects 21-cm signals. Therefore, as a caveat, we must keep an eye on the degeneracy between $w$ and DM-DE interactions in following discussions.

Now we turn to IDE models. For simplicity, in this section, we mainly focus on Model I \& IV and present some of their qualitative results in contrast to the $\Lambda$CDM model. 
For Models II (III) we simply describe its feature in 21-cm signals in light of Model I (IV) but omit full illustrations due to their similarity. The fiducial cosmological parameters are kept the same (see Sect.~\ref{sec.intro}), except the DE EoS and interaction parameter which we slightly change to $w = -0.999, \lambda_1 = -0.001$ for Model I and $w = -1.001, \lambda = 0.001$ for Model IV, in order to properly appreciate the effect of those parameters under the IDE models.

\begin{figure*}
	\subfloat[]{\label{fig.IDE1_w-delta}
		\includegraphics[width=0.31\textwidth]{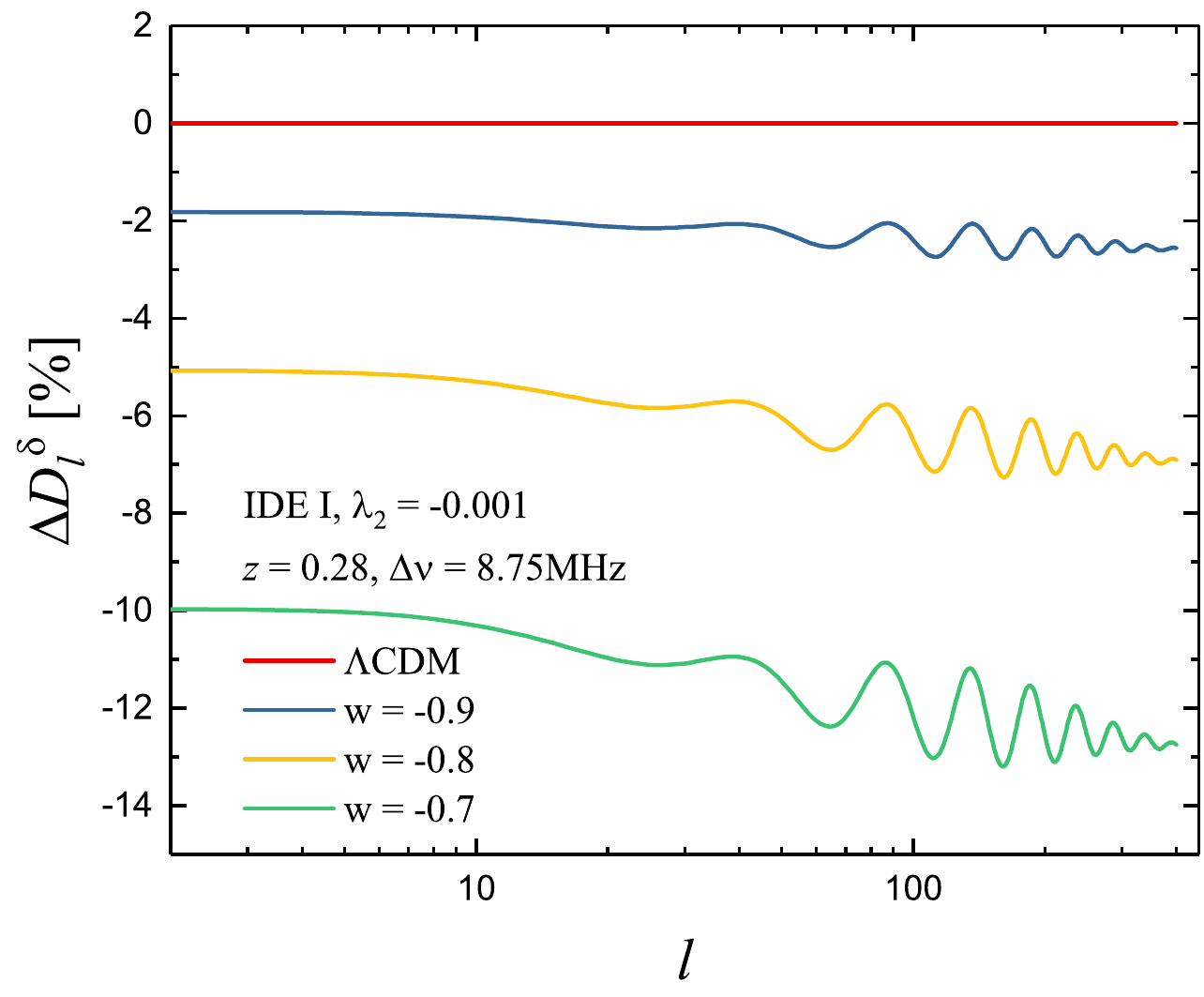}
	}
	\subfloat[]{\label{fig.IDE1_w-RSD}
		\includegraphics[width=0.31\textwidth]{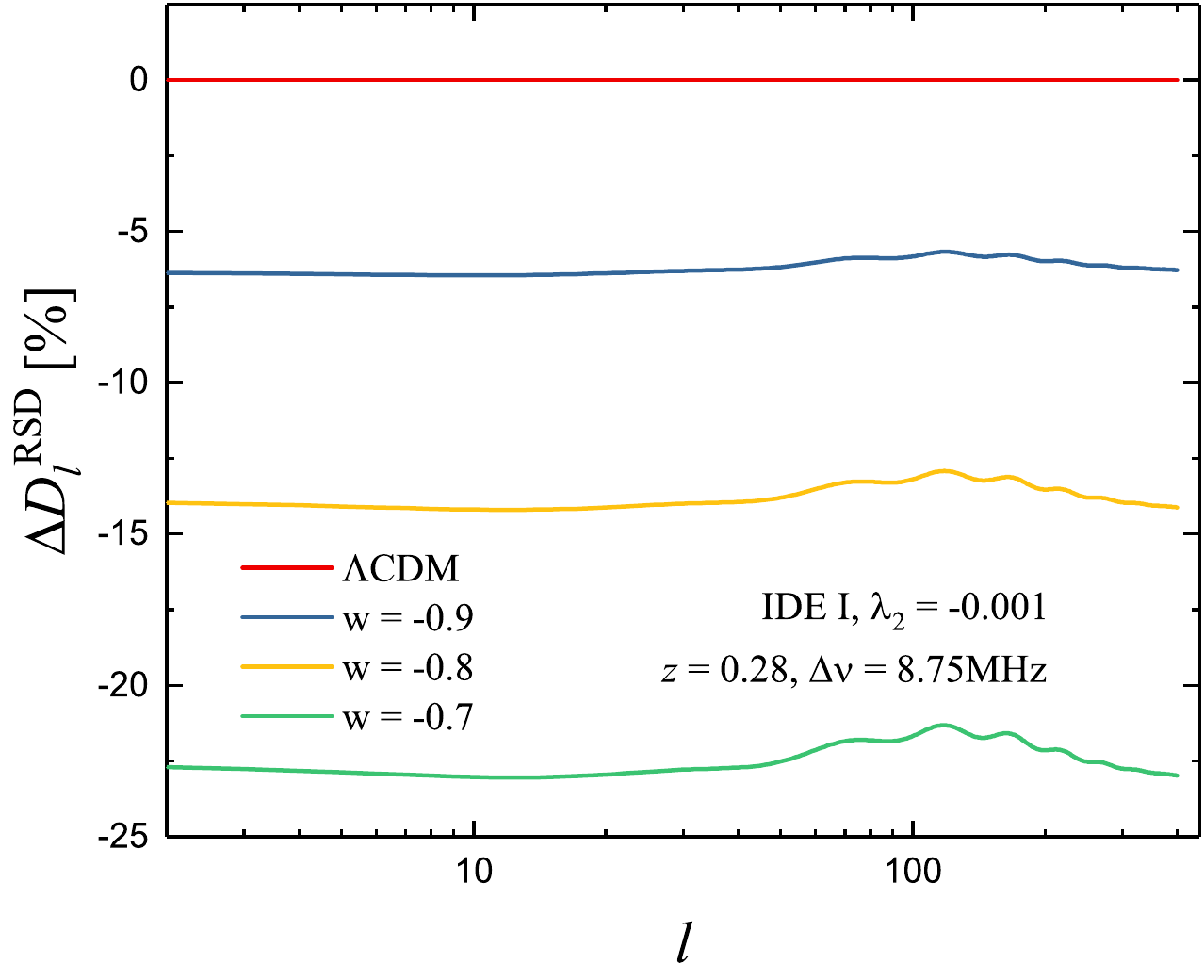}
	}
	\subfloat[]{\label{fig.IDE1_w-Pot}
		\includegraphics[width=0.31\textwidth]{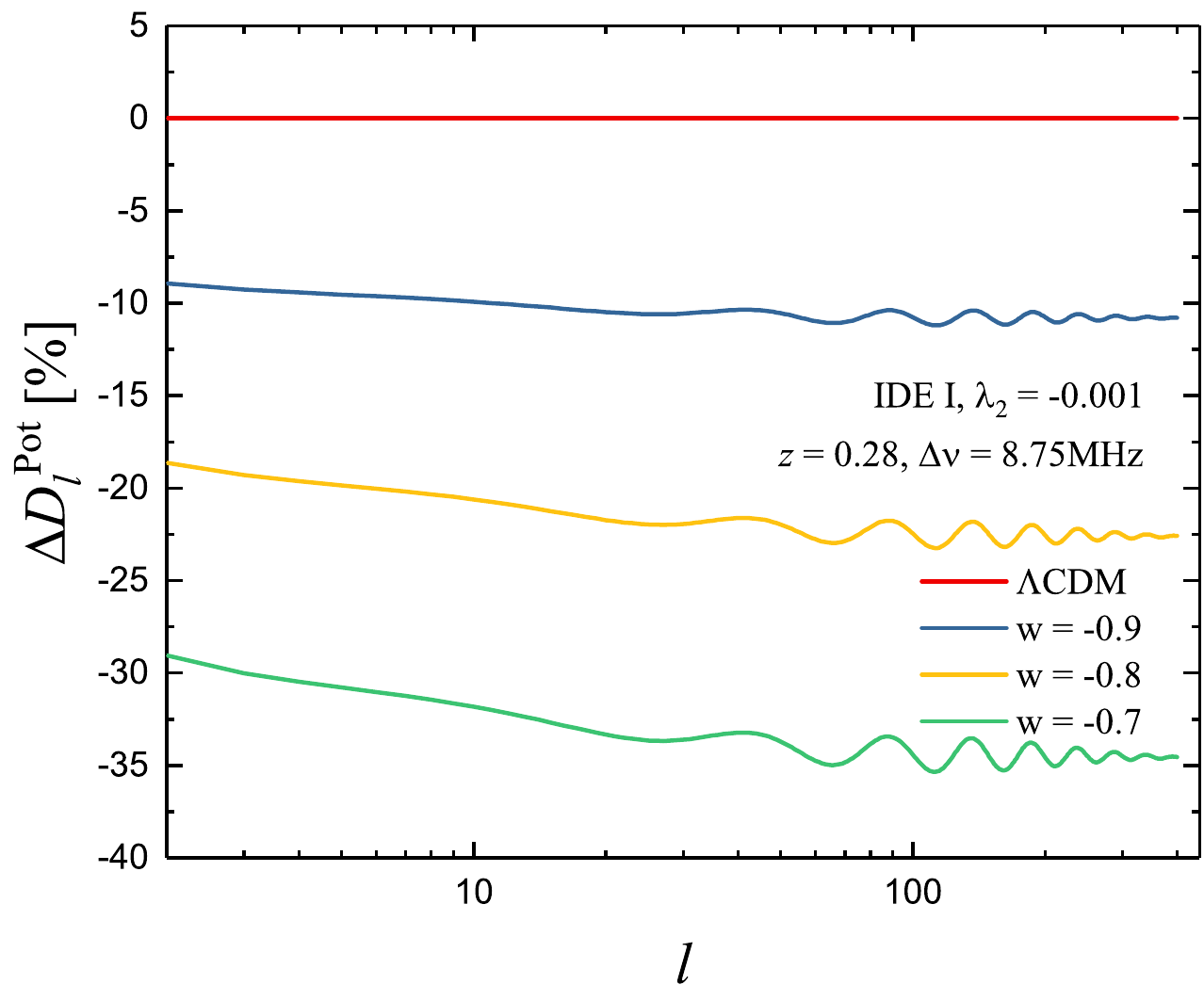}
	}
	
	\subfloat[]{\label{fig.IDE1_w-v}
		\includegraphics[width=0.31\textwidth]{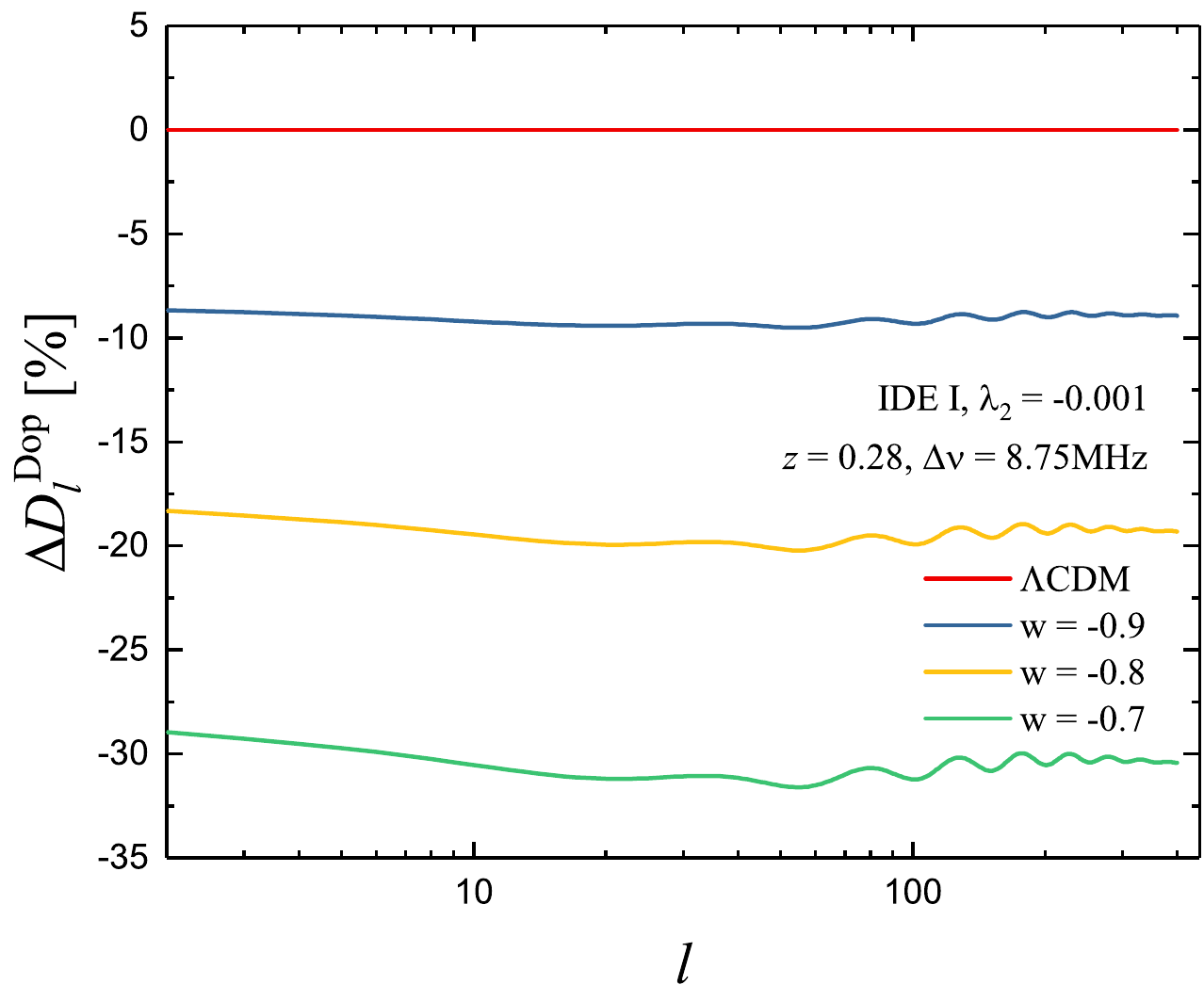}
	}
	\subfloat[]{\label{fig.IDE1_w-ISW}
		\includegraphics[width=0.31\textwidth]{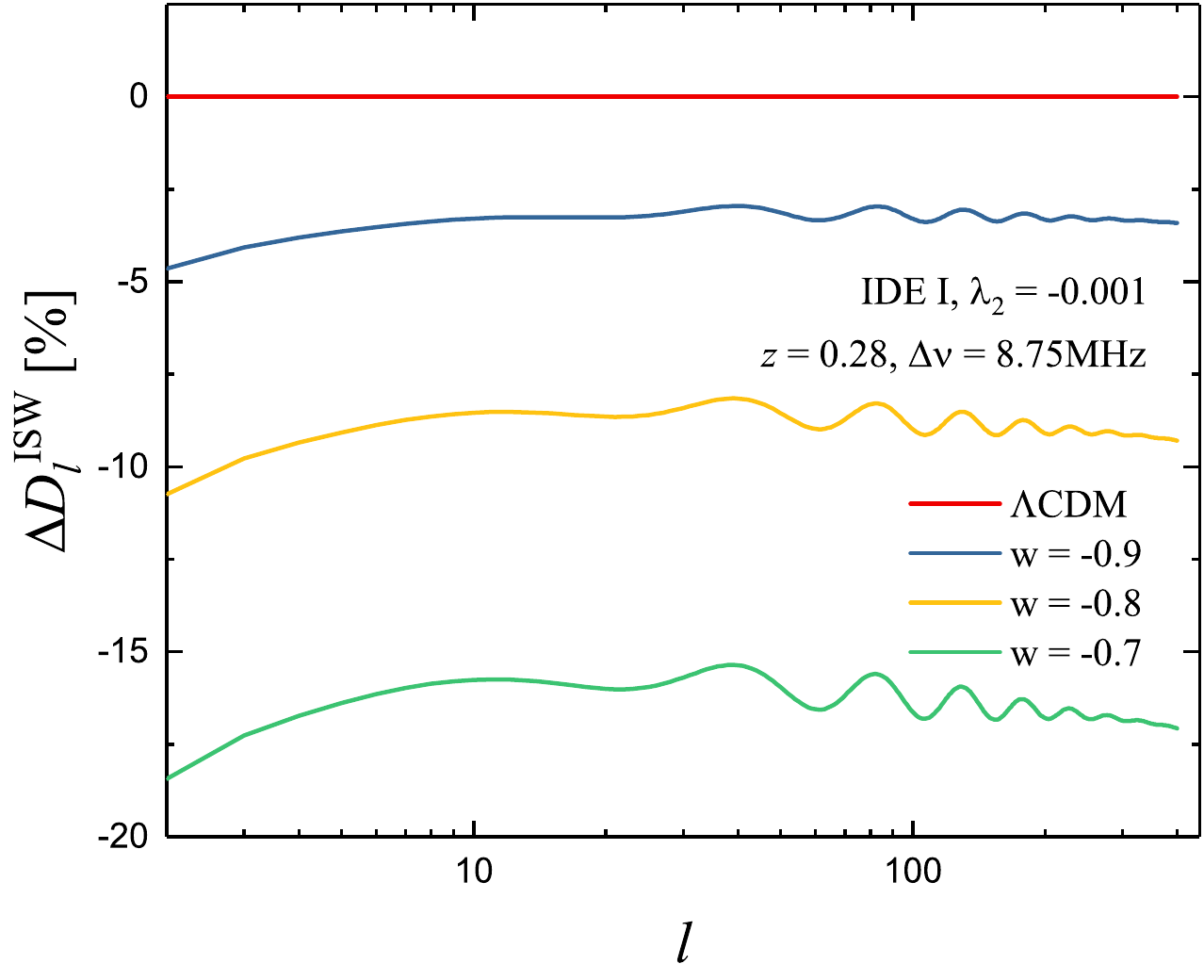}
	}
	\subfloat[]{\label{fig.IDE1_w-IDE}
		\includegraphics[width=0.31\textwidth]{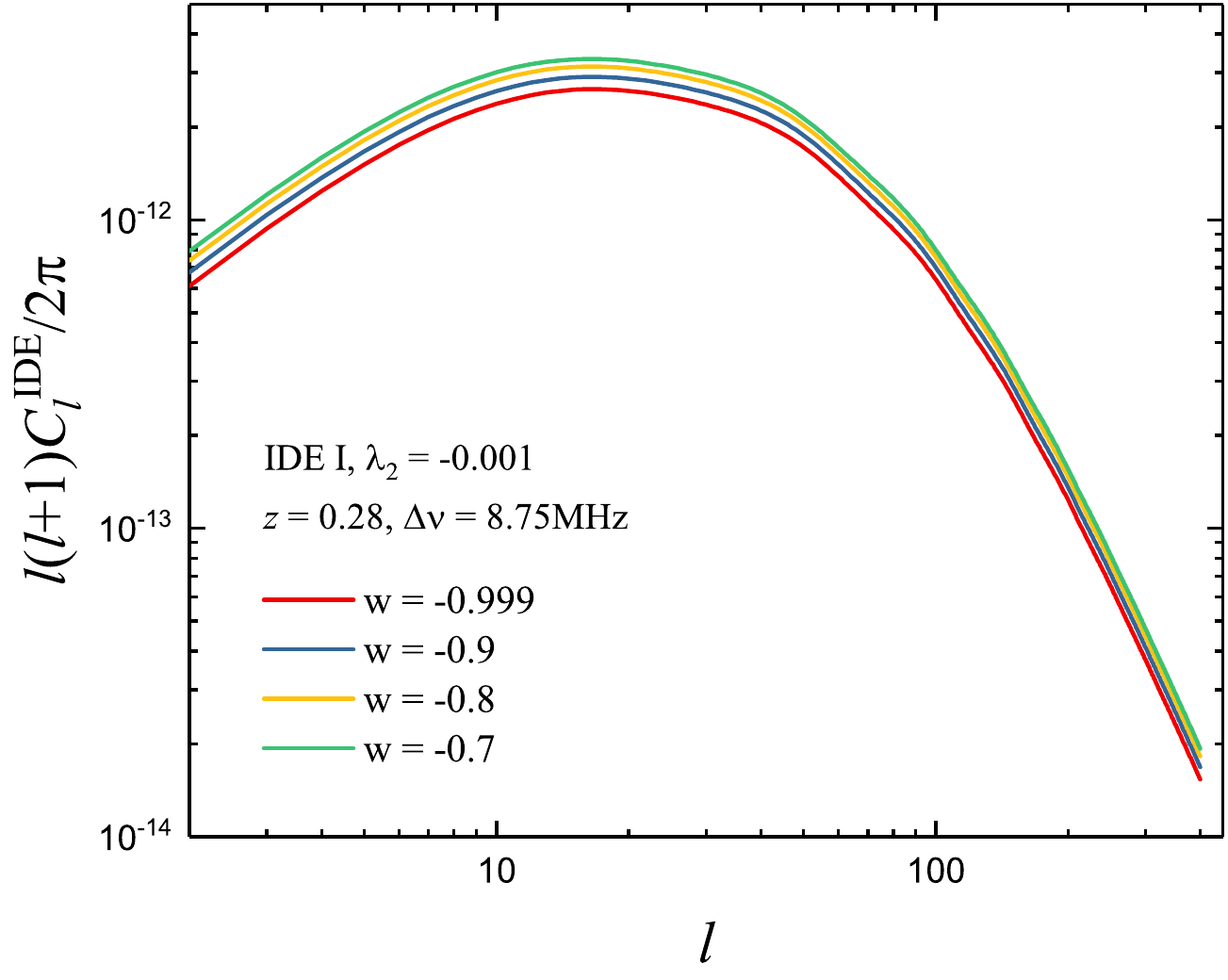}
	}
	\caption{\label{fig.IDE1_w} The $w$-varying fractional auto-spectra of each contribution from (a) overdensity $\delta$, (b) RSD, (c) potential terms, (d) Doppler effect and (e) ISW effect, respectively, for IDE Model I with respect to $\Lambda$CDM. Panel (f) is the auto-spectrum of the last term, the IDE-induced one, in Eq.~\ref{eq:perturb2}. A larger $w$ will suppress the signal for every contribution at all scales, except the extra IDE component. The ``mirror image" of these fractional angular power with respect to the zero axis can be approximately regarded as the illustrations of Model II.}
\end{figure*}

\begin{figure*}
	\subfloat[]{\label{fig.IDE1_lam2-delta}
		\includegraphics[width=0.31\textwidth]{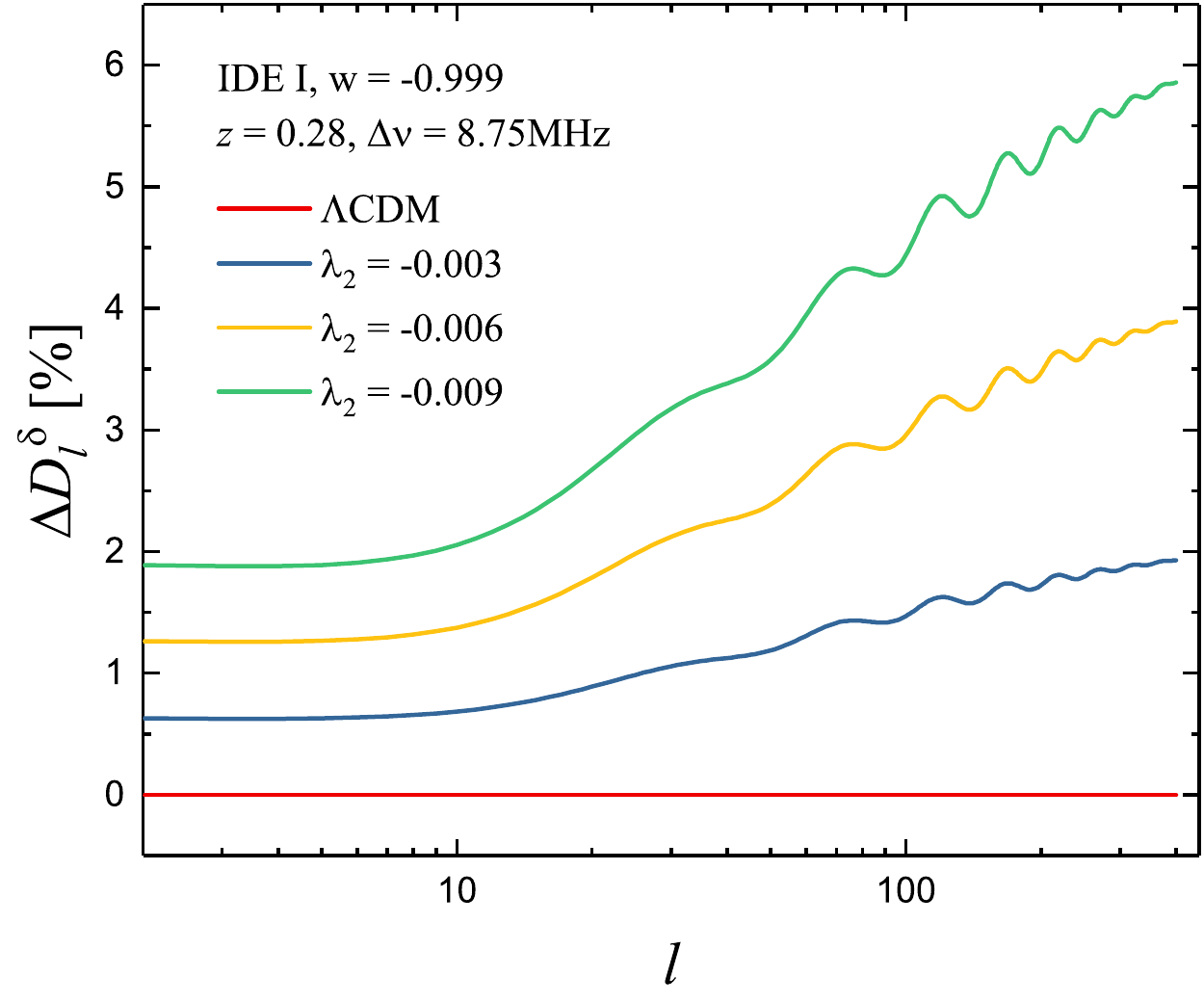}
	}
	\subfloat[]{\label{fig.IDE1_lam2-RSD}
		\includegraphics[width=0.31\textwidth]{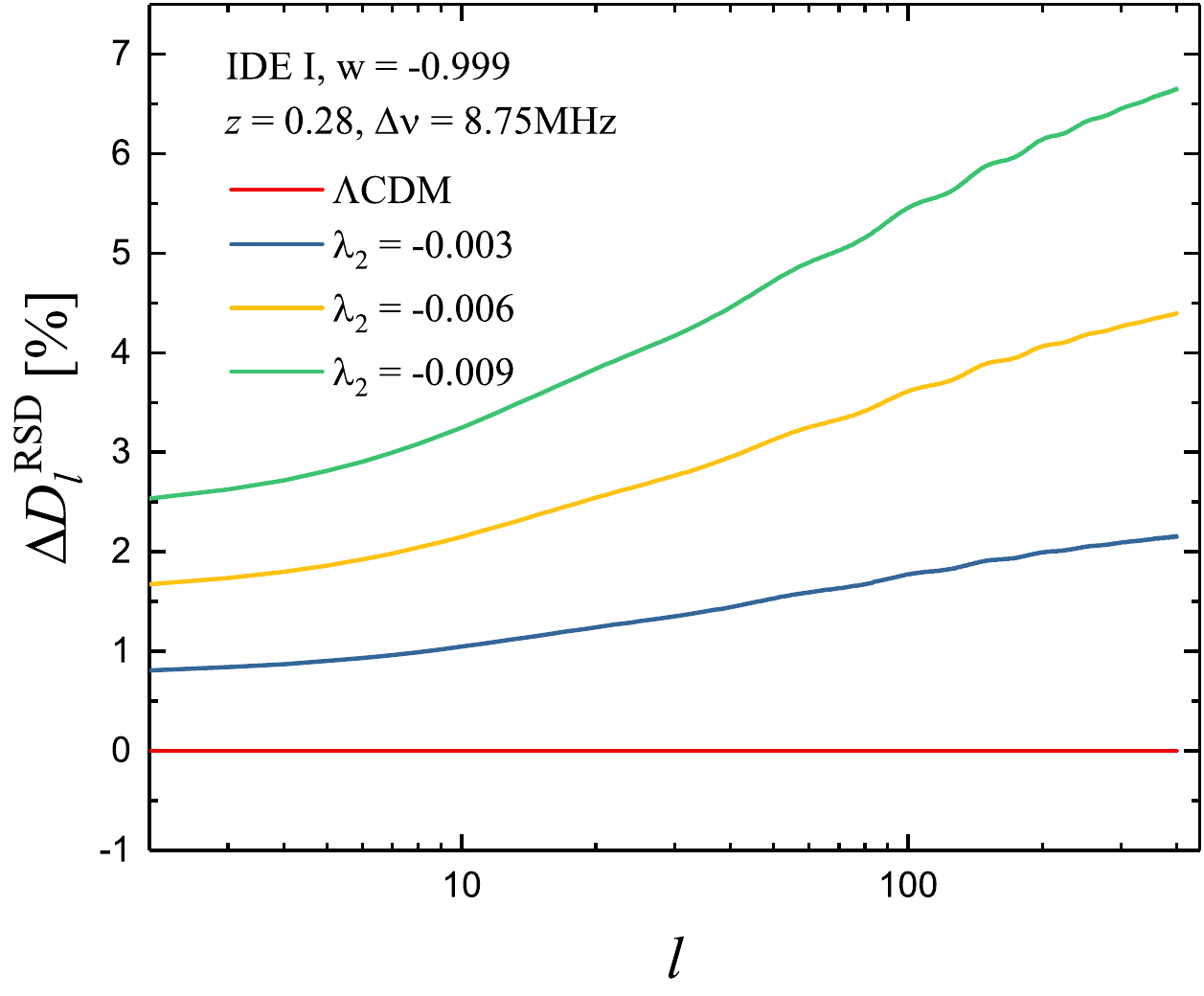}
	}
	\subfloat[]{\label{fig.IDE1_lam2-Pot}
		\includegraphics[width=0.31\textwidth]{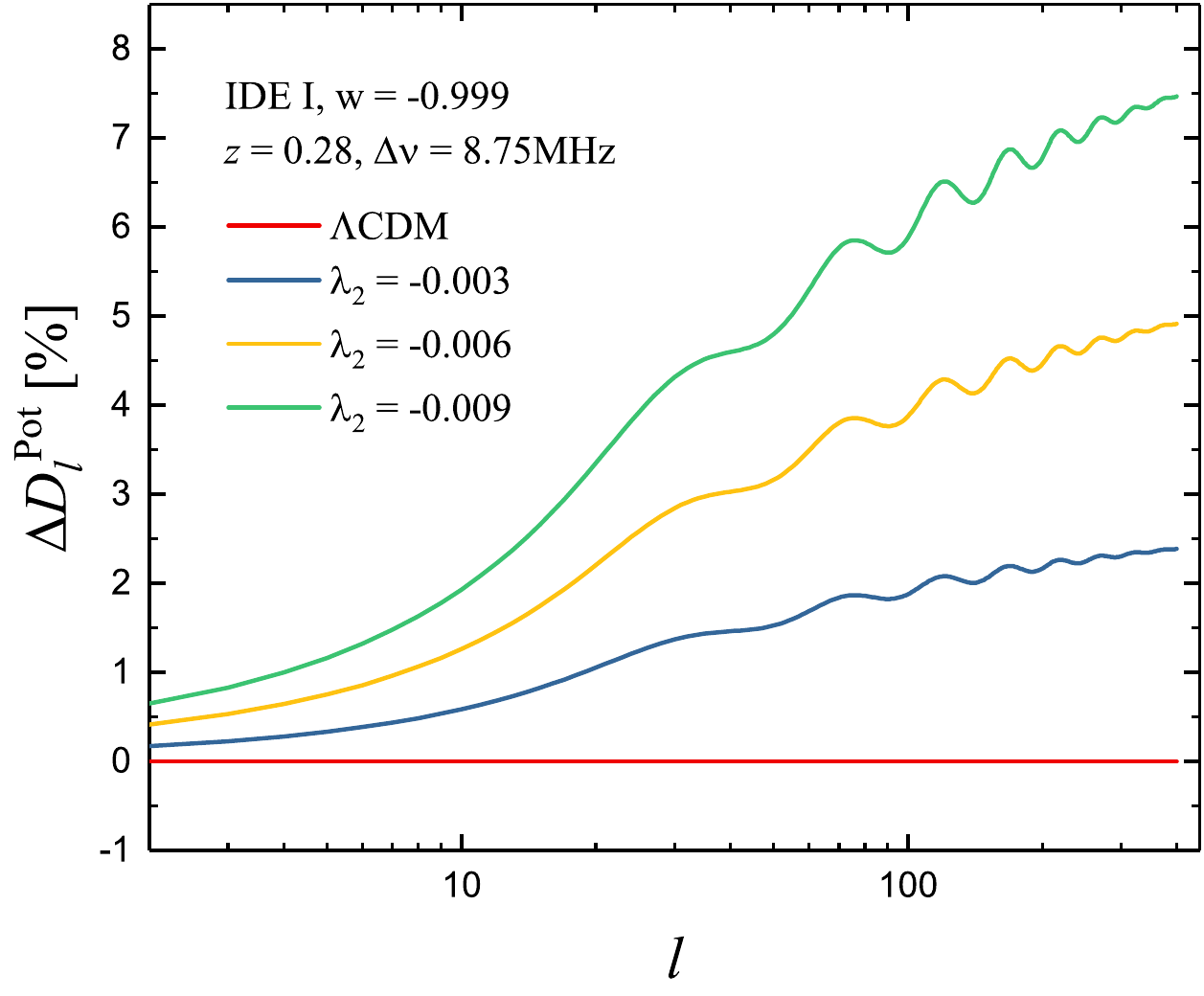}
	}
	
	\subfloat[]{\label{fig.IDE1_lam2-v}
		\includegraphics[width=0.31\textwidth]{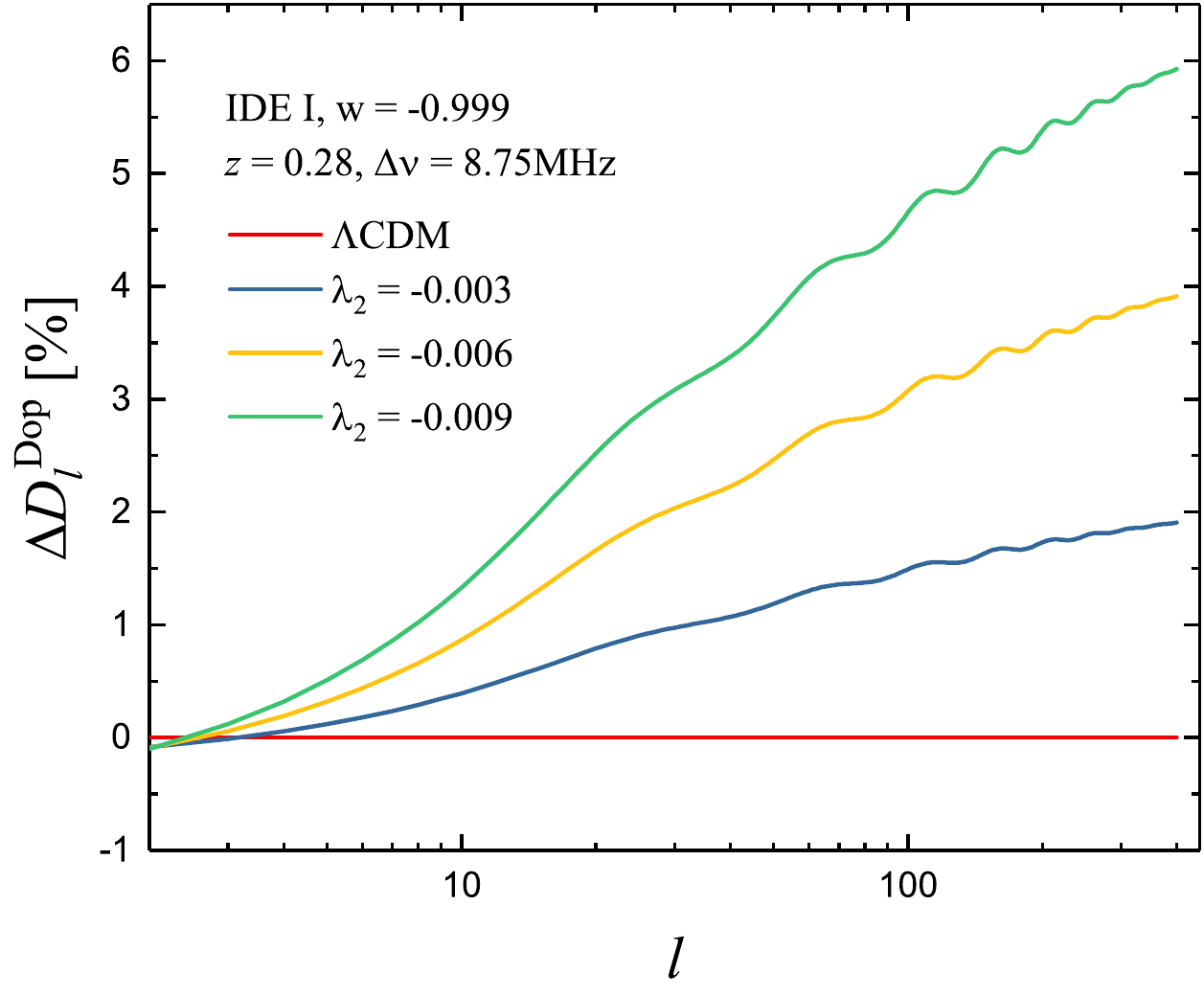}
	}
	\subfloat[]{\label{fig.IDE1_lam2-ISW}
		\includegraphics[width=0.31\textwidth]{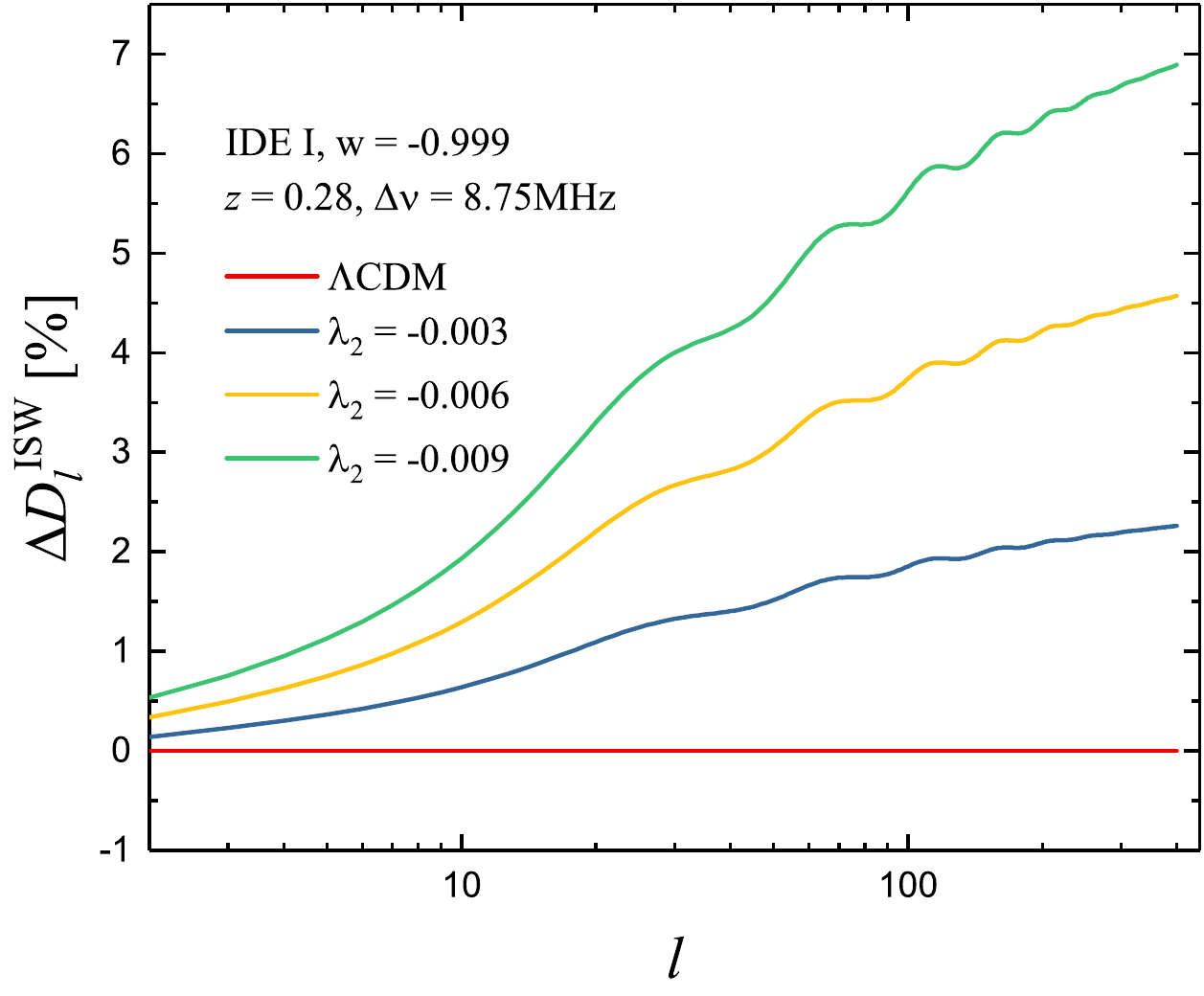}
	}
	\subfloat[]{\label{fig.IDE1_lam2-IDE}
		\includegraphics[width=0.31\textwidth]{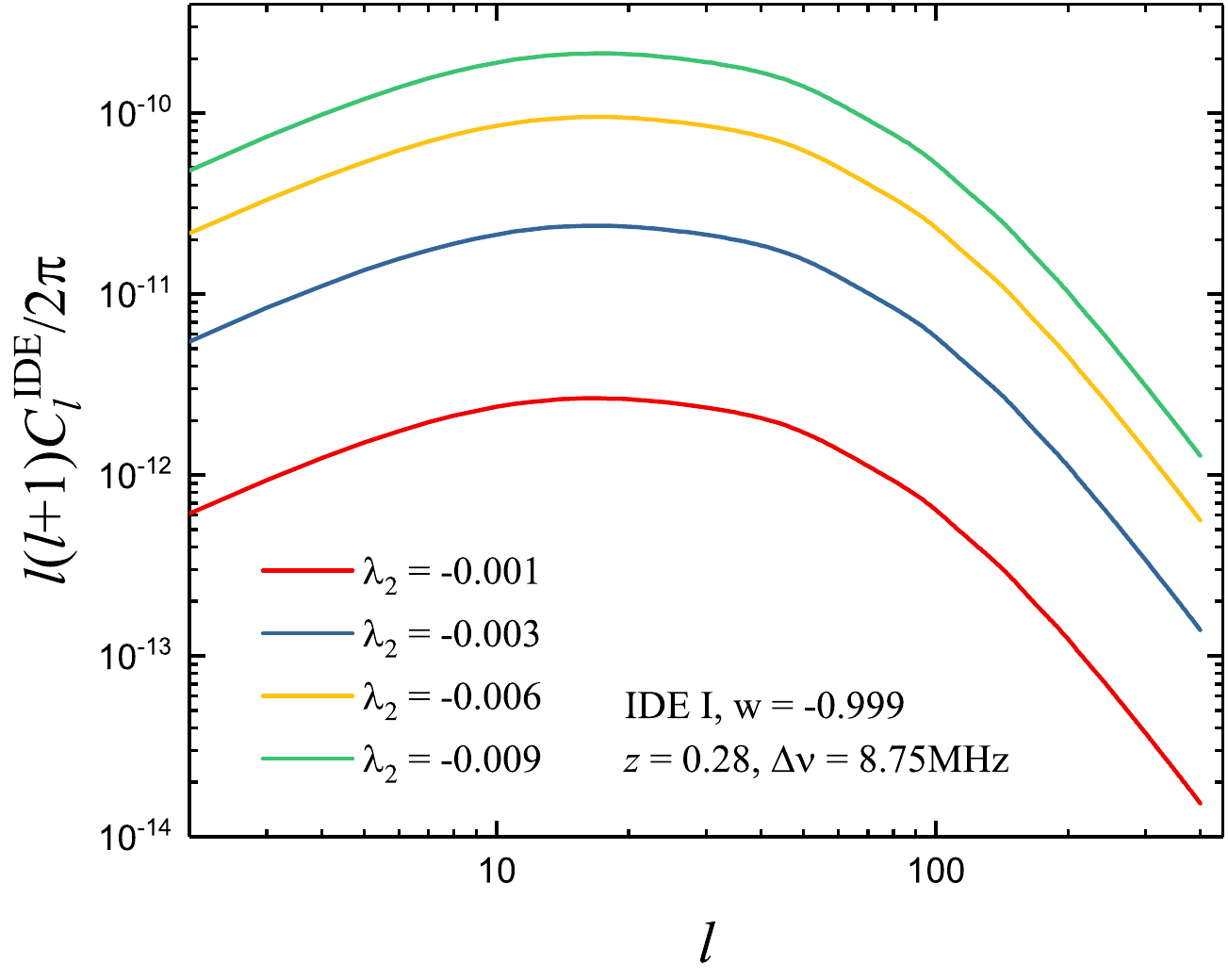}
	}
	\caption{\label{fig.IDE1_lam2} Same as in Fig.~\ref{fig.IDE1_w}, but for a varying interacting strength $\lambda_2$ in Model I. Except for the extra IDE contribution in panel (f), a DM-DE interaction mainly power boost the signal at small scales.}
\end{figure*}

Let us first consider the IDE Model I. In Fig.~\ref{fig.IDE1_w}, we plot the changes to the auto-spectra of each signal component induced by varying $w$. Except for the extra IDE term in Eq.~(\ref{eq:perturb2}), every other $D_{\ell}^{i}$ decreases with an increasing $w$. By comparing with Fig.~\ref{fig.cl_21}, we see the contribution from the extra IDE term (see Fig.~\ref{fig.IDE1_w-IDE}) is comparable to the ISW effect. Therefore, the total signal will be weakly affected and follow the pattern for $w >-1$ in Fig.~\ref{fig.cl_wCDM}. Here the coupling strength $\lambda_2$ has been assigned a very tiny value and, thus, the DM-DE interaction does not play a major role in the evolution of perturbations to the first order. Therefore, those nearly scale-independent power variations should be mainly attributed to the varying $w$, resembling the circumstance of $w$CDM. On the other hand, if we fix $w$ and vary $\lambda_2$, we find another story.  Taking for granted that similar behaviors appear in the background evolution by varying $w$ or $\lambda_2$, we anticipated a degeneracy between effects of $w$ and $\lambda_2$ in the 21-cm spectrum. An energy transfer from DM to DE, described by a negative $\lambda_2$ allowed in IDE Model I, which requires more DM and less DE in the past if the mean density of every cosmic component is fixed at nowadays. It seems that we ought to have deeper gravitational potentials, larger overdensities and in-fall velocities, hence correspondingly stronger 21-cm signals. Although this can be regarded as a physical interpretation to the similar qualitative influences of varying $w$ or $\lambda_2$, their behaviours on the perturbation level show different scale dependencies, as can be seen by comparing Fig.~\ref{fig.IDE1_w} and Fig.~\ref{fig.IDE1_lam2}. Increasing the interaction, the power of each contribution gets strong boost on small scales and  the extra IDE term is more sensitive to the interaction (i.e., $aQ$). This scale-dependent characteristic due to the variation of the interaction between dark sectors is clearly different from the influence given by the change of $w$, which can be used to  break the degeneracy between $w$ and $\lambda_2$ and  distinguish IDE from $\Lambda$CDM at high $\ell$s.

Referring to Table~\ref{tab.models}, it is reasonable to assert that Model II is a ``mirror image" of Model I and as a result, the former scenario should behave in an opposed manner with respect to the later on 21-cm signals. In particular, the 21-cm $D_{\ell}^{\rm tot}$ will obtain an approximately scale-independent increase by a DE EoS $w < -1$ and otherwise, a larger $\lambda_2$ is dedicated to a global angular power loss, especially on small scales. The physical reason behind is of course their different stability requirements that is essentially, the opposite direction of energy transfer. We confirmed this feature by carrying out the same analysis as we did for Model I but to avoid repetition, we skip over the illustration of Model II and go forward to Model III. 
	
\begin{figure*}
	\subfloat[]{\label{fig.IDE4_w-delta}
		\includegraphics[width=0.31\textwidth]{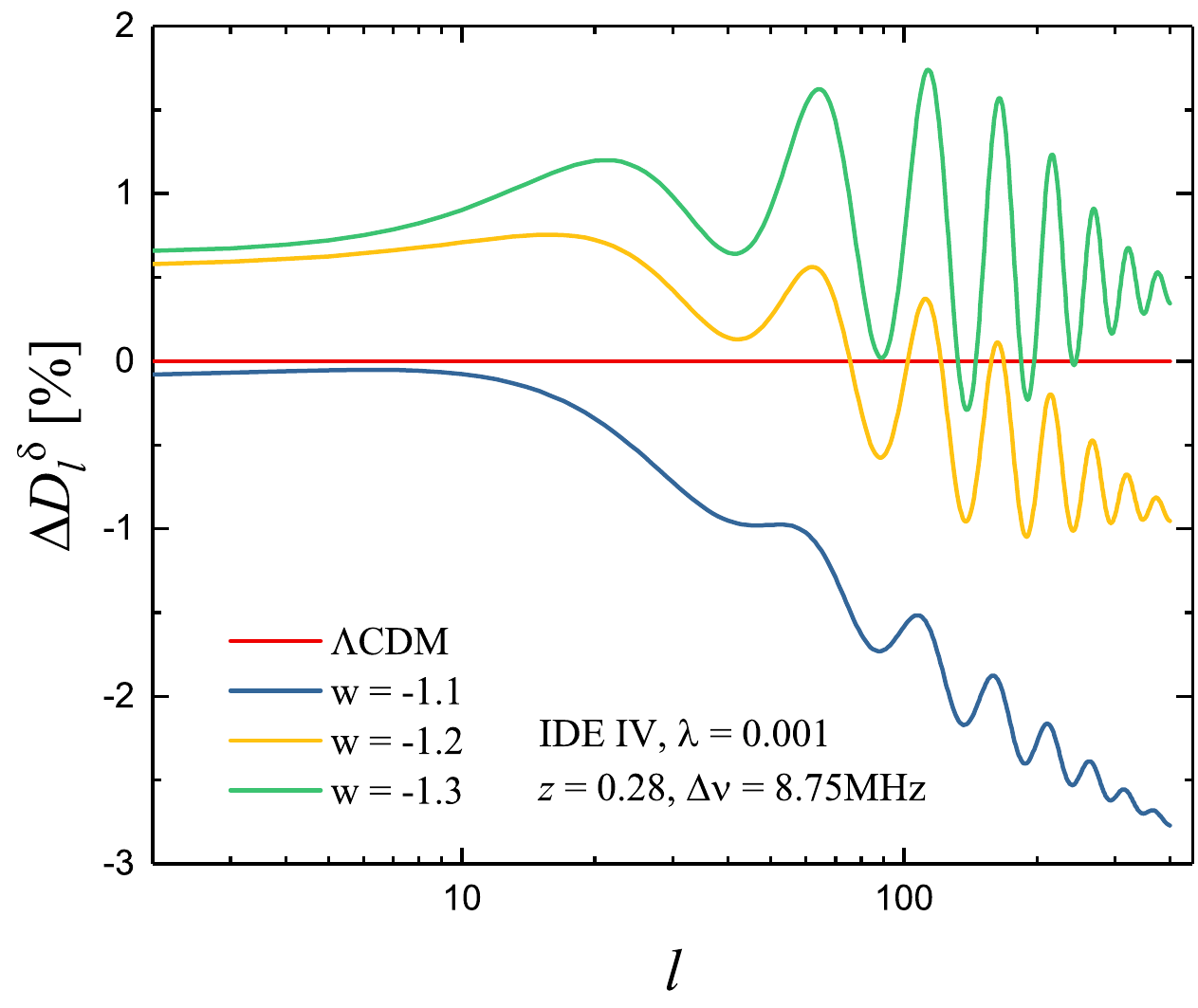}
	}
	\subfloat[]{\label{fig.IDE4_w-RSD}
		\includegraphics[width=0.31\textwidth]{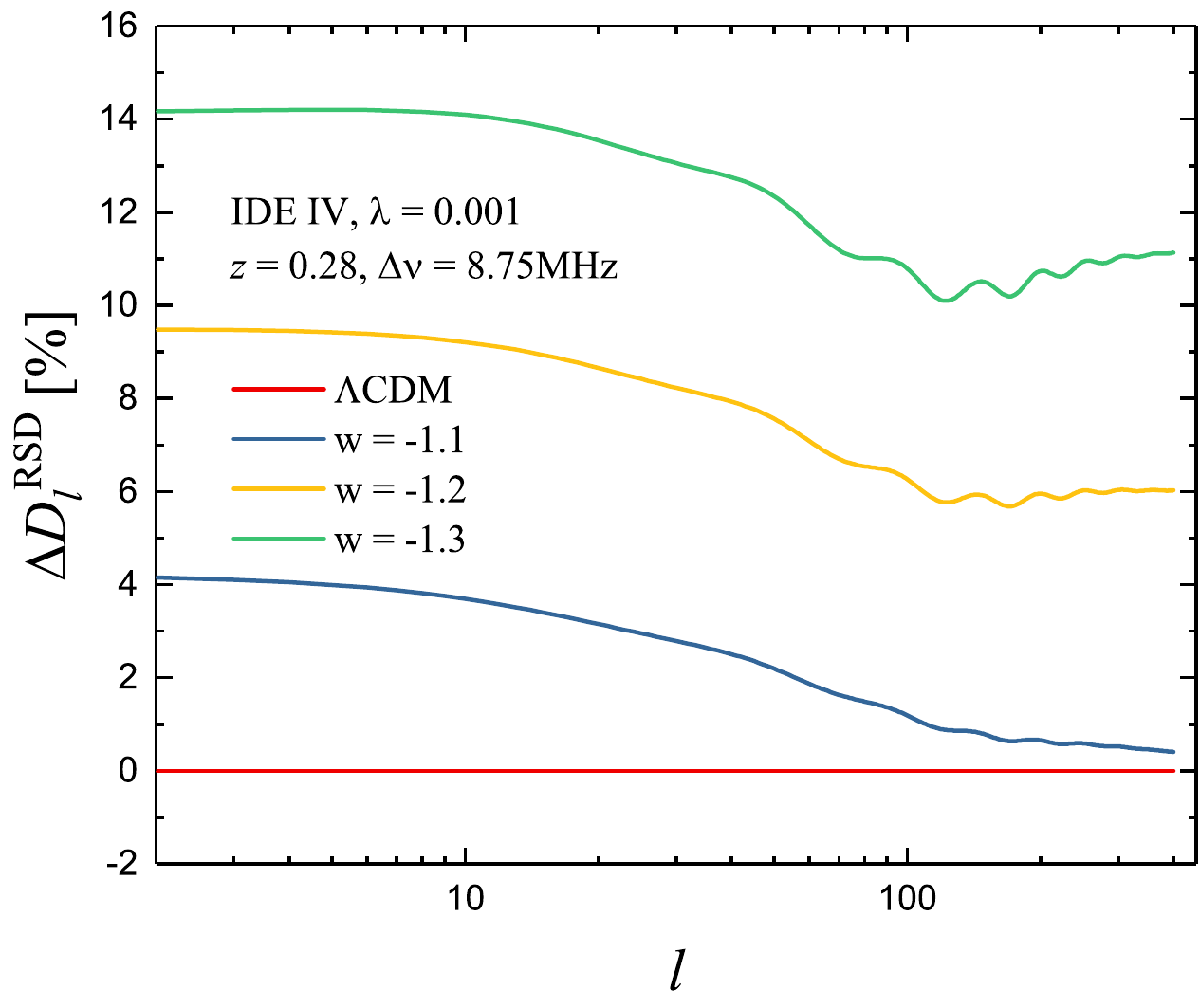}
	}
	\subfloat[]{\label{fig.IDE4_w-Pot}
		\includegraphics[width=0.31\textwidth]{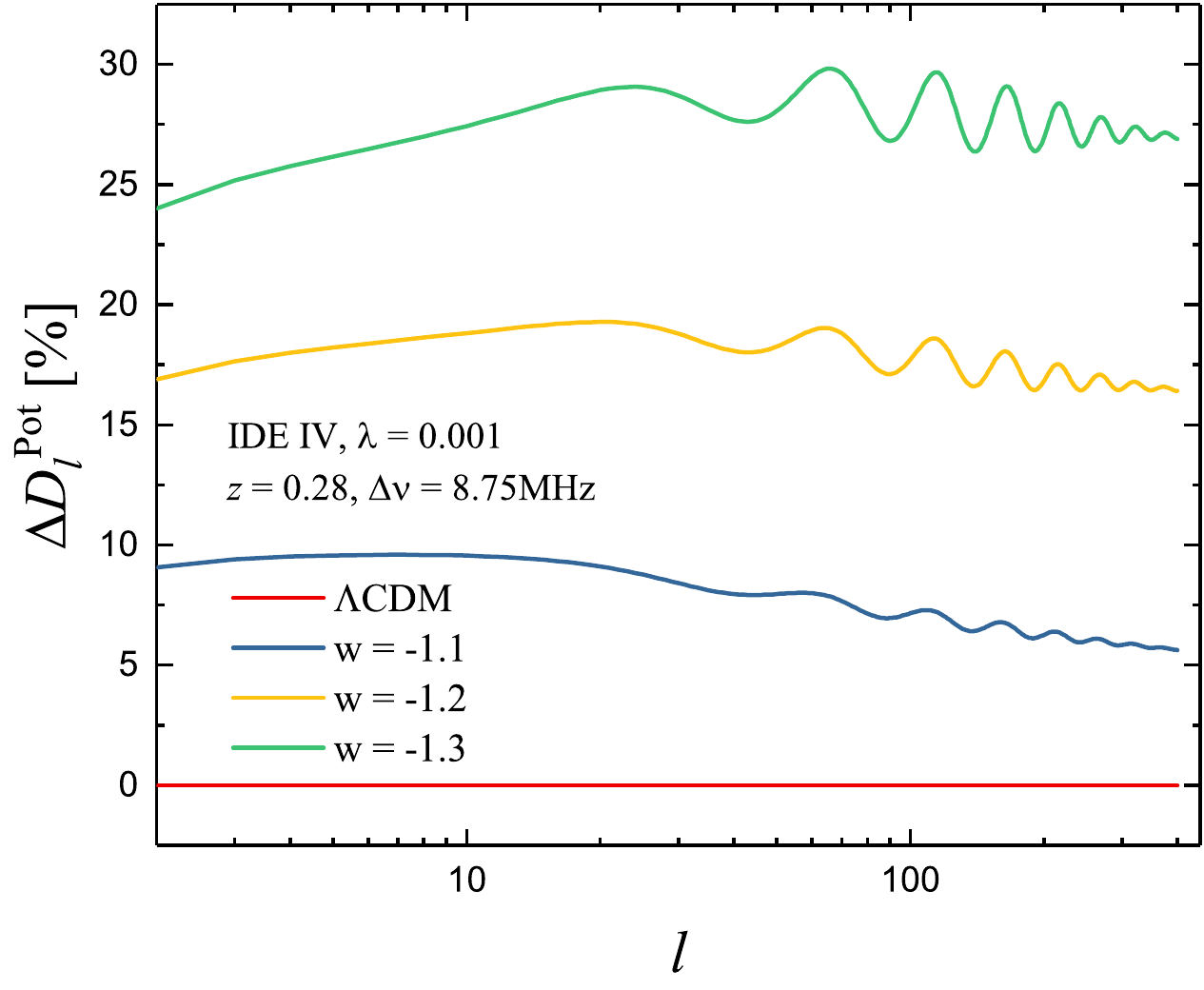}
	}
	
	\subfloat[]{\label{fig.IDE4_w-v}
		\includegraphics[width=0.31\textwidth]{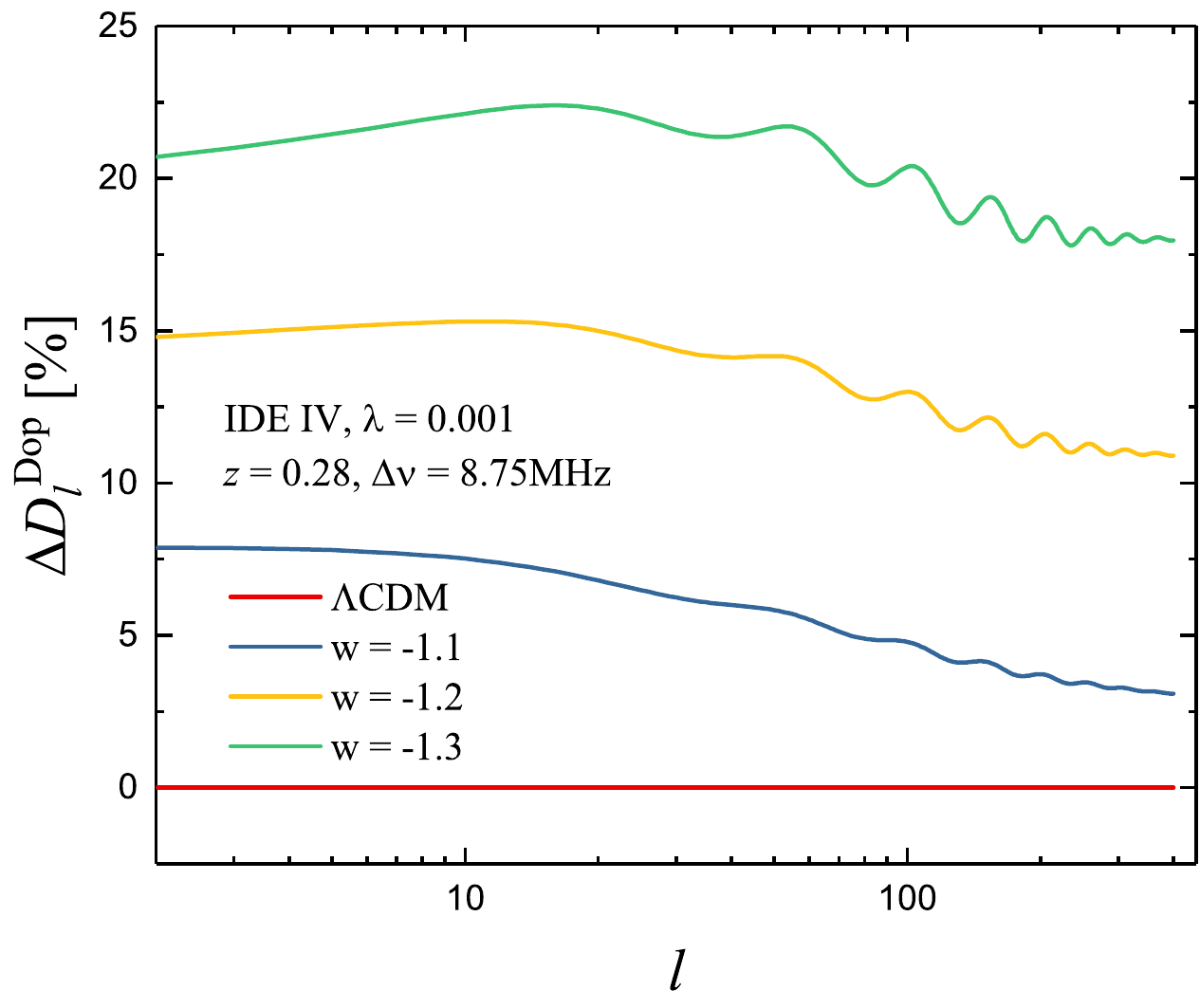}
	}
	\subfloat[]{\label{fig.IDE4_w-ISW}
		\includegraphics[width=0.31\textwidth]{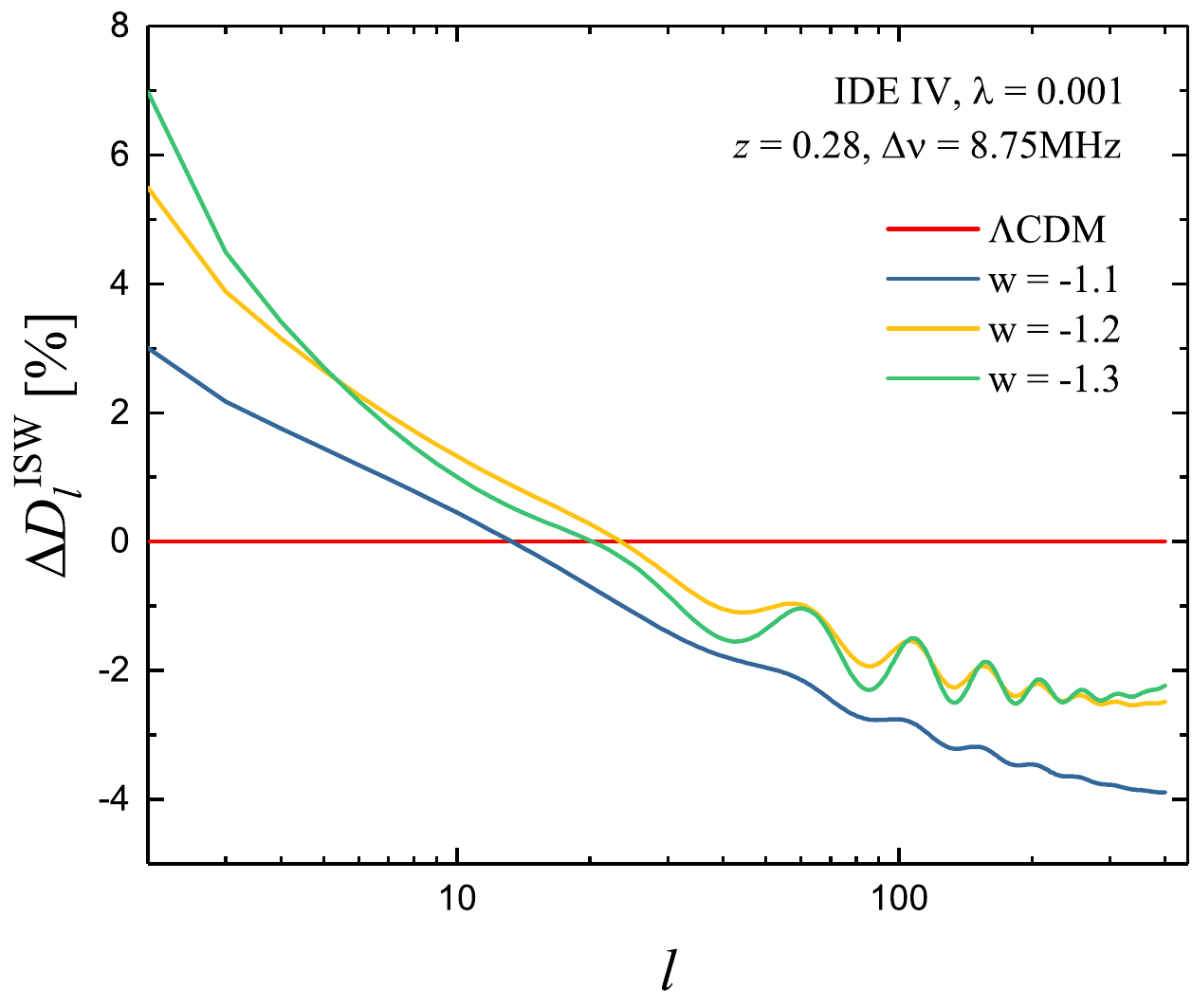}
	}
	\subfloat[]{\label{fig.IDE4_w-IDE}
		\includegraphics[width=0.31\textwidth]{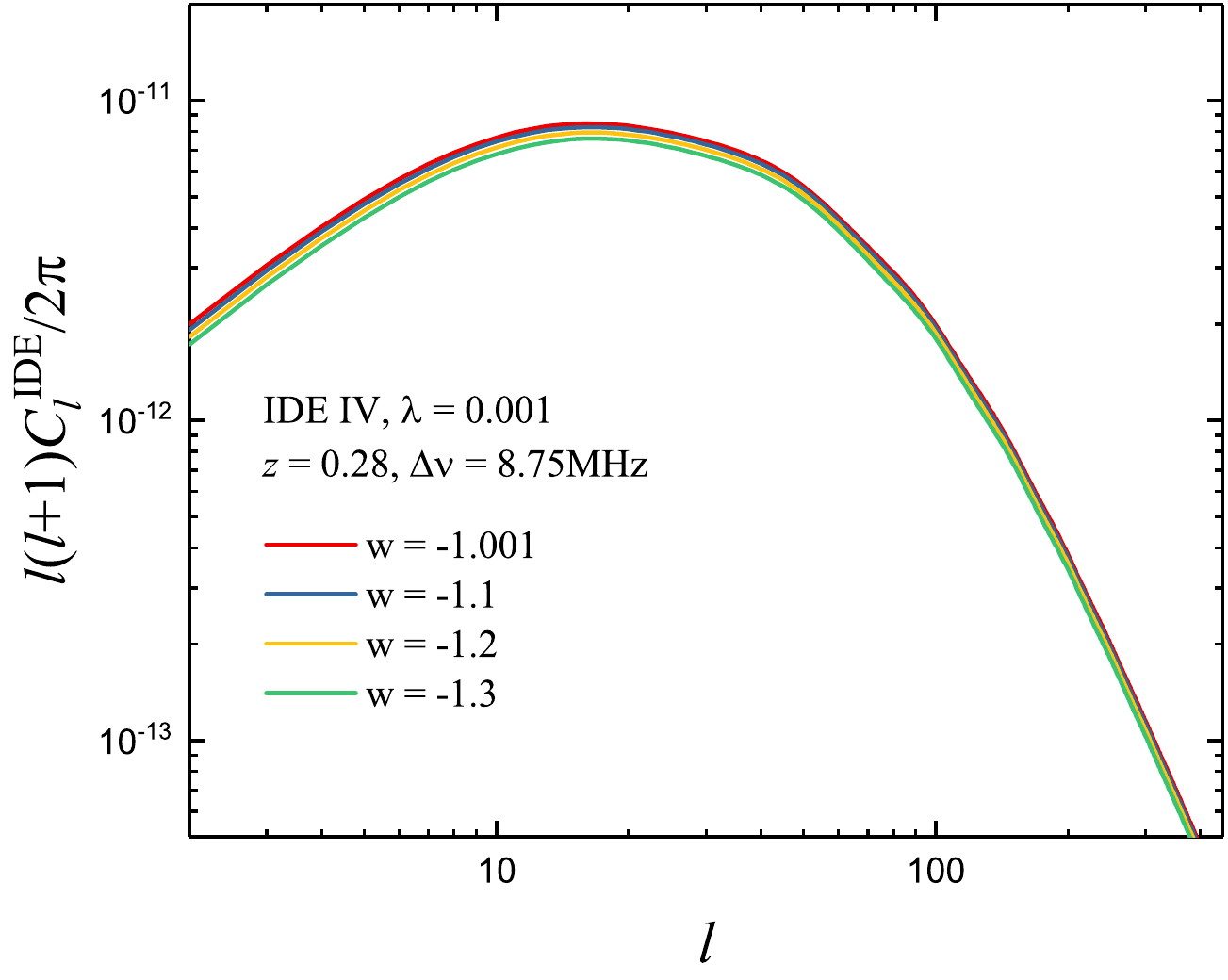}
	}
	\caption{\label{fig.IDE4_w} Same as in Fig.~\ref{fig.IDE1_w}, but for a varying equation of state $w$ in Model IV. We employ these fractional angular power patterns further as an alternative to the illustrations of Model III, by substituting $\lambda$ for $\lambda_1$.}
\end{figure*}

\begin{figure*}
	\subfloat[]{\label{fig.IDE4_lam-delta}
		\includegraphics[width=0.31\textwidth]{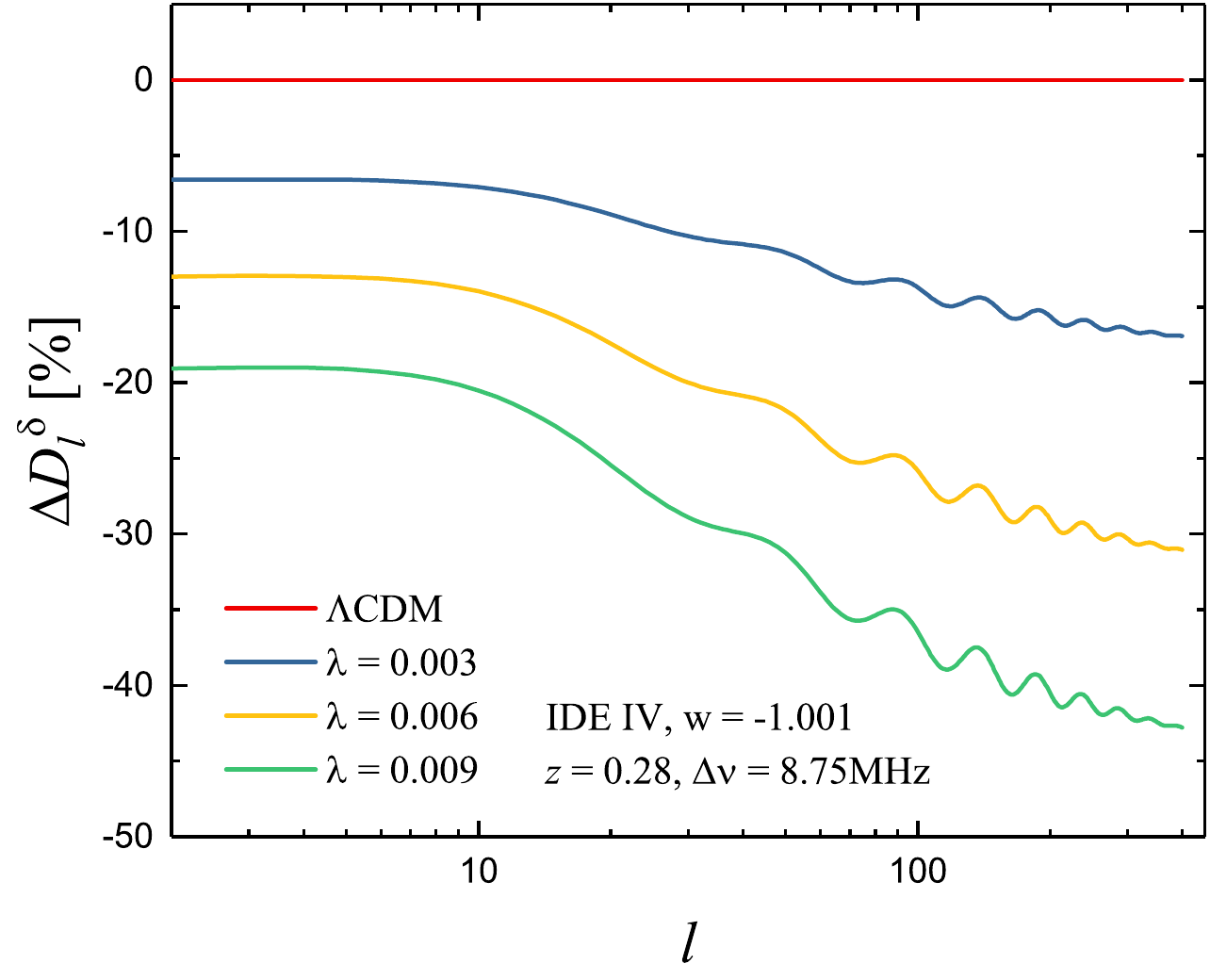}
	}
	\subfloat[]{\label{fig.IDE4_lam-RSD}
		\includegraphics[width=0.31\textwidth]{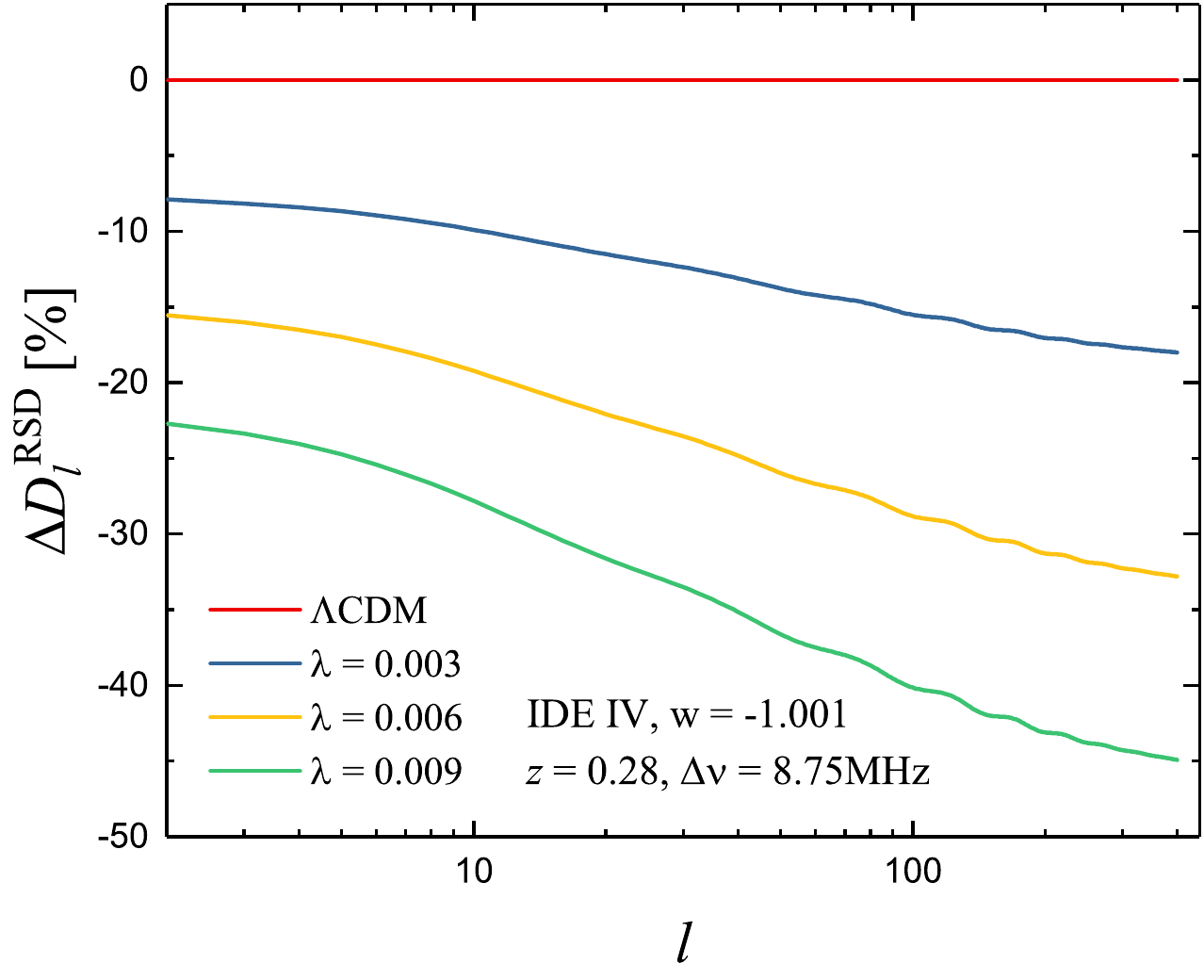}
	}
	\subfloat[]{\label{fig.IDE4_lam-Pot}
		\includegraphics[width=0.31\textwidth]{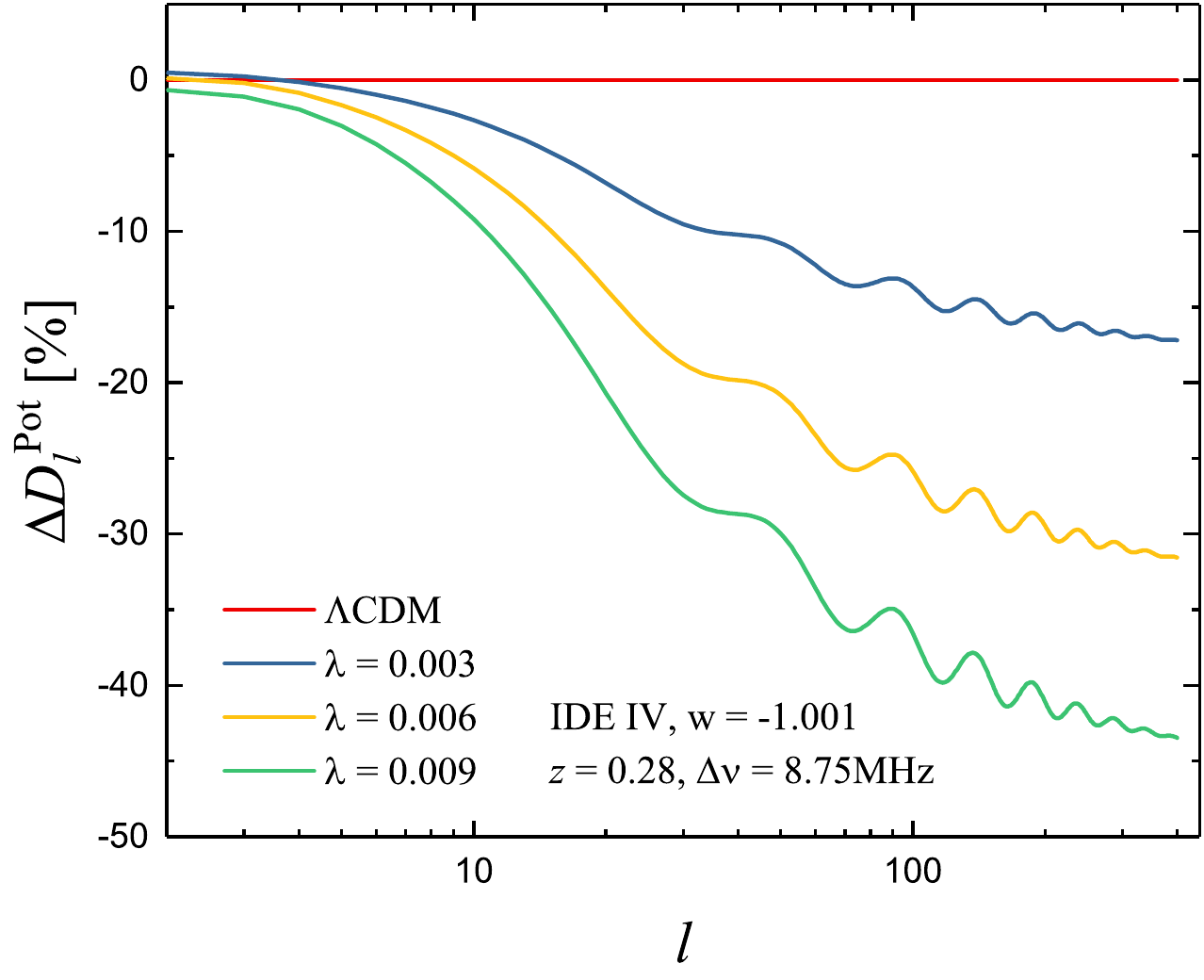}
	}
	
	\subfloat[]{\label{fig.IDE4_lam-v}
		\includegraphics[width=0.31\textwidth]{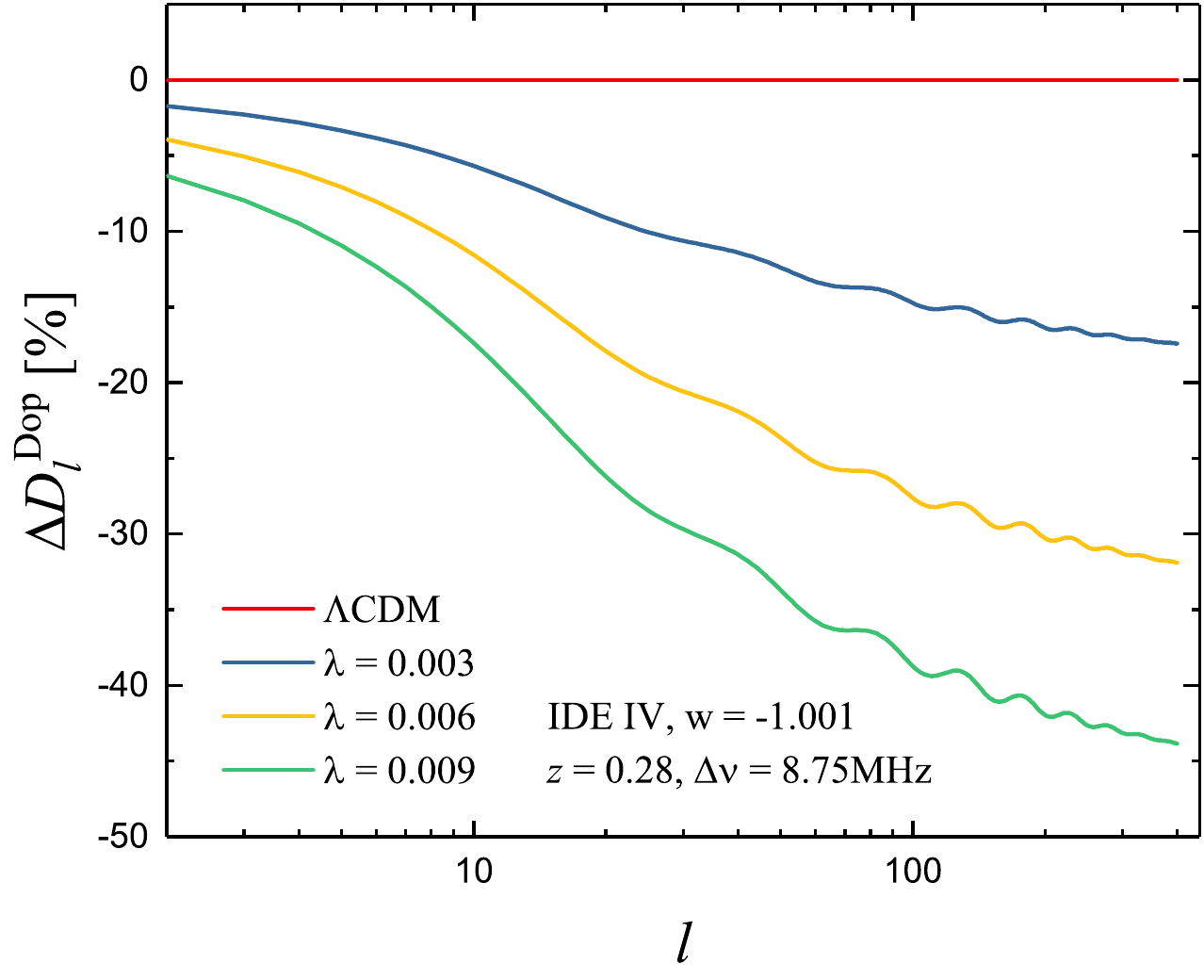}
	}
	\subfloat[]{\label{fig.IDE4_lam-ISW}
		\includegraphics[width=0.31\textwidth]{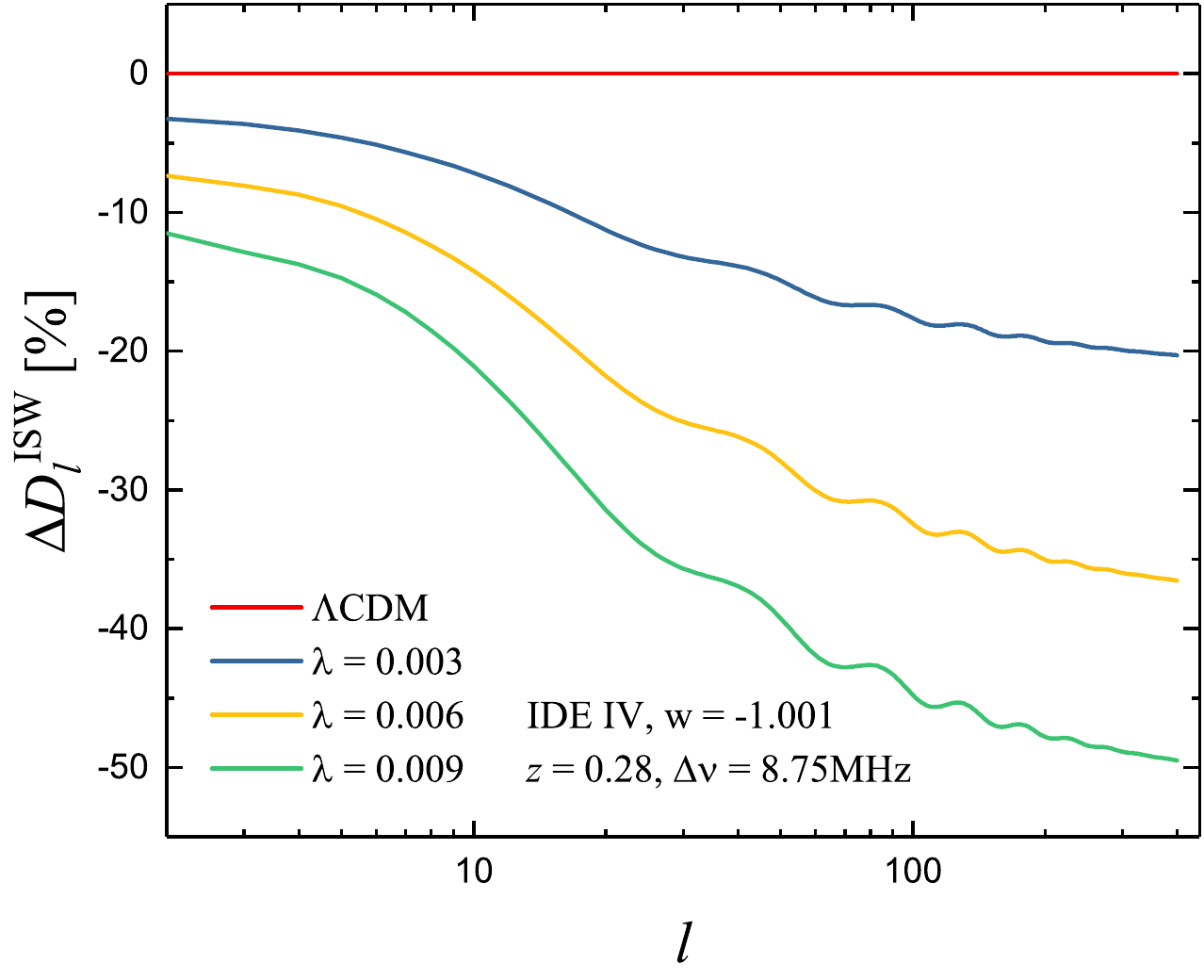}
	}
	\subfloat[]{\label{fig.IDE4_lam-IDE}
		\includegraphics[width=0.31\textwidth]{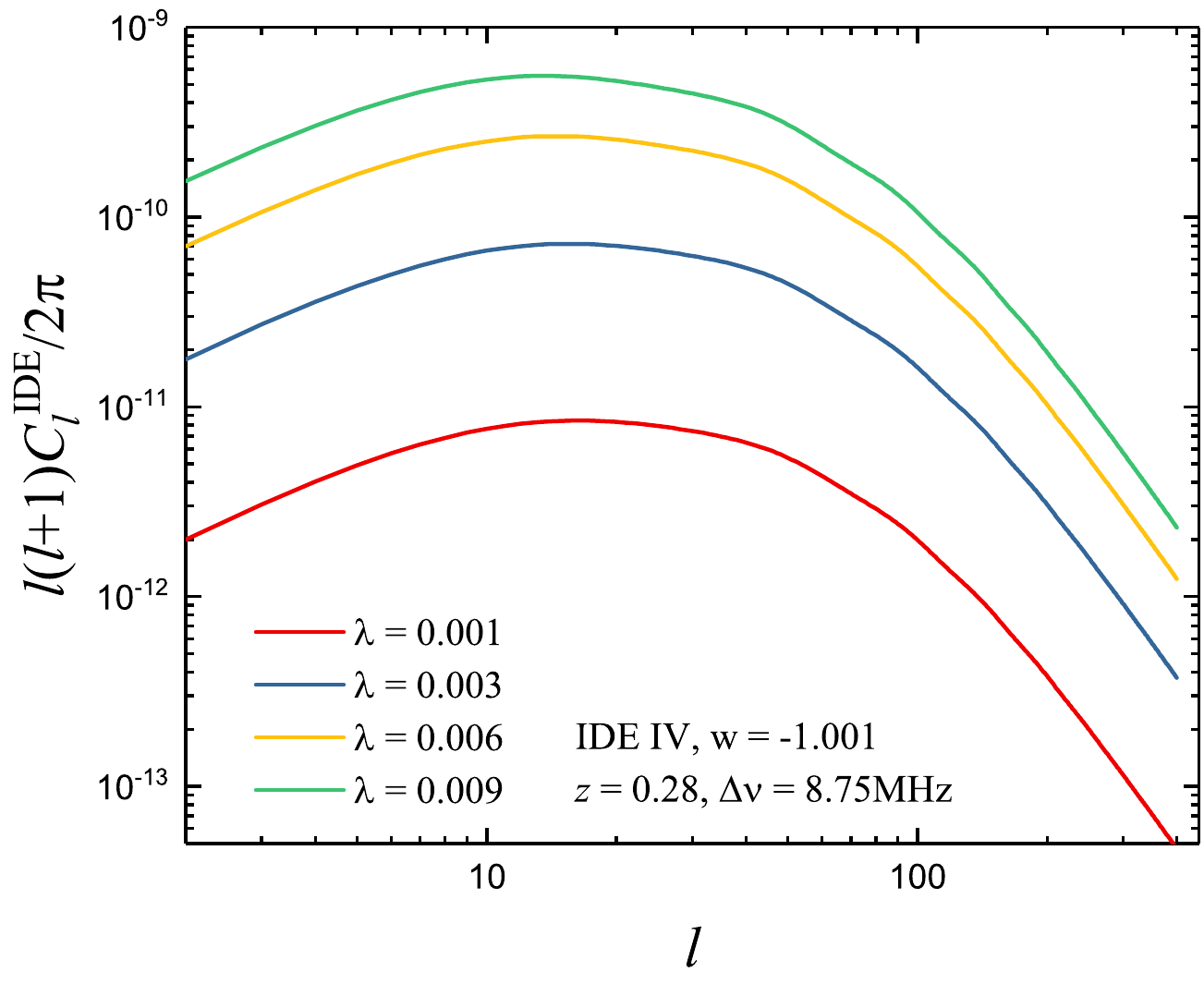}
	}
	\caption{\label{fig.IDE4_lam} Same as in Fig.~\ref{fig.IDE4_w}, but for a varying interacting strength $\lambda$ in Model IV.}
\end{figure*}

In contrast to previous two scenarios, Model III works with an interacting term $Q \propto \rho_c$. Such interaction shall play a major role in the matter dominant era, which is much earlier than those proportional to the DE energy density in Model I and II. The accumulated influence drawn from relatively more considerable energy transfer and longer dominating time span shall bring significant discrepancy on 21-cm signals from the $\Lambda$CDM prediction, making it easier to be detected. This high sensitivity to the interacting parameter, $\lambda_1$,  shows its appearance not merely in 21-cm measurements, but also in other cosmological observations, illustrated as an example in~\citet{Costa:2016tpb} that the CMB dataset is able to lay tighter constraints on Model III and IV. From the side of physics, the energy transfer direction in Model III is from DE to DM, indicating that a stronger interacting strength prefers to decrease angular power but instead, $w < -1$ will still support for the power rise as before. This is akin to what we see in the following discussion of Model IV and certainly has been confirmed by our repetitive analysis. Consequently, we again leave out the illustration of Model III and the reader may turn to Fig.~\ref{fig.IDE4_w} -~\ref{fig.IDE4_lam} as an alternative reference.

As we did for Model I, every signal component of Model IV is separately illustrated in Fig.~\ref{fig.IDE4_w} and \ref{fig.IDE4_lam}. Due to its interaction term of $Q \propto (\rho_c + \rho_d)$, Model IV is expected to be an updated IDE scenario by mixing Model II \& III together. The interaction again brings in a signal suppression across the whole multipole range but much severer at high $\ell$s, which is able to cancel off the enhancement by $w < -1$, especially in the patterns of the overdensity, RSD and ISW effects. The extra IDE contribution in Model IV is not the leading term, which was found similarly in Model I, but the interaction between dark sectors can still leave clear imprints in other terms contributing to the 21-cm angular power spectrum.  

\section{Forecast}\label{sec.forecast}
In this section, we first review the Fisher matrix method and cosmological parameters used in our analysis in Sect.~\ref{sec.Fisher}. Then, we present our forecast results in Sect.~\ref{sec.results}, encompassing the signal contributions from the overdensity and RSD components. After that, we further discuss in Sect.~\ref{sec.extension} the impact on the parameter constraints from different redshift binning schemes and including or not the RSD effect. The interference from other two systematics, foreground residuals as well as the uncertainty of $\Omega_{\mathrm{HI}}(z)$ both on its magnitude and redshift dependence, are well analyzed in Sect.~\ref{sec.residual} and \ref{sec.Omhi}, respectively.

\subsection{The method of Fisher Matrix Analysis}\label{sec.Fisher}
The Fisher matrix is frequently used to forecast the cosmological parameter constraints~\citep[e.g.,][]{Dodelson2003,Asorey:2012rd,Hall:2012wd}. Given a set of cosmological parameters, the Fisher matrix $\mathbf{F}$ yields the smallest error bars with which the parameters can be measured with some specific data set. $\mathbf{F}^{-1}$ can be thought as the best possible covariance matrix for the constraints on the parameters~\citep{Tegmark:1996qt}. Elements with higher absolute values in the Fisher matrix correspond to higher precision in the measured parameters. In this section, we perform a forecast for BINGO and SKA1-MID via a Fisher matrix analysis, such that we can inspect the ability of HI IM in constraining the IDE model.

The Fisher matrix for the parameters $\theta_i$ within a model $\mathcal{M}$ is the ensemble average of the Hessian matrix of the log-likelihood. When Gaussian fields with zero mean are assumed, each element in the Fisher matrix for HI IM surveys reads~\citep{Dodelson2003,Asorey:2012rd}
\begin{align}
F_{ij} \equiv \left\langle-\frac{\partial\ln \mathcal{L}}{\partial\theta_i \, \partial\theta_j}\right\rangle = \frac{1}{2}\mathrm{Tr}\left[\mathrm{\bf C}^{-1}\frac{\partial\mathrm{\bf C}}{\partial\theta_i}\mathrm{\bf C}^{-1}\frac{\partial\mathrm{\bf C}}{\partial\theta_j}\right] \;.
\label{eq:fish}
\end{align}
The covariance $\mathrm{\bf C}$ comprising both signals and noises is given by\footnote{In order to properly take into account the effect of $\Omega_{\mathrm{HI}}$ in our estimation, we will use the dimensional angular spectra, which can be calculated recasting Eq.~(\ref{eq:window}) as ${\Delta}_{T_{\rm b},\ell}^W(\mathbf{k}) = \int_0^\infty {\rm d} z \bar{T}_{\mathrm{b}}(z) W(z)
\Delta_{T_{\rm b},\ell}(\mathbf{k},z)$. Our analyses on dimensional $\Delta D_{\ell}^{i}$ show that it will not significantly change the features discussed in Sect.~\ref{sec.analyses}.}

\begin{align}
\mathrm{\bf C} & = {C}_{\ell}^{\rm HI}(z_i,z_j) + \delta_{ij}C_{\ell}^{\rm shot}(z_i,z_j) + N_{\ell}(z_i,z_j)B_{\ell}(z_i, z_j) \nonumber \\
& + C_l^{\rm FG}(z_i, z_j)\,.
\label{eq:Cl_tot}
\end{align}

These 21-cm $C_{\ell}$s encompass information mapping a 3D volume, which is intrinsic different from CMB measurements to a fixed redshift. Consequently, we extend the original CMB diagonal matrix into a diagonal block matrix
\begin{align}
\mathrm{\bf C} &=
\begin{bmatrix}
A_{\ell = 2} & 0 & ... & 0 \\
0 & A_3 & ... & 0 \\
\vdots & \vdots & ... & \vdots \\
0 & 0& ... & A_n
\end{bmatrix}
\,, \qquad \\ 
\text{where} \nonumber\\
A_\ell &= (2\ell + 1)
\begin{bmatrix}
C_\ell(z_1,z_1) & C_\ell(z_1,z_2) & ... & C_\ell(z_1,z_n) \\
C_\ell(z_2,z_1) & C_\ell(z_2,z_2) & ... & C_\ell(z_2,z_n) \\
\vdots & \vdots & ... & \vdots \\
C_\ell(z_n,z_1) & C_\ell(z_n,z_2)& ... & C_\ell(z_n,z_n)
\end{bmatrix}
\,.
\label{equcov}
\end{align}

The set of cosmological parameters for the IDE models we will consider in our forecast is
\begin{equation}
\boldsymbol{\theta} = \{\Omega_{\mathrm{b}}h^2, \Omega_{\mathrm{c}}h^2, w, h, n_s, \mathrm{log}(10^{10}A_s), b_{\rm HI}, \lambda_1, \lambda_2\}\,.
\label{basepar}
\end{equation}
We will assume the fiducial value for the parameters as $w = -0.999$ for Model I, $w = -1.001$ for Model II $\sim$ IV, $b_{\rm HI} = 1$ and $\lambda_1 = \lambda_2 = 0$. Note that in Model IV $\lambda \equiv \lambda_1 = \lambda_2$. The other parameters' fiducial values follow the {\it Planck} best-fit values listed in Sect.~\ref{sec.intro}. We numerically calculate the partial derivative of HI power spectrum with respect to each cosmological parameter in Eq.~(\ref{eq:fish}). The value of $\Delta\theta$ should be carefully modulated to avoid miscalculating the derivative or introducing numerical errors. We set $\Delta\theta=0.5\%\times\theta$. Due to their stability conditions, the derivatives with respect to the interacting strengths or EoS are limited to one side, hence, we employ second-order difference for high numerical accuracy. The $1\sigma$ uncertainty in each parameter is obtained from the inverse of the Fisher matrix in Eq.~(\ref{eq:fish}), after being marginalized over other parameters.
 
In our fiducial Fisher analysis (i.e., consisting of Sect.~\ref{sec.results} $\sim$~\ref{sec.extension}), we work in an optimistic situation with no foreground residual and a fixed $\Omega_{\mathrm{HI}} = 6.2 \times 10^{-4}$. In Sect.~\ref{sec.residual}, we extend the fiducial analysis by incorporating foreground residuals, whereas the analysis in Sect.~\ref{sec.Omhi} is focusing on the uncertainty from $\Omega_{\mathrm{HI}}$.

\subsection{Forecast Results}\label{sec.results}
In this subsection, we gather the projected constraints on the parameter set $\boldsymbol{\theta}$ for three HI IM projects: BINGO, SKA1-MID Band\,1 and Band\,2, with survey configurations listed in Table~\ref{tab.survey_parameter}. We also introduce the covariance matrices for IDE models with the {\it Planck} 2018 dataset as in~\cite{Bachega:2019fki}, such that a joint analysis of HI IM and CMB measurement is accessible. Therefore, we can compare the constraints from HI IM to those from CMB, as well as combine those observations focusing on different physical processes to help tightening the cosmological constraints.

We observe in Fig.~\ref{fig.cl_21} that  velocity, potentials and ISW contributions are negligible  to the total signal. Thereby, in order to improve the computer performance, we present the projected constraints based on the total 21-cm signal including contributions from $\delta_n$ and RSD only. Here, we take into account the shot noise and thermal noise (see  Sect.~\ref{sec.surveys} for more details), but exclude the foreground residuals.

\begin{figure*}
	\subfloat[]{\label{fig.IDE1_lam2_h}
		\includegraphics[width=0.33\textwidth]{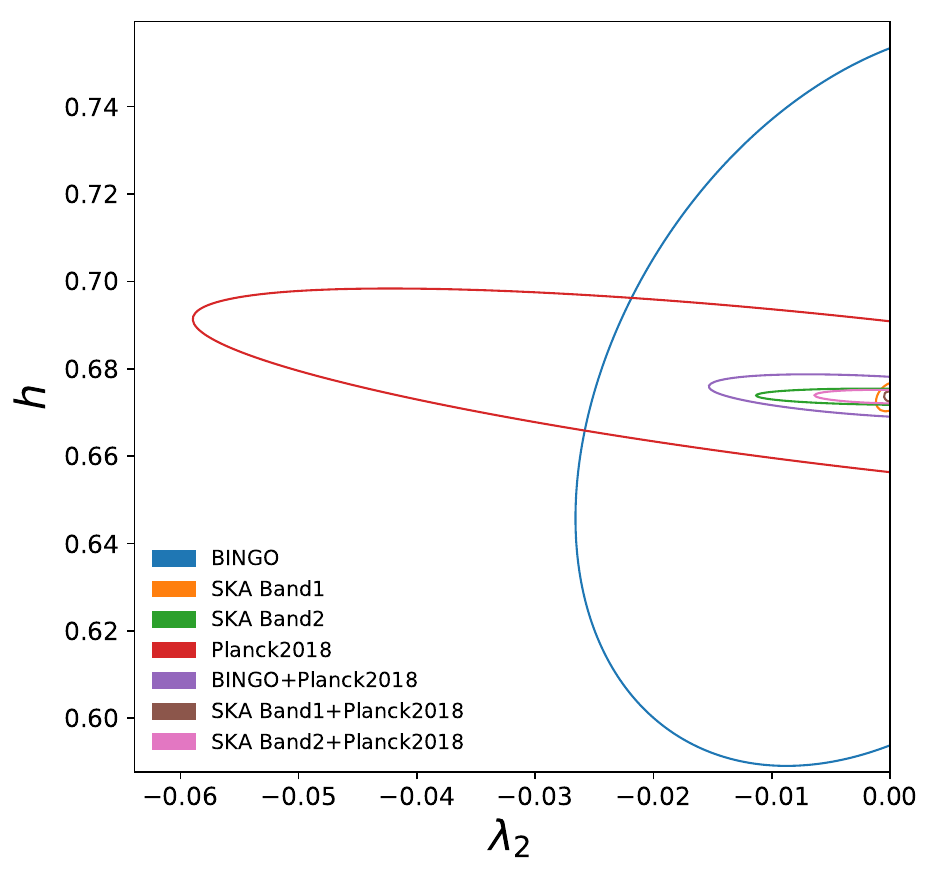}
	}
	\subfloat[]{\label{fig.IDE1_w_lam}
		\includegraphics[width=0.33\textwidth]{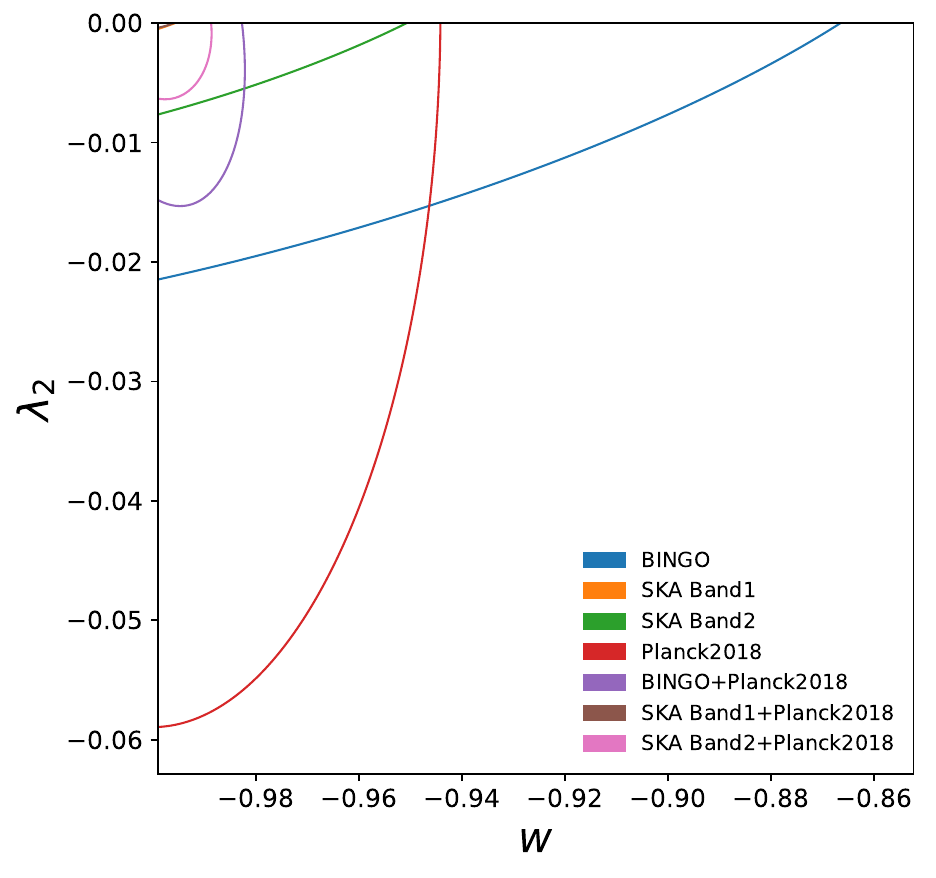}
	}
	\subfloat[]{\label{fig.IDE1_w_h}
		\includegraphics[width=0.33\textwidth]{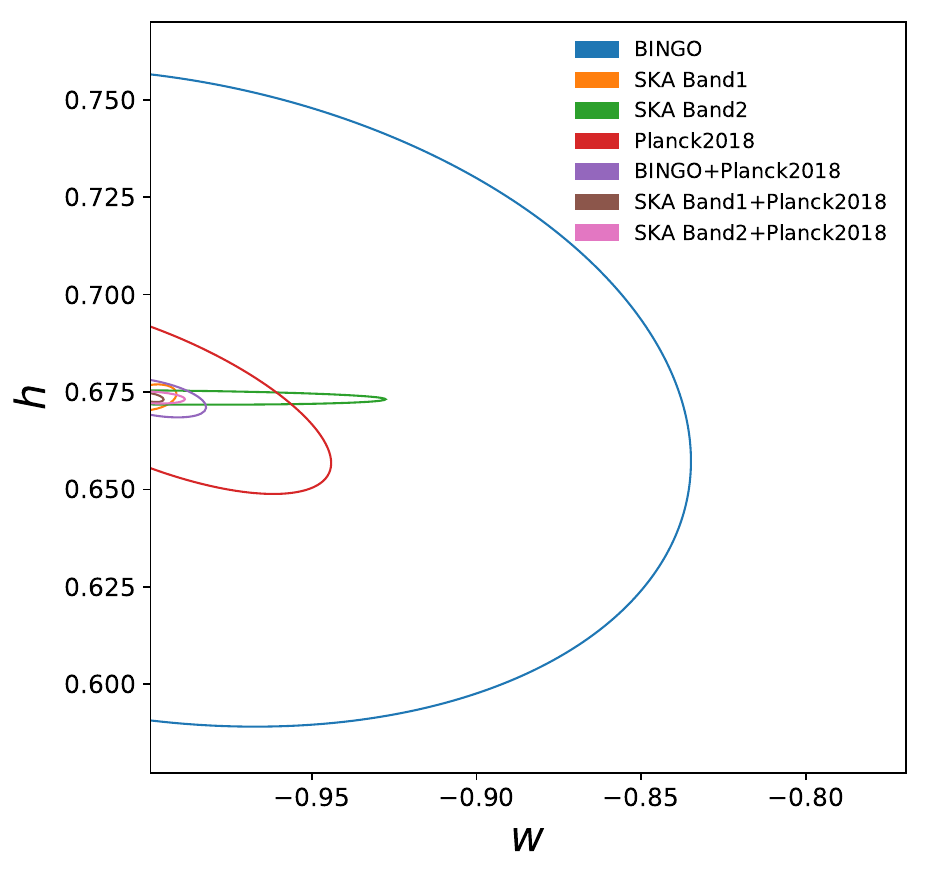}
	}	
	\caption{\label{fig.IDE1_2D_1D} The forecasted 2D distributions for $\lambda_2$, $w$ and $h$ in case of Model I. SKA1-MID shows its remarkable strength in parameter constraints and both HI IM projects have advantages on laying bounds to the interacting strength as well as $h$ over {\it Planck} 2018.}
\end{figure*}

We present the forecasted distributions for the three parameters more related to low-redshift measurements, $\lambda_2, w$ and $h$, within Model I in Fig.~\ref{fig.IDE1_2D_1D}, and a complete result summary at $1\sigma$ confidence level is found in Table~\ref{tab.IDE1}. Although our constraints should be Gaussian distributed around their fiducial values, we cut off those areas not allowed by the stability conditions in the IDE models. Two conclusions are inferred from Fig.~\ref{fig.IDE1_2D_1D}: 1) the two SKA1-MID Bands are expected to have huge potential in constraining $\lambda_2$, $w$ and $h$, whose abilities are basically above the level of {\it Planck} 2018; 2) all three HI IM projects can lay tighter constraints on the interacting strength than the CMB measurement to date. This can be explained by the wide observational redshift range of HI IM projects, which can break the degeneracy between $w$ and $h$ resided in {\it Planck} data and improve the measurement of IDE. Nevertheless, Table~\ref{tab.IDE1} shows that {\it Planck} 2018 still holds advantage on restricting early-Universe parameters, i.e., $A_s$ and $n_s$. In practice, the projected $1\sigma$ uncertainties from SKA1-MID Band\,1 are substantially the same order as those of {\it Planck} 2018 from~\cite{Bachega:2019fki}, whose advantage compared to Band\,2 is less than one order of magnitude, whereas BINGO is $\sim 1$ order of magnitude below. Another sparkle worth mentioning is the strict bound on $h$ laid by SKA1-MID Band\,2, outperforming all other projects considered here. Besides, we readily notice that all three HI IM surveys perform better than, or at least as good as, {\it Planck} 2018 does on constraining $\Omega_{\mathrm{c}}h^2$.

Fixing the binning scheme ($\delta \nu = 8.75$\,MHz), it is not hard to understand the performance differences between the three HI IM surveys. Both the 21-cm signal and shot noise are determined by the given cosmology and redshift range, the remaining contributions to the projected uncertainties are the thermal noise characterized by $\sigma_{\rm T}$ in Eq.~(\ref{eq:radiometer}) and the beam resolution in Eq.~(\ref{eq.B_l}). BINGO suffers from more thermal noise than SKA1-MID Band\,2 given its higher $T_{\rm sys}$ matched with lower $n_{\rm d}\times n_{\rm{beam}}$ and $\theta_{\rm FWHM}$, albeit across a similar observational redshift range and reduced sky coverage. Compared with Band\,1, Band\,2 encounters less thermal noise owing to its good control of $T_{\rm sys}$ and $\sigma_{\rm T}$. However, a higher shot noise level, arisen by the low redshifts, erodes its potential in measurements. Besides reducing the noise level, more cosmological information can be extracted by increasing the tomographic samples, namely a larger $N_{\rm bin}$. In this regard, Band\,1 has potential to become a sensation among these three projects.

Although SKA1-MID Band\,1 alone can provide better constraints than {\it Planck} 2018 on several parameters, combining multiple observations together can further improve the measurements. Adding the inverse of the {\it Planck} covariance matrix taken from {\bf \textsc{CosmoMC}} for Model I into the IM Fisher matrix analysis, the poor constraint by BINGO alone on $\Omega_{\mathrm{b}}h^2$ with $\approx 49.17\%$ accuracy is significantly improved to the level of $\approx 0.58\%$, and the bound on $\lambda_2$ is also narrowed by a factor of $1.75$. The improvements in SKA constraints are not as pronounced as for BINGO, however, they also worth attention. For example, the constraint on $h$ is further improved from $\approx 0.65\%$ to $\approx 0.24\%$ for Band\,1, whilst the uncertainty of $\lambda_2$ in Band\,2 is upgraded by a factor of $1.79$. The optimal constraints are laid by Band\,1+Band\,2+{\it Planck}\footnote{Band\,1 \& 2 in practise are not thoroughly independent, since there is a small overlap in redshift range, we however ignore this effect in our analysis.}, albeit a mild improvement relative to Band\,1+{\it Planck}. All projected $1\sigma$ uncertainties for the cosmological parameters in Model I are summarized in Table~\ref{tab.IDE1}.

\begin{table*}
	\centering
	\begin{tabular}{l|l|l|l|l|l|l|l|l}
		\hline
		\hline
		Parameters & $\Omega_{\mathrm{b}}h^2$ & $\Omega_{\mathrm{c}}h^2$ & $w$ & $\ln (10^{10}A_s)$ & $n_s$ & $\lambda_2$ & $h$ & $b_{\rm HI}$ \\
		&  [0.02237] & [0.12] & [-0.999] & [3.044] & [0.9649] & [0.00] & [0.6736] & [1.00]  \\
		\hline
		BINGO alone  & $\pm0.011$  & $\pm0.053$  & $\pm0.22$ & $\pm0.73$ & $\pm0.12$ & $\pm0.035$ & $\pm0.11$ & $\pm0.066$ \\ \hline
		SKA B\,1 alone  & $\pm0.00069$  & $\pm0.0029$  & $\pm0.010$ & $\pm0.036$ & $\pm0.011$ & $\pm0.0016$ & $\pm0.0044$ & $\pm0.0075$ \\ \hline
		SKA B\,2 alone & $\pm0.0015$ & $\pm0.0084$ & $\pm0.094$ & $\pm0.12$ & $\pm0.017$ & $\pm0.015$ & $\pm0.0025$ & $\pm0.026$ \\ \hline
		{\it Planck} & $\pm0.00015$ & $\pm0.033$  & $\pm0.072$  & $\pm0.016$ & $\pm0.0045$ &   $\pm0.078$ & $\pm0.033$ & $\dots$ \\ \hline
		BINGO+{\it Planck}  & $\pm0.00013$ & $\pm0.0079$ & $\pm0.022$ & $\pm0.016$ & $\pm0.0039$& $\pm0.020$ & $\pm0.0067$ & $\pm0.031$ \\ \hline
		SKA B\,1+{\it Planck}  & $\pm0.00012$ & $\pm0.00082$ & $\pm0.0052$ & $\pm0.012$ & $\pm0.0025$& $\pm0.00068$ & $\pm0.0016$ & $\pm0.0065$ \\ \hline
		SKA B\,2+{\it Planck} & $\pm0.00013$ & $\pm0.0037$  & $\pm0.014$ & $\pm0.016$ & $\pm0.0035$   & $\pm0.0084$ & $\pm0.0021$ & $\pm0.013$ \\ \hline
		SKA B\,1+SKA B\,2+{\it Planck} & $\pm0.00011$ & $\pm0.00071$  & $\pm0.0046$ & $\pm0.011$ & $\pm0.0022$   & $\pm0.00065$ & $\pm0.0013$ & $\pm0.0058$ \\ \hline
		\hline
	\end{tabular}
	\caption{The projected $1\sigma$ uncertainties for Model I from BINGO, SKA1-MID Band\,1 and Band\,2, respectively, via the Fisher matrix forecast and {\it Planck} 2018 MCMC. Also their joint results by adding up each Fisher matrices. The square brackets in the {\it 1st row} are the parameter fiducial values declared in Sect.~\ref{sec.intro}.}
	\label{tab.IDE1}
\end{table*}

\begin{figure*}
	\subfloat[]{\label{fig.IDE2_lam2_h}
		\includegraphics[width=0.33\textwidth]{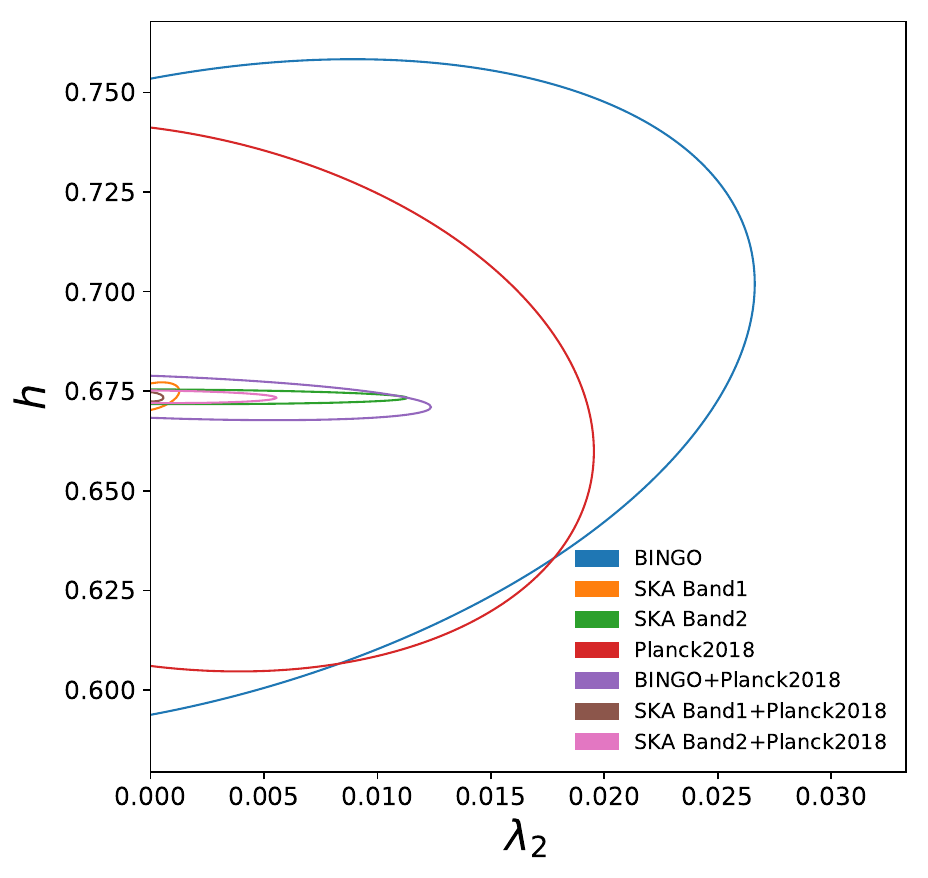}
	}
	\subfloat[]{\label{fig.IDE2_w_lam2}
		\includegraphics[width=0.33\textwidth]{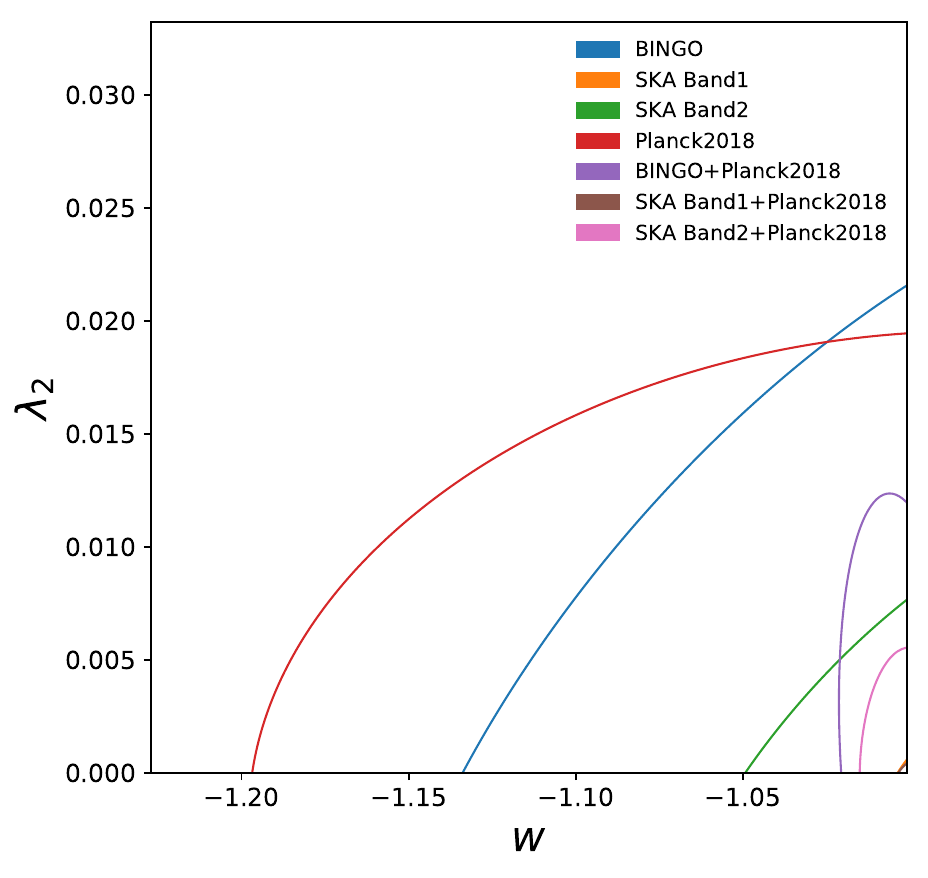}
	}
	\subfloat[]{\label{fig.IDE2_w_h}
		\includegraphics[width=0.33\textwidth]{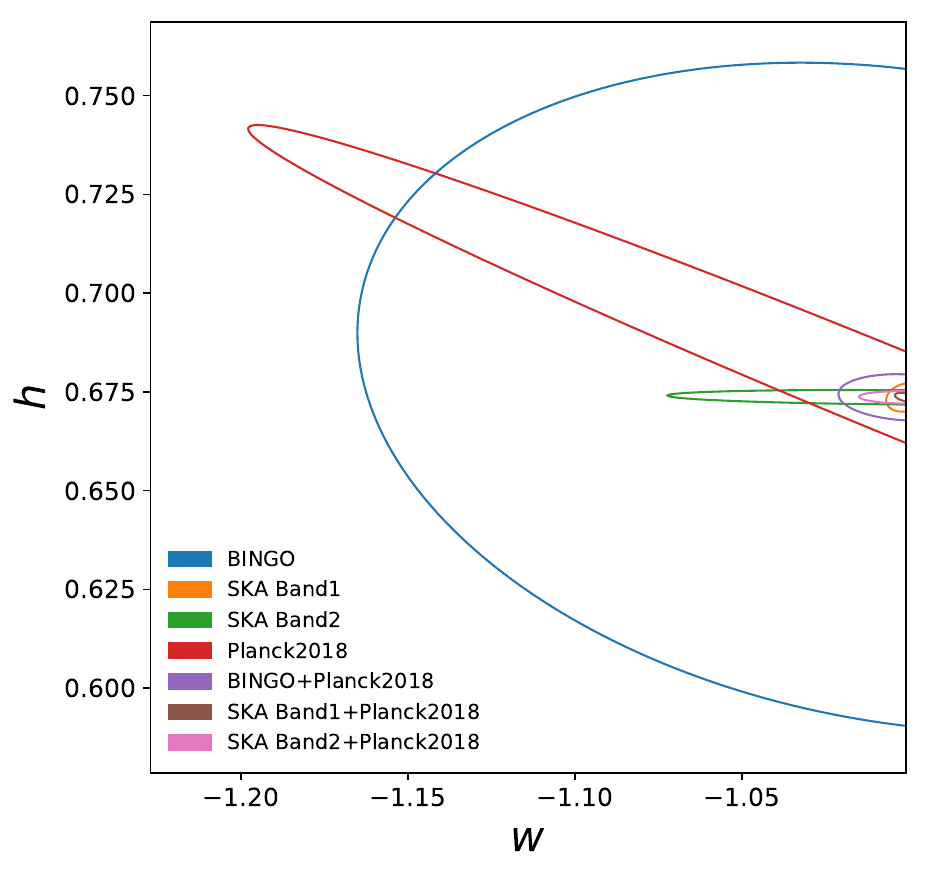}
	}
	\caption{\label{fig.IDE2_2D_1D} Same as Fig.~\ref{fig.IDE1_2D_1D}, but for Model II.}
\end{figure*}

Similar analyses have been carried out for Model II $\sim$ IV, the complete results are summarized in Table~\ref{tab.IDE2} $\sim$ \ref{tab.IDE4} and Fig.~\ref{fig.IDE2_2D_1D} $\sim$ \ref{fig.IDE4_2D_1D} illustrate the resulting contours. In principle, we can infer that the ability to constrain any of the four IDE scenarios with one of the three HI IM projects alone are on the same level, except for discrepancies in the interacting strengths and EoS. In terms of our Fisher matrix analyses, this is a natural result since we get derivatives with respect to one parameter by fixing others to their fiducial values, while for the interacting strengths, the discrepancies are directly attributed to differences in the $Q$ terms in those IDE scenarios. In addition, the strong degeneracy between the EoS and the interacting parameter will affect mostly the constraints in those two parameters. After a horizontal comparison, we further perceive that in terms of those predicted uncertainties by HI IM projects, Model I is highly in line with Model II, whereas Model III well resembles Model IV. This is consistent with our qualitative analysis aforementioned.

We also compare the results from the HI IM surveys with those from {\it Planck}. Although {\it Planck} data set is better in put constraints on $A_s$ and $n_s$, SKA1-MID provides much better results on $w$ and $h$, whereas BINGO yields similar values. In particular, the projected uncertainties on $w$ and $h$ for Model II with SKA1-MID Band\,2 alone are, respectively, 2.77 and 36.4 times smaller than with {\it Planck}. In the case of Model III (Model IV), those differences are increased to 2.93 (2.79) times for $w$ and 45.83 (45.83) times for $h$. SKA1-MID Band\,1 provides even superior results on $w$ but two-times larger uncertainty on $h$ relative to Band\,2. Whilst BINGO is not as impressive as SKA is, its performance is fairly close to {\it Planck}'s but better on determining $w$ in Model III and IV. Nevertheless, {\it Planck} puts better constraints in the interacting strength than BINGO in Model II $\sim$ IV and even SKA1-MID Band\,2 in Model III and IV.

After assessing the ability of a single observation in parameter constraints, we redo the joint analysis as we did for Model I. The results are clearly summarized in the lower halves of Table~\ref{tab.IDE2} $\sim$ \ref{tab.IDE4}. As expected the resulting constraints are better. Another key point we want to reiterate here is the intrinsic difference between Model II and Model III in their physical backgrounds, manifesting in two distinct constraints on $\lambda_2$ and $\lambda_1$. As aforementioned in Sect.~\ref{sec.analyses} that, $Q \propto \rho_{c}$ can easily reduce the signal to a lower level (demonstrated by Fig.~\ref{fig.IDE4_w} $\sim$~\ref{fig.IDE4_lam} as a good alternative, thanks to the similarity between Model III \& IV), which presumably lies in the fact that DM is far beyond DE in the time span of domination. Such inherent character of IDE models are also confirmed by other works, for example,~\cite{Costa:2019uvk} and~\cite{Bachega:2019fki}.

\begin{table*}
	\centering
	\begin{tabular}{l|l|l|l|l|l|l|l|l}
		\hline
		\hline
		Parameters & $\Omega_{\mathrm{b}}h^2$ & $\Omega_{\mathrm{c}}h^2$ & $w$ & $\ln (10^{10}A_s)$ & $n_s$ & $\lambda_2$ & $h$ & $b_{\rm HI}$ \\
		&  [0.02237] & [0.12] & [-1.001] & [3.044] & [0.9649] & [0.00] & [0.6736] & [1.00]  \\
		\hline
		BINGO alone  & $\pm0.011$  & $\pm0.053$  & $\pm0.22$ & $\pm0.73$ & $\pm0.12$ & $\pm0.035$ & $\pm0.11$ & $\pm0.066$ \\ \hline
		SKA B\,1 alone  & $\pm0.00066$  & $\pm0.0029$  & $\pm0.0079$ & $\pm0.037$ & $\pm0.011$ & $\pm0.0017$ & $\pm0.0048$ & $\pm0.0074$ \\ \hline
		SKA B\,2 alone & $\pm0.0015$ & $\pm0.0084$ & $\pm0.094$ & $\pm0.12$ & $\pm0.017$ & $\pm0.015$ & $\pm0.0025$ & $\pm0.026$ \\ \hline
		{\it Planck} & $\pm0.00015$ & $\pm0.0090$  & $\pm0.26$  & $\pm0.016$ & $\pm0.0043$ &   $\pm0.026$ & $\pm0.091$ & $\dots$ \\ \hline
		BINGO+{\it Planck}  & $\pm0.00014$ & $\pm0.0058$ & $\pm0.027$ & $\pm0.016$ & $\pm0.0039$& $\pm0.016$ & $\pm0.0077$ & $\pm0.030$ \\ \hline
		SKA B\,1+{\it Planck}  & $\pm0.00012$ & $\pm0.00079$ & $\pm0.0045$ & $\pm0.012$ & $\pm0.0026$& $\pm0.00075$ & $\pm0.0015$ & $\pm0.0064$ \\ \hline
		SKA B\,2+{\it Planck} & $\pm0.00013$ & $\pm0.0029$  & $\pm0.019$ & $\pm0.015$ & $\pm0.0033$   & $\pm0.0073$ & $\pm0.0021$ & $\pm0.013$ \\ \hline
		SKA B\,1+SKA B\,2+{\it Planck} & $\pm0.00012$ & $\pm0.00072$  & $\pm0.0041$ & $\pm0.011$ & $\pm0.0023$   & $\pm0.00073$ & $\pm0.0013$ & $\pm0.0057$ \\ \hline
		\hline
	\end{tabular}
	\caption{Same as the projected uncertainties listed in Table~\ref{tab.IDE1}, but for Model II.}
	\label{tab.IDE2}
\end{table*}

\begin{figure*}
	\subfloat[]{\label{fig.IDE3_lam1_h}
		\includegraphics[width=0.33\textwidth]{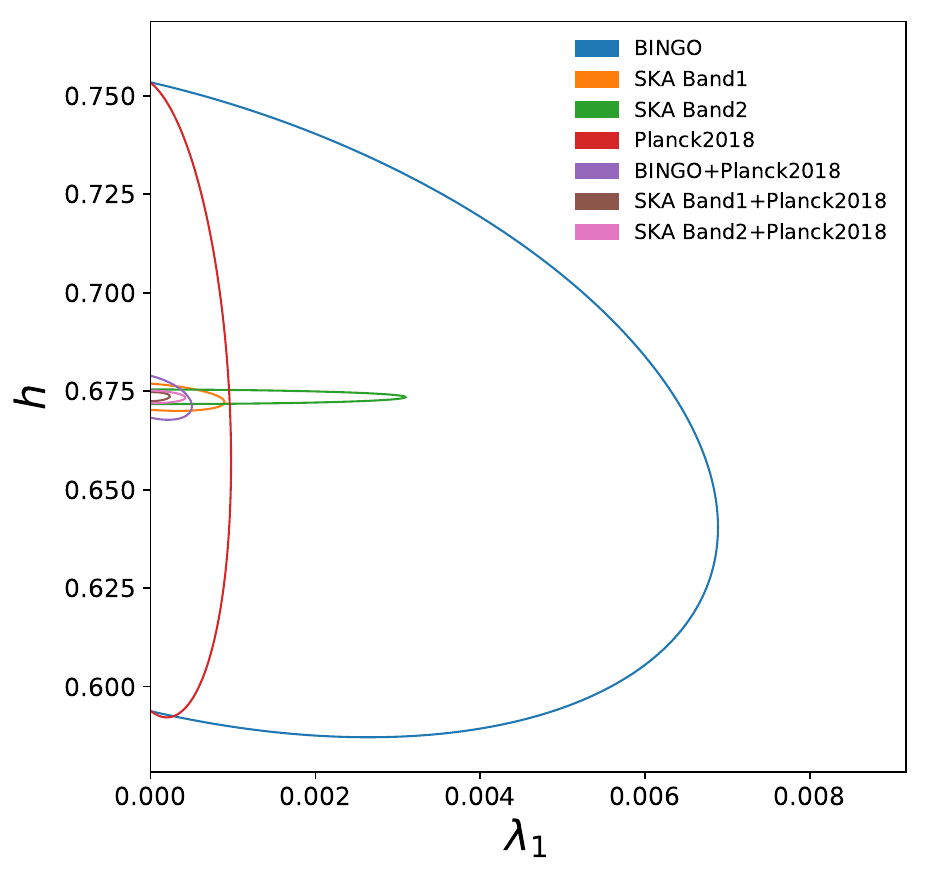}
	}
	\subfloat[]{\label{fig.IDE3_w_lam1}
		\includegraphics[width=0.33\textwidth]{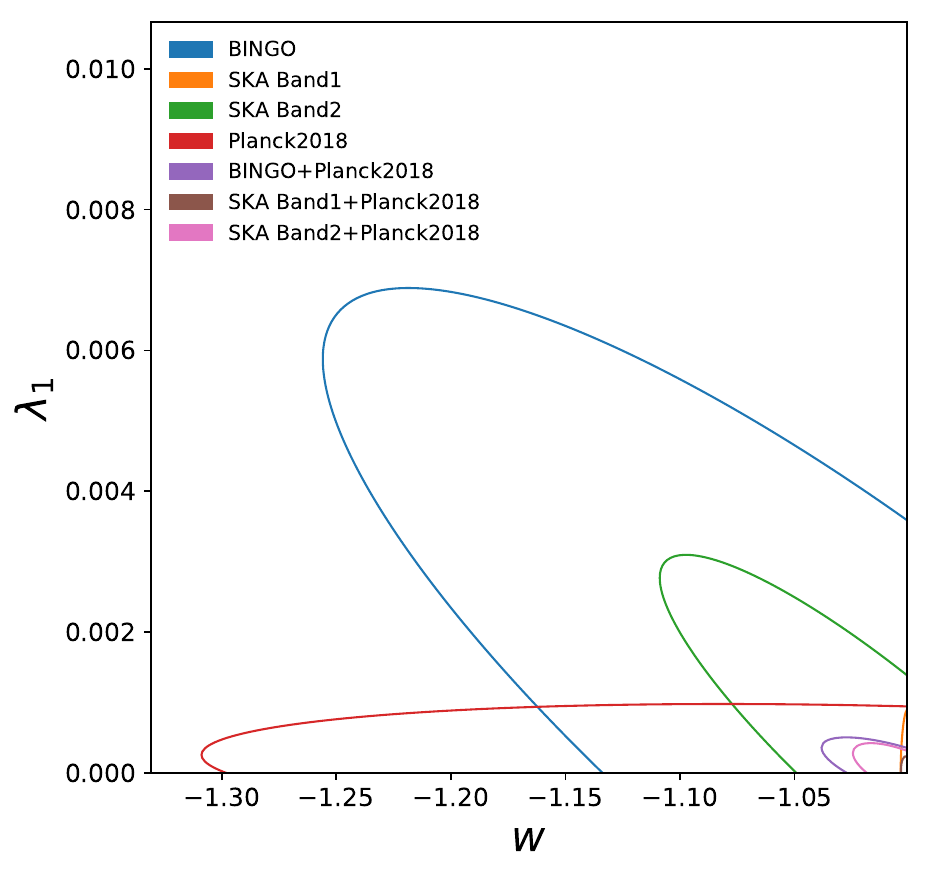}
	}
	\subfloat[]{\label{fig.IDE3_w_h}
		\includegraphics[width=0.33\textwidth]{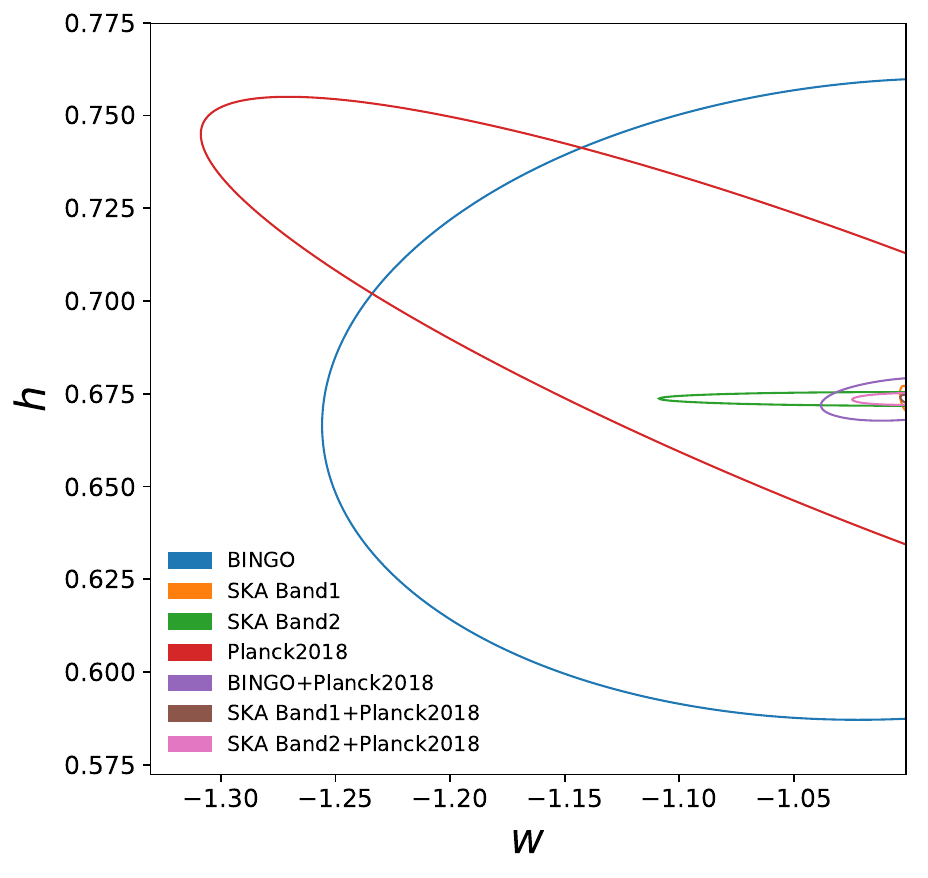}
	}
	\caption{\label{fig.IDE3_2D_1D} Same as Fig.~\ref{fig.IDE2_2D_1D}, but for Model III.}
\end{figure*}

\begin{table*}
	\centering
	\begin{tabular}{l|l|l|l|l|l|l|l|l}
		\hline
		\hline
		Parameters & $\Omega_{\mathrm{b}}h^2$ & $\Omega_{\mathrm{c}}h^2$ & $w$ & $\ln (10^{10}A_s)$ & $n_s$ & $\lambda_1$ & $h$ & $b_{\rm HI}$ \\
		&  [0.02237] & [0.12] & [-1.001] & [3.044] & [0.9649] & [0.00] & [0.6736] & [1.00]  \\
		\hline
		BINGO alone  & $\pm0.012$  & $\pm0.041$  & $\pm0.34$ & $\pm0.73$ & $\pm0.12$ & $\pm0.0091$ & $\pm0.11$ & $\pm0.12$ \\ \hline
		SKA B\,1 alone  & $\pm0.00036$  & $\pm0.0028$  & $\pm0.0036$ & $\pm0.057$ & $\pm0.013$ & $\pm0.0012$ & $\pm0.0047$ & $\pm0.0076$ \\ \hline
		SKA B\,2 alone & $\pm0.0015$ & $\pm0.0082$ & $\pm0.14$ & $\pm0.12$ & $\pm0.017$ & $\pm0.0041$ & $\pm0.0024$ & $\pm0.055$ \\ \hline
		{\it Planck} & $\pm0.00018$ & $\pm0.0036$  & $\pm0.41$  & $\pm0.016$ & $\pm0.0049$ &   $\pm0.0013$ & $\pm0.11$ & $\dots$ \\ \hline
		BINGO+{\it Planck}  & $\pm0.00017$ & $\pm0.0015$ & $\pm0.049$ & $\pm0.016$ & $\pm0.0039$& $\pm0.00066$ & $\pm0.0077$ & $\pm0.022$ \\ \hline
		SKA B\,1+{\it Planck}  & $\pm0.00012$ & $\pm0.00095$ & $\pm0.0034$ & $\pm0.013$ & $\pm0.0032$& $\pm0.00031$ & $\pm0.0015$ & $\pm0.0067$ \\ \hline
		SKA B\,2+{\it Planck} & $\pm0.00016$ & $\pm0.0012$  & $\pm0.031$ & $\pm0.015$ & $\pm0.0037$   & $\pm0.00055$ & $\pm0.0021$ & $\pm0.012$ \\ \hline
		SKA B\,1+SKA B\,2+{\it Planck} & $\pm0.00012$ & $\pm0.00085$  & $\pm0.0032$ & $\pm0.012$ & $\pm0.0027$   & $\pm0.00030$ & $\pm0.0013$ & $\pm0.0058$ \\ \hline
		\hline
	\end{tabular}
	\caption{Same as the projected uncertainties given in Table~\ref{tab.IDE2}, but for Model III.}
	\label{tab.IDE3}
\end{table*}

\begin{figure*}
	\subfloat[]{\label{fig.IDE4_lam_h}
		\includegraphics[width=0.33\textwidth]{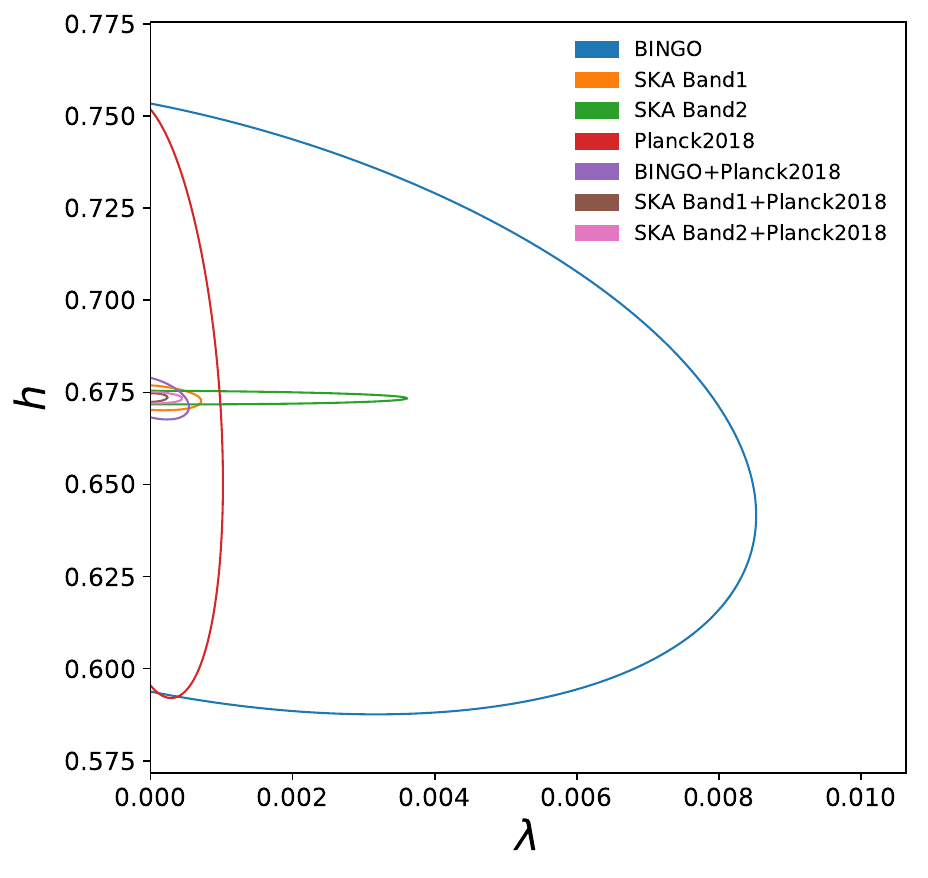}
	}
	\subfloat[]{\label{fig.IDE4_w_lam}
		\includegraphics[width=0.33\textwidth]{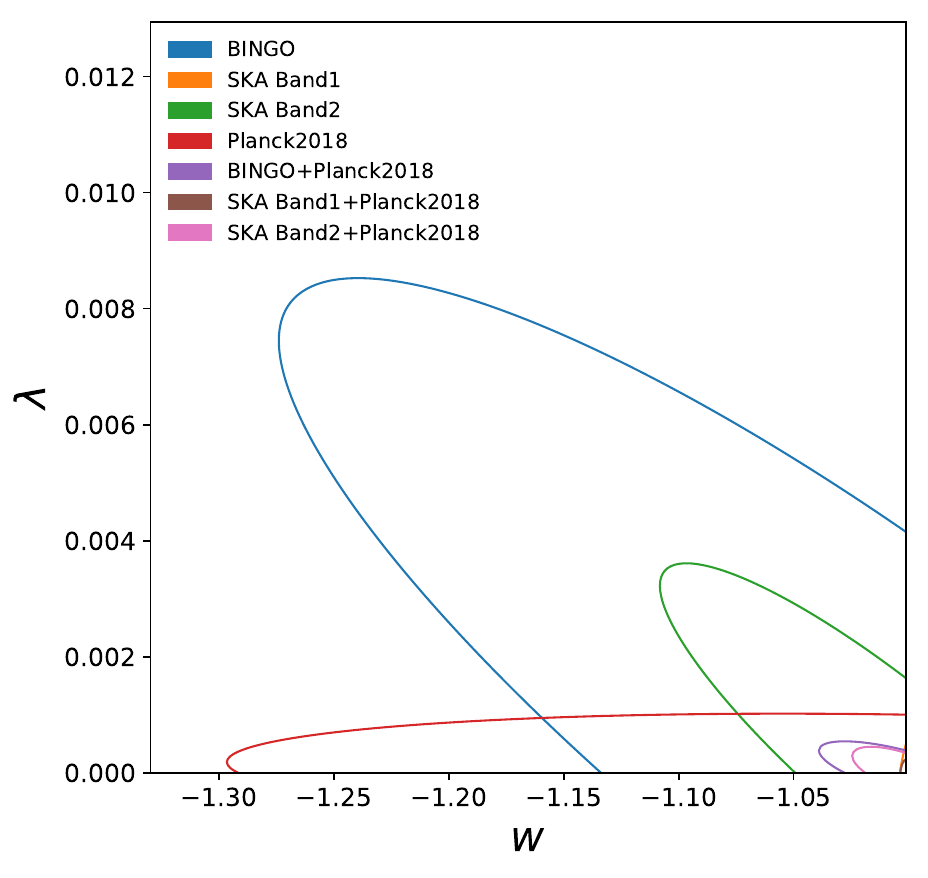}
	}
	\subfloat[]{\label{fig.IDE4_w_h}
		\includegraphics[width=0.33\textwidth]{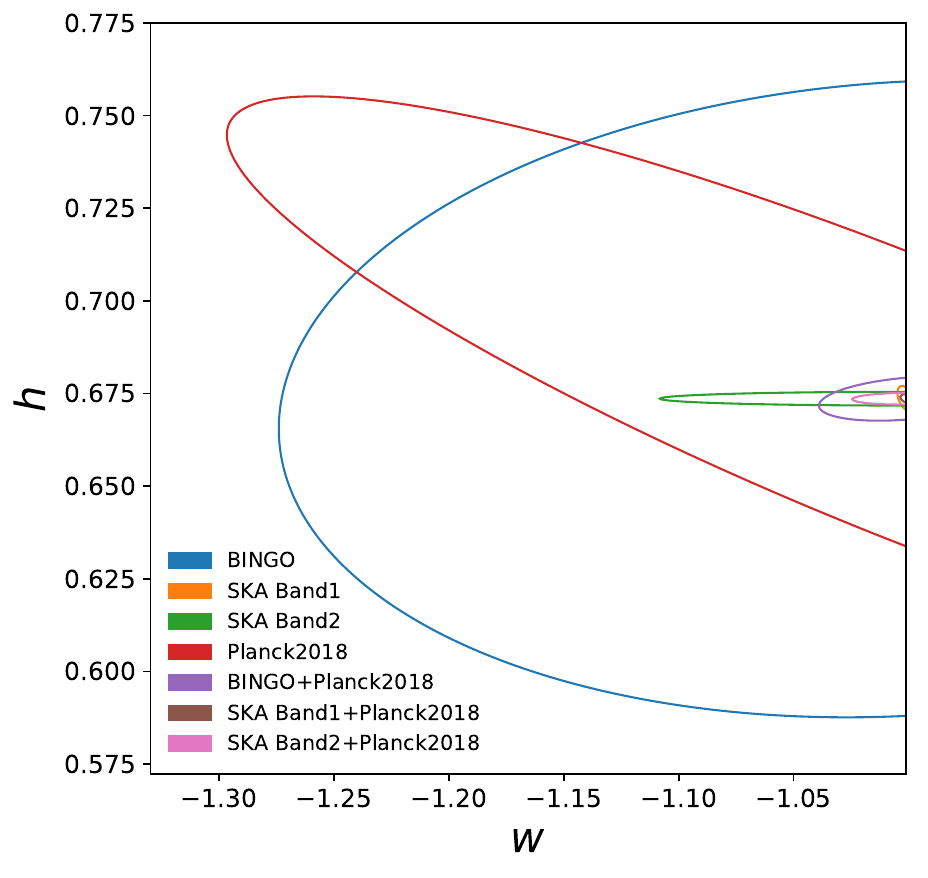}
	}
	\caption{\label{fig.IDE4_2D_1D} Same as Fig.~\ref{fig.IDE3_2D_1D}, but for Model IV.}
\end{figure*}

\begin{table*}
	\centering
	\begin{tabular}{l|l|l|l|l|l|l|l|l}
		\hline
		\hline
		Parameters & $\Omega_{\mathrm{b}}h^2$ & $\Omega_{\mathrm{c}}h^2$ & $w$ & $\log (10^{10}A_s)$ & $n_s$ & $\lambda$ & $h$ & $b_{\rm HI}$ \\
		&  [0.02237] & [0.12] & [-1.001] & [3.044] & [0.9649] & [0.00] & [0.6736] & [1.00]  \\
		\hline
		BINGO alone  & $\pm0.012$  & $\pm0.042$  & $\pm0.36$ & $\pm0.73$ & $\pm0.12$ & $\pm0.011$ & $\pm0.11$ & $\pm0.15$ \\ \hline
		SKA B\,1 alone  & $\pm0.00037$  & $\pm0.0030$  & $\pm0.0051$ & $\pm0.054$ & $\pm0.012$ & $\pm0.00094$ & $\pm0.0046$ & $\pm0.0076$ \\ \hline
		SKA B\,2 alone & $\pm0.0015$ & $\pm0.012$ & $\pm0.14$ & $\pm0.11$ & $\pm0.017$ & $\pm0.0048$ & $\pm0.0024$ & $\pm0.062$ \\ \hline
		{\it Planck} & $\pm0.00019$ & $\pm0.0040$  & $\pm0.39$  & $\pm0.017$ & $\pm0.0050$ &   $\pm0.0013$ & $\pm0.11$ & $\dots$ \\ \hline
		BINGO+{\it Planck}  & $\pm0.00017$ & $\pm0.0017$ & $\pm0.050$ & $\pm0.016$ & $\pm0.0040$& $\pm0.00071$ & $\pm0.0078$ & $\pm0.023$ \\ \hline
		SKA B\,1+{\it Planck}  & $\pm0.00013$ & $\pm0.0011$ & $\pm0.0034$ & $\pm0.013$ & $\pm0.0033$& $\pm0.00031$ & $\pm0.0015$ & $\pm0.0067$ \\ \hline
		SKA B\,2+{\it Planck} & $\pm0.00016$ & $\pm0.0014$  & $\pm0.031$ & $\pm0.016$ & $\pm0.0038$   & $\pm0.00059$ & $\pm0.0021$ & $\pm0.013$ \\ \hline
		SKA B\,1+SKA B\,2+{\it Planck} & $\pm0.00012$ & $\pm0.0010$  & $\pm0.0032$ & $\pm0.013$ & $\pm0.0028$   & $\pm0.00031$ & $\pm0.0013$ & $\pm0.0059$ \\ \hline
		\hline
	\end{tabular}
	\caption{Same as the projected uncertainties shown in Table~\ref{tab.IDE3}, but for Model IV.}
	\label{tab.IDE4}
\end{table*}

\subsection{Impacts of $N_{\rm bin}$ and RSD} \label{sec.extension}
Our forecast results presented before are based on a fixed channel bandwidth of $8.75$\,MHz and by including $\delta_n$ and RSD contributions to the signals. However, different binning schemes or contributions to the signals will indeed affect the projected uncertainties by altering the signal and noise level alone or simultaneously. In this subsection, we extend our discussion to the impacts of the number of frequency channels $N_{\rm bin}$ and RSD on the Fisher forecast. For simplicity, we merely focus on BINGO and one may turn to~\cite{Chen:2019jms} for a similar discussion on SKA.

\begin{figure*}
	\subfloat[]{\label{fig.IDE2_Nz}
		\includegraphics[width=0.51\textwidth]{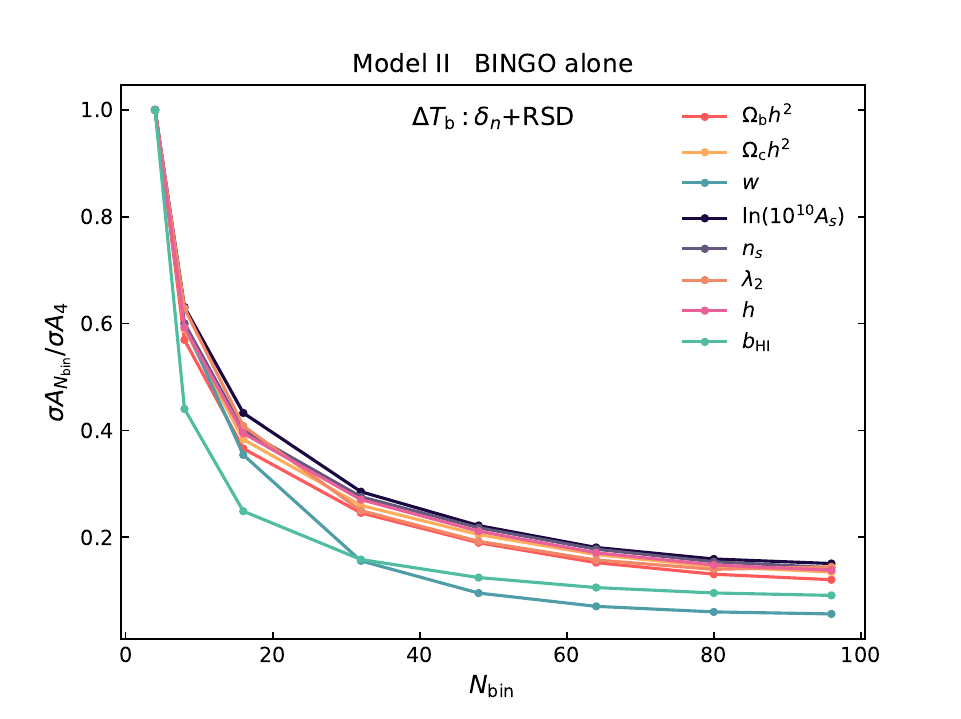}
	}
	\subfloat[]{\label{fig.IDE3_Nz}
		\includegraphics[width=0.51\textwidth]{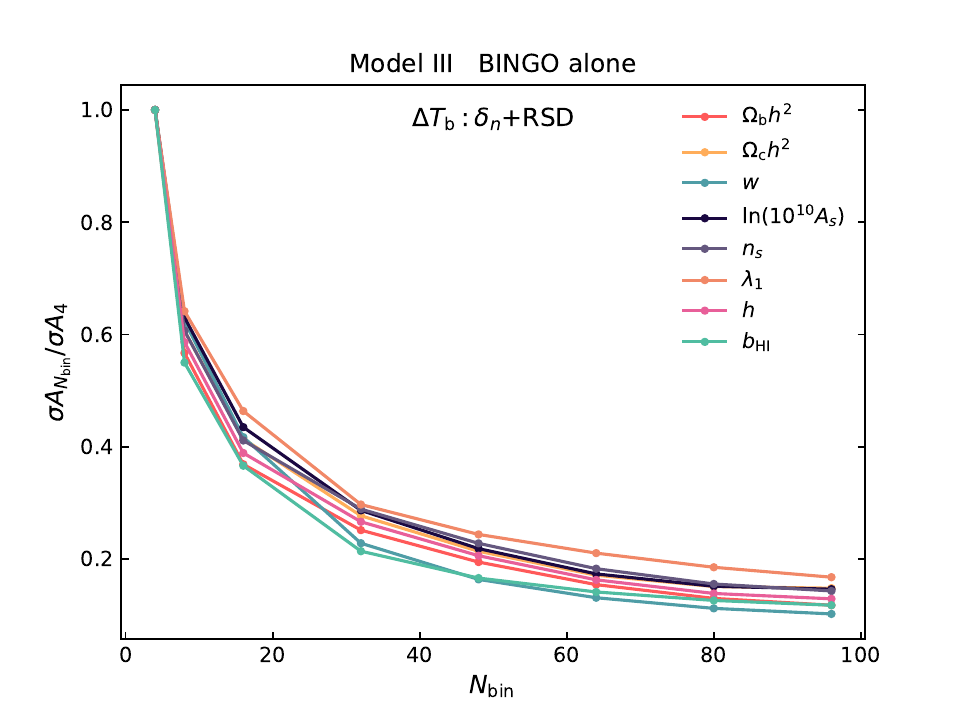}
	}
	\caption{\label{fig.Nz} The projected uncertainties from BINGO alone relative to those with $N_{\rm bin} = 4$ as a function of the number of frequency channels for IDE Model II \& III, respectively. The ratios shrink quickly as $N_{\rm bin}$ grows and flatten when $N_{\rm bin} \gtrsim 80$.}
\end{figure*}

A careful choice of binning scheme is of fundamental importance to a successful HI IM operation. Once we have specified the observed frequency range, the channel bandwidth $\Delta\nu$ is determined by the number of frequency channels $N_{\rm bin}$. As depicted in Fig.~\ref{fig.BINGO_noise}, both signal and noise levels get enhanced upon narrowing $\Delta\nu$ or equivalently by increasing $N_{\rm bin}$. This competing relationship raises a question to the existence of an optimal $\Delta\nu$ or $N_{\rm bin}$. On one hand, increased tomographic slices will accommodate more cosmological information, especially in time evolution and small-scale structures. On the other hand, the signal to noise ratio at $\ell \lesssim 200$ is, in fact, reduced due to higher increments in both shot and thermal noises. In an attempt to answer that question, we explore a wide range of $N_{\rm bin}$ from $4\sim96$ and their corresponding constraints on $\boldsymbol{\theta}$. We plot the ratios of these projected uncertainties relative to those with $N_{\rm bin} = 4$ in Fig.~\ref{fig.Nz} for Model II \& III, respectively, but neglect those for Model I \& IV due to their similarities. It is evident that the ratios shrink very quickly until $N_{\rm bin} \simeq 30$, and then they level off and eventually asymptotic to constants when $N_{\rm bin} \geq 80$, suggesting little or no additional information over there. Such kind of downward trend applies to every IDE scenario, without any exception. In this sense, setting $N_{\rm bin} = 32$ in this work as the fiducial value of BINGO configuration does not lose significant information and, furthermore, it meets the requirement of an efficient computation.

\begin{figure*}
	\subfloat[]{\label{fig.IDE2_RSD}
		\includegraphics[width=0.51\textwidth]{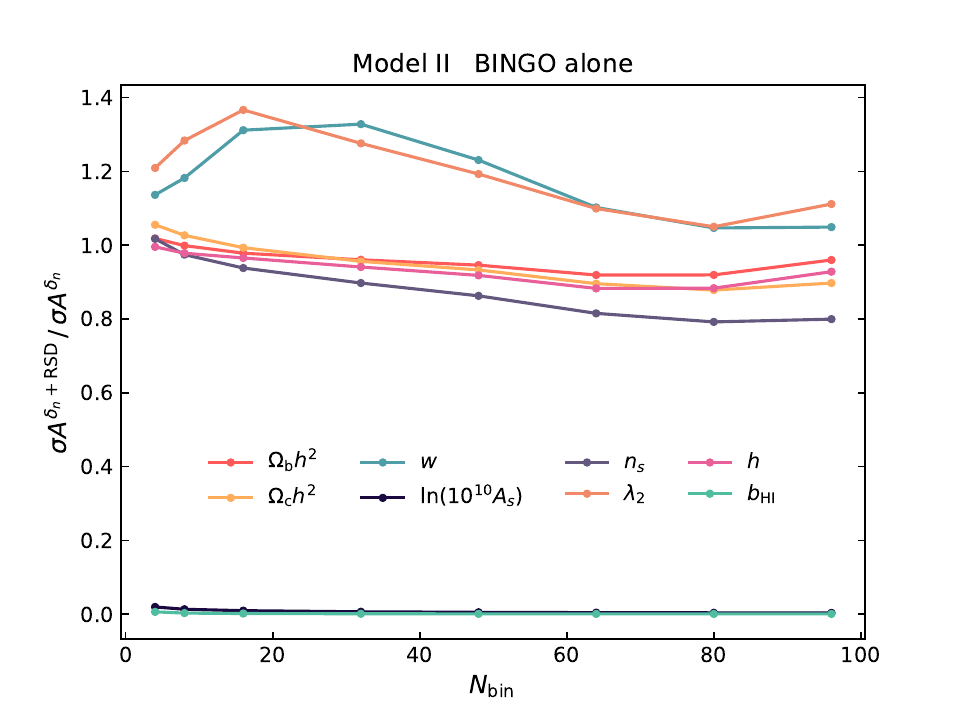}
	}
	\subfloat[]{\label{fig.IDE3_RSD}
		\includegraphics[width=0.51\textwidth]{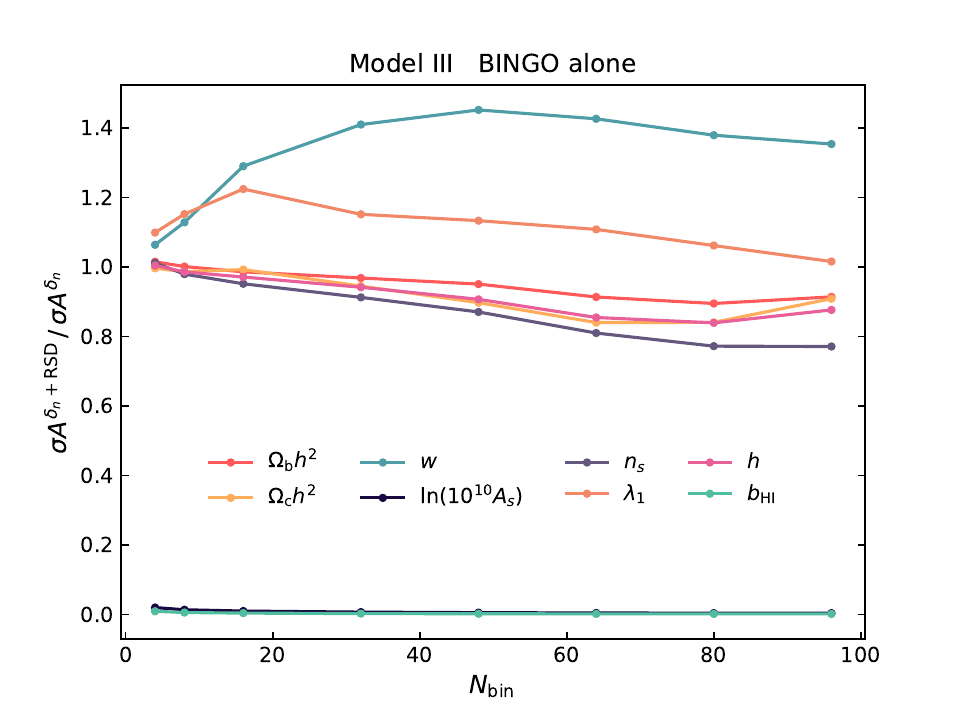}
	}
	\caption{\label{fig.RSD} The projected uncertainties under BINGO configurations for $\delta_n +$RSD relative to those with $\delta_n$ alone within a range of $N_{\rm bin}$ from $4 \sim 96$ for Model II \& III, respectively. The inclusion of RSD will basically facilitate the measurements of our parameters, albeit degrading the constraints on $w$ and the interacting strength.}
\end{figure*}

The term $b_{\rm HI}\delta^{\rm syn}$ in Eq.~(\ref{eq.bias}) manifests a complete degeneracy between $b_{\rm HI}$ and $A_s$, if we consider solely $\delta_n$ contributing to the 21-cm signal. A proper way to break such degeneracy is to include one or more other contributions, for example, the RSD component. As illustrated in Fig.~\ref{fig.IDE1_w} $\sim$ \ref{fig.IDE4_lam}, the RSD component deviates even more than the $\delta_n$ in affecting the HI angular power spectra, such that we do further expect it to tighten the parameters' constraints. In order to evaluate the degree to which RSD can improve the forecast constraints and its relationship to the binning scheme, we repeat the analysis carried out before for $N_{\rm bin}$, by depicting the ratios of projected uncertainties using the base angular spectra with $\delta_n$ + RSD relative to those without RSD for Model II \& III in Fig.~\ref{fig.RSD}. Again, Model I (IV) well resembles Model II (III) in the illustration here. Two lines at the bottom of each figure confirm that RSD can indeed break the degeneracy between $b_{\rm HI}$ and $A_s$. For the other parameters excluding DE EoS and the interacting strength, our results show that the participation of RSD is able to facilitate the measurements to a maximum amount of $\sim 20\%$. In contrast, the constraints on $w$ and $\lambda_{2}$ ($\lambda_{1}$) are fairly hindered by the accession of RSD, which is anchored on the degeneracy of these two parameters lurking in convolving the peculiar velocity of matter (i.e., the term $\lambda_{1} + \lambda_{2}/r$ in Eq.~(\ref{linear_pert_3})) to $\bar{T}_{\mathrm{b}}(z)$. 
Nonetheless, the downward trend of $\sigma_w$ and $\sigma_{\lambda_{2}}$ shown in Fig.~\ref{fig.IDE2_RSD} indicates the effectiveness of increasing $N_{\rm bin}$ partially in mitigating such impact, albeit a marginal reduction of $\sigma_w$ as we see in Fig.~\ref{fig.IDE3_RSD}. Although we can better recover the statistical properties of large scale structures with a thinner frequency bin, it is still worth emphasizing that, the amount of information laid in linear region is limited and the noise level also enhances with $N_{\rm bin}$. Consequently, if $N_{\rm bin}$ is up to 80 or higher, the information carried by a linear-modelled RSD is close to saturation, generating the requirement for a sophisticated approach to nonlinearity.

\subsection{Impacts of Foreground Residuals} \label{sec.residual}
Our fiducial Fisher analysis has no consideration of foreground residuals and, as mentioned in Sect.~\ref{sec.foreground}, an accurate interpretation of cosmological information from a 21-cm survey is subject to its foreground subtraction, indicating the necessity to scrutinise how foreground residuals will impact parameter determinations, especially in IDE scenarios. The analysis is carried out by varying the efficiency of foreground removal, $\epsilon_{\rm FG}$, in light of the foreground residual model, Eq.~(\ref{eq:cl_fg}). We vary $\epsilon_{\rm FG}$ from $10^{-2}$ to $10^{-6}$ and demonstrate the projected uncertainties by calculating their ratios to the fiducial predictions shown before.

\begin{figure*}
	\subfloat[]{\label{fig.IDE2_FG}
		\includegraphics[width=0.51\textwidth]{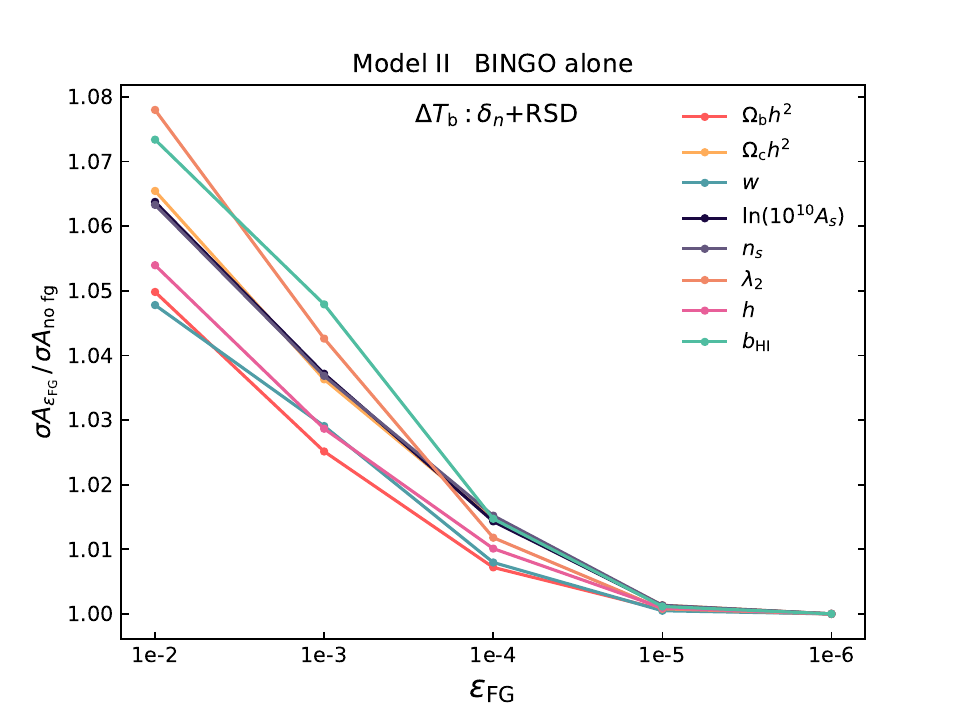}
	}
	\subfloat[]{\label{fig.IDE3_FG}
		\includegraphics[width=0.51\textwidth]{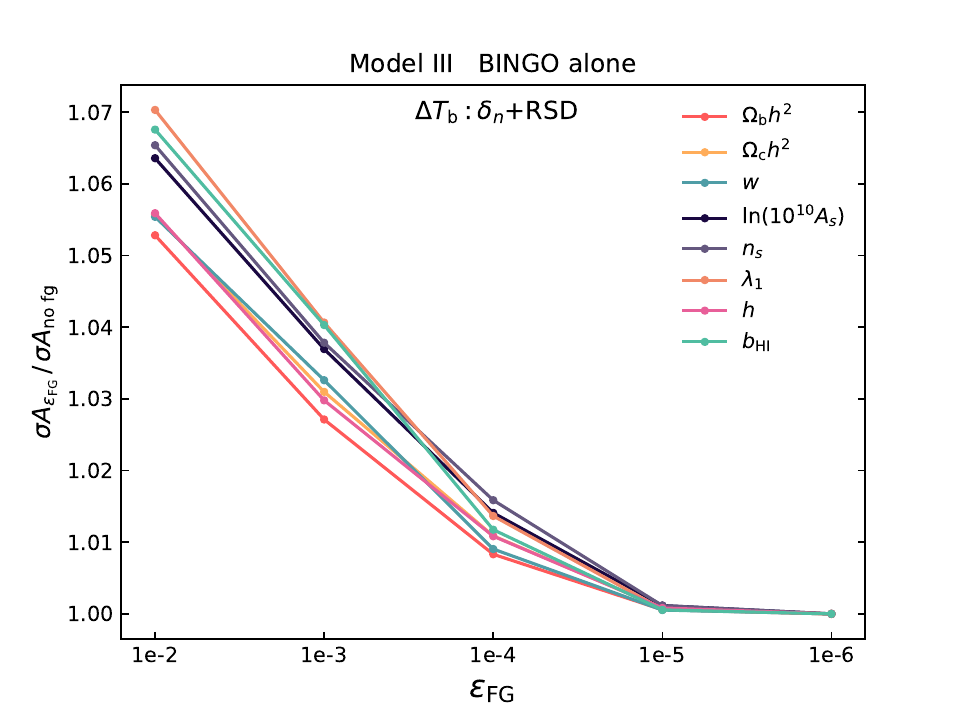}
	}
	\caption{\label{fig.FG} The projected uncertainties for $\delta_n +$RSD laid by BINGO in the presence of foreground residuals relative to those with the fiducial configuration as a function of foreground removal efficiency $\epsilon_{\rm FG}$, respectively for Model II \& III. Those Gaussian-distributed assumed residuals are able to degrade parameter determinations by a level of $\gtrsim 5\%$ when $\epsilon_{\rm FG} \geq 10^{-3}$, while for $\epsilon_{\rm FG} \leq 10^{-5}$ the interference is negligible, demonstrating the effectiveness of foreground subtraction.}
\end{figure*}

In Fig.~\ref{fig.FG}, we can explicitly figure out that the interference coming from foreground residuals act almost fairly to both scenarios of $Q \propto \rho_d$ and $\propto \rho_c$. Residuals modeled by $\epsilon_{\rm FG} = 10^{-2}$ in Eq.~(\ref{eq:cl_fg}) result in an extra uncertainty on the parameter forecasts which are $\lesssim 10\%$ with respect to the fiducial BINGO configuration and, in line with our expectation, such uncertainty decreases with a higher foreground removal efficiency, and is finally asymptotic to the optimistic situation of no foreground contamination, where the threshold value characterizing an ideal foreground subtraction is $\epsilon_{\rm FG} \lesssim 10^{-5}$. In spite of such clear pattern, we must point out that our estimation of residual interference is for reference only owing to the rough modelling and, in reality, the foreground reduction is still a challenging technique as we have briefly mentioned in Sect.~\ref{sec.foreground}. Additionally, our calculation has demonstrated that the residual interference to the parameter determinations from the two SKA1-MID bands are rather close to the BINGO case and, to avoid redundancy, we skip over their illustrations.

\subsection{Impacts of $\Omega_{\mathrm{HI}}$} \label{sec.Omhi}
Previously we had fixed $\Omega_{\mathrm{HI}} = 6.2 \times 10^{-4}$  throughout the forecast. From the observational side, however, there is a considerable uncertainty on this HI fraction, both on its magnitude and redshift dependence. This parameter is completely degenerate with the primordial amplitude, $A_s$, as manifested by Eq.~(\ref{eq:Tb}) and~(\ref{eq:window}) $\sim$~(\ref{eq:Cl}) for the dimensional $C_\ell$. Such degeneracy is inherent to parameter measurements from 21-cm intensity mapping and can only be broken by joint analyses with observations on other tracers. To this end, the Fisher calculation we perform in this section is combined with a prior from {\it Planck} 2018 and our analysis is separated in two steps. First, we neglect the redshift evolution of HI fraction and solely consider its magnitude variation affecting the parameter forecasts, dubbed as Case I. Second, we go further accounting for both magnitude and redshift uncertainties and model the dependence on $\Omega_{\mathrm{HI}}$ by following~\citet{Crighton:2015pza}, $\Omega_{\mathrm{HI}}(z) = A(1+z)^{\gamma}$, which constrained those parameters as $A = (4.00 \pm 0.24) \times 10^{-4}$ and $\gamma = 0.60 \pm 0.05$ by fitting such power law to a compilation of $\Omega_{\mathrm{HI}}$ measurements residing in $0 \lesssim z \lesssim 5$. Rightfully, the typical value we take in the follow-up Fisher study is $A = 4.00 \times 10^{-4}$ and $\gamma = 0.60$, which are consistent with our fiducial choice, $\Omega_{\mathrm{HI}} = 6.2 \times 10^{-4}$, when $z \lesssim 1$. We abbreviate this second step analysis as Case II. Given the very tiny value of $\Omega_{\mathrm{HI}}\,(A)$, we will follow the same treatment as for $A_s$ and calculate the derivatives with respect to $\ln (10^{4}\Omega_{\mathrm{HI}})$ or $(\ln (10^{4}A))$.

We summarize the forecast results of Case I in Table~\ref{tab.Omhi_1}. Compared to Tables~\ref{tab.IDE1} $\sim$~\ref{tab.IDE4}, we observe that the variation of $\Omega_{\mathrm{HI}}$ will degrade all parameter measurements. Specifically, the maximum degradation appearing in Mode II is for $h$, which is $\sim 3$ times under SKA1-MID Band\,2 configuration, the next most sensitive parameter is $b_{\rm HI}$ in Model III constrained by BINGO, whose projected uncertainty is degraded by a factor of $1.77$. While for other parameters as $\Omega_{\mathrm{b}}h^2$ or $n_s$, the degradation is rather limited. These impacts can be explained from two aspects: on one hand, a new degree of freedom raised by $\Omega_{\mathrm{HI}}$ joining in the Fisher matrix should be in charge of a global degradation; On the other hand, some parameters are mostly constrained by {\it Planck} data and will not change significantly with that additional HI parameter. Our results demonstrate a bright prospect for HI IM + {\it Planck} 2018 in measuring the HI fraction, yielding an uncertainty of $\sim 5.2\%$ for BINGO and $\lesssim 1.5\%$ for SKA1-MID\footnote{Notice our percentage error of the HI fraction is shown with respect to $\Omega_{\mathrm{HI}}\,(A)$ but not to its ln correspondence.}.

The predicted $1\sigma$ constraints for Case II are entirely collected in Table~\ref{tab.Omhi_2}. As anticipated, a comprehensive modelling of $\Omega_{\mathrm{HI}}$ by considering its time evolution will further deteriorate the forecast results. Aside from a new degree of freedom, the participation of $\gamma$ will introduce an additional degeneracy to $A$, inducing a maximum $\sigma_h$ of $\sim 3.8$ times larger than our fiducial constraint by SKA1-MID Band\,2+{\it Planck} laid in Model II. Also, the maximum degradation to $\sigma_w$ is around a factor of $2.5$ for each IDE scenario. The price for introducing $\gamma$ in bounding $A$ is not considerable yet, as the uncertainty on $A$ is of $\lesssim 8.0\%$ for BINGO+{\it Planck} and $\lesssim 1.9\%$ for SKA1-MID+{\it Planck}, slightly larger than what we see in Table~\ref{tab.Omhi_1}. In general, SKA1-MID Band\,1+{\it Planck} outperforms the other two in constraining $\sigma_{\gamma}$ to a level of $1.1\%$, whereas Band\,2+{\it Planck} has advantage in restricting the magnitude of $\Omega_{\mathrm{HI}}$ no matter in Case I or II. In addition, the constraints predicted by Band\,1+{\it Planck} are less sensitive to the inclusion of $\gamma$ in relative to others, and of course, the optimal bounds are reported by combining {\it Planck} with two SKA1-MID bands, especially on $A$ and $\gamma$ at a level of $\lesssim 0.46\%$ and $\lesssim 0.83\%$, respectively.

\begin{table*}
	\centering
	\begin{tabular}{l|l|l|l|l|l|l|l|l|l}
		\hline
		\hline
		Model I & $\Omega_{\mathrm{b}}h^2$ & $\Omega_{\mathrm{c}}h^2$ & $w$ & $\ln (10^{10}A_s)$ & $n_s$ & $\lambda_2$ & $h$ & $b_{\rm HI}$ & $\ln (10^{4}\Omega_{\mathrm{HI}})$ \\
		&  [0.02237] & [0.12] & [-0.999] & [3.044] & [0.9649] & [0.00] & [0.6736] & [1.00] & [1.825]  \\
		\hline
		BINGO+{\it Planck}  & $\pm0.00014$ & $\pm0.0079$ & $\pm0.038$ & $\pm0.016$ & $\pm0.0040$& $\pm0.021$ & $\pm0.010$ & $\pm0.035$ & $\pm0.051$ \\ \hline
		SKA B\,1+{\it Planck}  & $\pm0.00012$ & $\pm0.00084$ & $\pm0.0058$ & $\pm0.016$ & $\pm0.0025$& $\pm0.00070$ & $\pm0.0019$ & $\pm0.0074$ & $\pm0.012$ \\ \hline
		SKA B\,2+{\it Planck} & $\pm0.00013$ & $\pm0.0041$  & $\pm0.017$ & $\pm0.016$ & $\pm0.0036$   & $\pm0.010$ & $\pm0.0058$ & $\pm0.018$ & $\pm0.0092$ \\ \hline
		SKA B\,1+SKA B\,2+{\it Planck} & $\pm0.00011$ & $\pm0.00077$  & $\pm0.0049$ & $\pm0.011$ & $\pm0.0022$   & $\pm0.00066$ & $\pm0.0018$ & $\pm0.0058$ & $\pm0.0041$ \\ \hline
		\hline
		\multicolumn{10}{c}{} \\
		\hline
		\hline
		Model II & $\Omega_{\mathrm{b}}h^2$ & $\Omega_{\mathrm{c}}h^2$ & $w$ & $\ln (10^{10}A_s)$ & $n_s$ & $\lambda_2$ & $h$ & $b_{\rm HI}$ & $\ln (10^{4}\Omega_{\mathrm{HI}})$ \\
		&  [0.02237] & [0.12] & [-1.001] & [3.044] & [0.9649] & [0.00] & [0.6736] & [1.00] & [1.825]  \\
		\hline
		BINGO+{\it Planck}  & $\pm0.00014$ & $\pm0.0059$ & $\pm0.042$ & $\pm0.016$ & $\pm0.0040$& $\pm0.017$ & $\pm0.011$ & $\pm0.034$ & $\pm0.048$ \\ \hline
		SKA B\,1+{\it Planck}  & $\pm0.00012$ & $\pm0.00082$ & $\pm0.0049$ & $\pm0.016$ & $\pm0.0027$& $\pm0.00077$ & $\pm0.0019$ & $\pm0.0074$ & $\pm0.012$ \\ \hline
		SKA B\,2+{\it Planck} & $\pm0.00013$ & $\pm0.0035$  & $\pm0.020$ & $\pm0.016$ & $\pm0.0035$   & $\pm0.0096$ & $\pm0.0064$ & $\pm0.017$ & $\pm0.0099$ \\ \hline
		SKA B\,1+SKA B\,2+{\it Planck} & $\pm0.00012$ & $\pm0.00078$  & $\pm0.0044$ & $\pm0.011$ & $\pm0.0023$   & $\pm0.00074$ & $\pm0.0017$ & $\pm0.0057$ & $\pm0.0041$ \\ \hline
		\hline
		\multicolumn{10}{c}{} \\
		\hline
		\hline
		Model III & $\Omega_{\mathrm{b}}h^2$ & $\Omega_{\mathrm{c}}h^2$ & $w$ & $\ln (10^{10}A_s)$ & $n_s$ & $\lambda_1$ & $h$ & $b_{\rm HI}$ & $\ln (10^{4}\Omega_{\mathrm{HI}})$ \\
		&  [0.02237] & [0.12] & [-1.001] & [3.044] & [0.9649] & [0.00] & [0.6736] & [1.00] & [1.825]  \\
		\hline
		BINGO+{\it Planck}  & $\pm0.00017$ & $\pm0.0018$ & $\pm0.084$ & $\pm0.016$ & $\pm0.0040$& $\pm0.00072$ & $\pm0.0085$ & $\pm0.039$ & $\pm0.055$ \\ \hline
		SKA B\,1+{\it Planck}  & $\pm0.00013$ & $\pm0.00096$ & $\pm0.0035$ & $\pm0.015$ & $\pm0.0036$& $\pm0.00037$ & $\pm0.0020$ & $\pm0.0074$ & $\pm0.014$ \\ \hline
		SKA B\,2+{\it Planck} & $\pm0.00016$ & $\pm0.0012$  & $\pm0.031$ & $\pm0.015$ & $\pm0.0037$   & $\pm0.00057$ & $\pm0.0045$ & $\pm0.013$ & $\pm0.0069$ \\ \hline
		SKA B\,1+SKA B\,2+{\it Planck} & $\pm0.00012$ & $\pm0.00094$  & $\pm0.0033$ & $\pm0.012$ & $\pm0.0028$   & $\pm0.00030$ & $\pm0.0017$ & $\pm0.0058$ & $\pm0.0040$ \\ \hline
		\hline
		\multicolumn{10}{c}{} \\
		\hline
		\hline
		Model IV & $\Omega_{\mathrm{b}}h^2$ & $\Omega_{\mathrm{c}}h^2$ & $w$ & $\ln (10^{10}A_s)$ & $n_s$ & $\lambda$ & $h$ & $b_{\rm HI}$ & $\ln (10^{4}\Omega_{\mathrm{HI}})$ \\
		&  [0.02237] & [0.12] & [-1.001] & [3.044] & [0.9649] & [0.00] & [0.6736] & [1.00] & [1.825]  \\
		\hline
		BINGO+{\it Planck}  & $\pm0.00017$ & $\pm0.0020$ & $\pm0.083$ & $\pm0.016$ & $\pm0.0041$& $\pm0.00077$ & $\pm0.0086$ & $\pm0.040$ & $\pm0.054$ \\ \hline
		SKA B\,1+{\it Planck}  & $\pm0.00013$ & $\pm0.0011$ & $\pm0.0039$ & $\pm0.016$ & $\pm0.0038$& $\pm0.00039$ & $\pm0.0021$ & $\pm0.0074$ & $\pm0.015$ \\ \hline
		SKA B\,2+{\it Planck} & $\pm0.00017$ & $\pm0.0014$  & $\pm0.032$ & $\pm0.016$ & $\pm0.0039$   & $\pm0.00062$ & $\pm0.0047$ & $\pm0.013$ & $\pm0.0071$ \\ \hline
		SKA B\,1+SKA B\,2+{\it Planck} & $\pm0.00012$ & $\pm0.0011$  & $\pm0.0033$ & $\pm0.013$ & $\pm0.0029$   & $\pm0.00031$ & $\pm0.0017$ & $\pm0.0059$ & $\pm0.0040$ \\ \hline
		\hline
	\end{tabular}
	\caption{The projected $1\sigma$ uncertainties from joint analyses of {\it Planck} 2018 $+$ HI IM surveys for four IDE scenarios, respectively. As expected, the inclusion of $\Omega_{\mathrm{HI}}$ to the Fisher analysis will degrade the projected constraints, affecting most on $h$ but least on $n_s$, in comparison to Table~\ref{tab.IDE1} $\sim$~\ref{tab.IDE4}. Considering to the tiny value of HI fraction, in our Fisher study the derivative in fact is with respect to $\ln (10^{4}\Omega_{\mathrm{HI}})$ other than $\Omega_{\mathrm{HI}}$, and thereby, the uncertainties listed in the last column are referring to the former.}
	\label{tab.Omhi_1}
\end{table*}

\begin{table*}
	\centering
	\begin{tabular}{l|l|l|l|l|l|l|l|l|l|l}
		\hline
		\hline
		Model I & $\Omega_{\mathrm{b}}h^2$ & $\Omega_{\mathrm{c}}h^2$ & $w$ & $\ln (10^{10}A_s)$ & $n_s$ & $\lambda_2$ & $h$ & $b_{\rm HI}$ & $\ln (10^{4}A)$ & $\gamma$ \\
		&  [0.02237] & [0.12] & [-0.999] & [3.044] & [0.9649] & [0.00] & [0.6736] & [1.00] & [1.386] & [0.60]  \\
		\hline
		BINGO+{\it Planck}  & $\pm0.00015$ & $\pm0.012$ & $\pm0.057$ & $\pm0.016$ & $\pm0.0043$& $\pm0.034$ & $\pm0.013$ & $\pm0.043$ & $\pm0.058$ & $\pm0.086$ \\ \hline
		SKA B\,1+{\it Planck}  & $\pm0.00012$ & $\pm0.00091$ & $\pm0.0058$ & $\pm0.016$ & $\pm0.0027$& $\pm0.00070$ & $\pm0.0021$ & $\pm0.0087$ & $\pm0.016$ & $\pm0.0068$ \\ \hline
		SKA B\,2+{\it Planck} & $\pm0.00013$ & $\pm0.0073$  & $\pm0.033$ & $\pm0.016$ & $\pm0.0037$   & $\pm0.020$ & $\pm0.0076$ & $\pm0.024$ & $\pm0.011$ & $\pm0.053$ \\ \hline
		SKA B\,1+SKA B\,2+{\it Planck} & $\pm0.00011$ & $\pm0.00082$  & $\pm0.0051$ & $\pm0.013$ & $\pm0.0024$   & $\pm0.00066$ & $\pm0.0018$ & $\pm0.0060$ & $\pm0.0046$ & $\pm0.0050$ \\ \hline
		\hline
		\multicolumn{11}{c}{} \\
		\hline
		\hline
		Model II & $\Omega_{\mathrm{b}}h^2$ & $\Omega_{\mathrm{c}}h^2$ & $w$ & $\ln (10^{10}A_s)$ & $n_s$ & $\lambda_2$ & $h$ & $b_{\rm HI}$ & $\ln (10^{4}A)$ & $\gamma$ \\
		&  [0.02237] & [0.12] & [-1.001] & [3.044] & [0.9649] & [0.00] & [0.6736] & [1.00] & [1.386] & [0.60]  \\
		\hline
		BINGO+{\it Planck}  & $\pm0.00015$ & $\pm0.0084$ & $\pm0.066$ & $\pm0.016$ & $\pm0.0041$& $\pm0.024$ & $\pm0.016$ & $\pm0.040$ & $\pm0.058$ & $\pm0.079$ \\ \hline
		SKA B\,1+{\it Planck}  & $\pm0.00012$ & $\pm0.00086$ & $\pm0.0049$ & $\pm0.016$ & $\pm0.0028$& $\pm0.00077$ & $\pm0.0020$ & $\pm0.0086$ & $\pm0.015$ & $\pm0.0065$ \\ \hline
		SKA B\,2+{\it Planck} & $\pm0.00013$ & $\pm0.0060$  & $\pm0.034$ & $\pm0.016$ & $\pm0.0035$   & $\pm0.017$ & $\pm0.0081$ & $\pm0.021$ & $\pm0.011$ & $\pm0.050$ \\ \hline
		SKA B\,1+SKA B\,2+{\it Planck} & $\pm0.00012$ & $\pm0.00081$  & $\pm0.0045$ & $\pm0.013$ & $\pm0.0025$   & $\pm0.00074$ & $\pm0.0017$ & $\pm0.0060$ & $\pm0.0045$ & $\pm0.0048$ \\ \hline
		\hline
		\multicolumn{11}{c}{} \\
		\hline
		\hline
		Model III & $\Omega_{\mathrm{b}}h^2$ & $\Omega_{\mathrm{c}}h^2$ & $w$ & $\ln (10^{10}A_s)$ & $n_s$ & $\lambda_1$ & $h$ & $b_{\rm HI}$ & $\ln (10^{4}A)$ & $\gamma$ \\
		&  [0.02237] & [0.12] & [-1.001] & [3.044] & [0.9649] & [0.00] & [0.6736] & [1.00] & [1.386] & [0.60]  \\
		\hline
		BINGO+{\it Planck}  & $\pm0.00017$ & $\pm0.0020$ & $\pm0.13$ & $\pm0.016$ & $\pm0.0041$& $\pm0.00079$ & $\pm0.016$ & $\pm0.048$ & $\pm0.080$ & $\pm0.066$ \\ \hline
		SKA B\,1+{\it Planck}  & $\pm0.00013$ & $\pm0.00097$ & $\pm0.0035$ & $\pm0.015$ & $\pm0.0036$& $\pm0.00038$ & $\pm0.0021$ & $\pm0.0086$ & $\pm0.018$ & $\pm0.0064$ \\ \hline
		SKA B\,2+{\it Planck} & $\pm0.00016$ & $\pm0.0012$  & $\pm0.040$ & $\pm0.016$ & $\pm0.0038$   & $\pm0.00058$ & $\pm0.0076$ & $\pm0.013$ & $\pm0.011$ & $\pm0.030$ \\ \hline
		SKA B\,1+SKA B\,2+{\it Planck} & $\pm0.00012$ & $\pm0.00096$  & $\pm0.0034$ & $\pm0.013$ & $\pm0.0031$   & $\pm0.00031$ & $\pm0.0017$ & $\pm0.0062$ & $\pm0.0043$ & $\pm0.0047$ \\ \hline
		\hline
		\multicolumn{11}{c}{} \\
		\hline
		\hline
		Model IV & $\Omega_{\mathrm{b}}h^2$ & $\Omega_{\mathrm{c}}h^2$ & $w$ & $\ln (10^{10}A_s)$ & $n_s$ & $\lambda$ & $h$ & $b_{\rm HI}$ & $\ln (10^{4}A)$ & $\gamma$ \\
		&  [0.02237] & [0.12] & [-1.001] & [3.044] & [0.9649] & [0.00] & [0.6736] & [1.00] & [1.386] & [0.60]  \\
		\hline
		BINGO+{\it Planck}  & $\pm0.00017$ & $\pm0.0023$ & $\pm0.13$ & $\pm0.016$ & $\pm0.0042$& $\pm0.00084$ & $\pm0.016$ & $\pm0.048$ & $\pm0.079$ & $\pm0.066$ \\ \hline
		SKA B\,1+{\it Planck}  & $\pm0.00014$ & $\pm0.0011$ & $\pm0.0039$ & $\pm0.016$ & $\pm0.0038$& $\pm0.00040$ & $\pm0.0022$ & $\pm0.0087$ & $\pm0.019$ & $\pm0.0065$ \\ \hline
		SKA B\,2+{\it Planck} & $\pm0.00017$ & $\pm0.0014$  & $\pm0.040$ & $\pm0.016$ & $\pm0.0039$   & $\pm0.00062$ & $\pm0.0077$ & $\pm0.014$ & $\pm0.011$ & $\pm0.030$ \\ \hline
		SKA B\,1+SKA B\,2+{\it Planck} & $\pm0.00012$ & $\pm0.0011$  & $\pm0.0034$ & $\pm0.013$ & $\pm0.0032$   & $\pm0.00032$ & $\pm0.0017$ & $\pm0.0063$ & $\pm0.0044$ & $\pm0.0048$ \\ \hline
		\hline
	\end{tabular}
	\caption{The projected $1\sigma$ uncertainties obtained by accounting for the time dependence of HI fraction with $\Omega_{\mathrm{HI}}(z) = A(1+z)^{\gamma}$, as an extension to Table~\ref{tab.Omhi_1}. Such $\Omega_{\mathrm{HI}}$ modelling will further degrade the constraints, especially for $h$ owing to its degeneracy to $\Omega_{\mathrm{HI}}$ drawn from Eq.~(\ref{eq:Tb}).}
	\label{tab.Omhi_2}
\end{table*}



\section{Conclusions}\label{sec.conclusions}
In this work, we estimate the capabilities of three upcoming HI IM surveys, BINGO, SKA1-MID Band\,1 and Band\,2, in constraining a beyond-standard cosmological model encompassing a phenomenologically inspired interaction between DM and DE. The projected uncertainties of cosmological parameters are obtained by employing a conventional forecast methodology using the Fisher matrix analysis.

We start with a simple review of this comprehensive model incorporating four specific interacting scenarios. Then, we redo the derivation of the 21-cm angular power spectrum in the context of our interacting DE models and perceive an extra contribution to the 21-cm signal induced by the interaction recasting the Euler equation of the bulk velocity of HI. After clarifying the fiducial survey configurations, we qualitatively discuss how interacting DE can leave imprints on 21-cm signals through the equation of state $w$ and two interacting strength parameters, $\lambda_1$ and $\lambda_2$. Regardless, the physical reason behind is not abstruse: more DM or less DE during the cosmic evolution is helpful to matter condensation and, then, resulting in a higher 21-cm signal. Assuming  an optimistic situation of no foreground contamination, we further illustrate the impacts on signals and two types of interference, the shot and thermal noises, by varying the channel bandwidth of frequency channels for the three HI IM surveys, respectively. We summarize three conclusions we have obtained: 1) The narrower the channel bandwidth is, the higher the levels of signal and noise are. 2) The rapid growth of thermal noise at $\ell \gtrsim 100$, a feature produced by the instrumental beam correction in essence, validates our assumption of ignoring nonlinear effects. 3) The position of signal-noise intersection in the multipole $\ell$ space is mildly shifted by the value of channel bandwidth.

Three HI IM projects: SKA1-MID Band\,1 , SKA1-MID Band\,2, BINGO are listed according to their capability in parameter constraints from strong to weak. Compared with {\it Planck} 2018, although HI IM surveys are weaker in measuring early-Universe parameters (i.e., $A_s$ and $n_s$), we readily find that they have great potential in bounding late-Universe parameters (i.e., $w$ and $h$) and the strength of DM-DE interactions. In particular, for each interacting DE scenario, barring the minimal projected $1\sigma$ uncertainty of $w$ as a credit to SKA1-MID Band\,1, the other two HI IM projects can also outperform/play a draw game against {\it Planck} in constraining the interacting strength, together with the tightest bound on $h$ given by SKA1-MID Band\,2. Among the four specific IDE scenarios, the corresponding optimal and worst constraints forecasted with HI IM only on $w$, $h$ and $\lambda_1$ or $\lambda_2$ are of magnitude $\sim 0.0036$ against $\sim 0.36$, $\sim 0.0024$ against $\sim 0.11$, and $\sim 0.001$ against $\sim 0.035$, respectively.

Another desirable feature of HI IM projects worth stressing is to measure the overdensity bias of HI gas from matter, $b_{\rm HI}$, up to an accuracy of ${\cal O}(10^{-2}$), which is inaccessible to CMB observations. Of course, by adding the inverse of covariances from {\it Planck} 2018 into the Fisher matrix of one specific HI IM survey, we obtain joint constraints with less uncertainties. The most improvement is dedicated to BINGO, whereas the minimum progress comes in SKA1-MID Band\,1. The most restricted uncertainties come from 'SKA1-MID Band\,1 + Band\,2 + {\it Planck} 2018', whose magnitudes are of $\sigma_w \sim 0.0032\,(0.32\%)$, $\sigma_h \sim 0.0013\,(0.19\%)$, $\sigma_{\lambda_1} = 0.00030$ in Model III, and $b_{\rm HI} = 0.0057\,(0.57\%)$ in Model II.

The forecasted constraints are strongly related to survey configurations and signal components and, thereby, we extend our discussion to the impacts of binning scheme and RSD on BINGO forecast. In a condition of a fixed range of observing frequency, we find that the projected uncertainties shrink with an increase of $N_{\rm bin}$, the number of frequency channels, until $N_{\rm bin} \gtrsim 80$, where the cosmological information from tomographic slices are close to saturation owing to well enhanced noise levels. Also, an inclusion of RSD contribution is fairly useful in breaking the complete degeneracy of $b_{\rm HI}$ and $A_s$ if the overdensity of HI is the only source of the signal. Putting aside its negative effects of impeding the measurements of $w$ and $\lambda_{2}$ as well as $\lambda_{1}$, RSD is able to update the projected constraints for our interacting DE scenarios by up to $\sim 20\%$, together with an increased $N_{\rm bin}$.

Besides, foreground residuals and the uncertainty on measured $\Omega_{\mathrm{HI}}(z)$ are two additional obstacles in reality to the full information extraction from 21-cm lines. Regarding to the former, we investigate the extent to which the projected constraints will be degraded via firstly a simple modelling of residuals by assuming Gaussian-distributed foregrounds manipulated by the efficiency of foreground removal, $\epsilon_{\rm FG}$, and thence for a set of varying $\epsilon_{\rm FG}$s, calculating the resulting ratios of projected uncertainties to the previous forecast results without foreground residual consideration. Our calculation manifests that the degradation to the parameter-determining power of three HI IM surveys alone are almost to the same level as $\lesssim 10\%$ when $\epsilon_{\rm FG} = 10^{-2}$. Also as anticipated, a reduction in uncertainty appears with a smaller $\epsilon_{\rm FG}$, referring to a better foreground subtraction, and a threshold value of $\epsilon_{\rm FG} \lesssim 10^{-5}$ is found for approaching to a nearly perfect situation of no foreground contamination. In terms of the later issue, we discuss the interference separately from the magnitude and time dependence of $\Omega_{\mathrm{HI}}$, after imposing a {\it Planck} 2018 prior as to break the degeneracy between HI fraction and $A_s$. Indeed, $\Omega_{\mathrm{HI}}$ as a new degree of freedom to the Fisher matrix will introduce a global degradation to the forecast results. Solely focusing on the magnitude variation, we find that two most affected parameters are $h$ and $b_{\rm HI}$. The former is caused by its degeneracy to $\Omega_{\mathrm{HI}}$ drawn from their product in the expression of $\bar{T}_{\mathrm{b}}(z)$, of which the degradation on $\sigma_h$ with respect to the fiducial prediction is up to $\sim 3$ times, while the later encounters a uncertainty increment of $1.77$ times. Despite that, this sensitivity to $\Omega_{\mathrm{HI}}$ provides the possibility to its measurement through HI IM surveys once those degeneracies are properly taken into account. We predict that a synergy of BINGO+{\it Planck} is able to measure $\Omega_{\mathrm{HI}}$ with an accuracy of $\sim 5.2\%$, and the cooperation of SKA1-MID+{\it Planck} can upgrade the precision even to the degree of $\lesssim 1.5\%$. Taking into account the time evolution, $\Omega_{\mathrm{HI}}(z) = A(1+z)^{\gamma}$, will further downgrade the constraining power of HI IM+{\it Planck}. Relative to the fiducial prediction, the maximum $\sigma_h$ is about $3.8$ times larger and the bound on $w$ also has been relaxed at most by a factor about $2.5$, regardless of the IDE scenarios. Still, the appearance of $\gamma$ will not severely hinder the detection of $A$, the amplitude of $\Omega_{\mathrm{HI}}$, and the best constraints given by SKA1-MID Band\,1+Band\,2+{\it Planck} on $A$ and $\gamma$ are of $\lesssim 0.46\%$ and $\lesssim 0.83\%$, respectively.

Although the detectability of IDE with future HI IM observations has been previously studied, our work constitutes a complementary extension to~\cite{Xu:2017rfo}, especially we have uncovered a new term in the brightness temperature coming from the interaction. We have also extended the physical analyses and quantitative estimation of RSD impacts along with two other systematics of foreground residuals and the uncertainty from $\Omega_{\mathrm{HI}}(z)$ measurements. However, we notice some differences with~\cite{Xu:2017rfo} in our fiducial projected variances. Taking Model I ($w > -1$) as an example, our $\sigma_w$ with BINGO configurations is about 5.5 times weaker than theirs, whereas for SKA1-MID Band\,1 their $\sigma_{\lambda_2}$ is in excess of a factor around 5.5 in relative to ours. These discrepancies may be related with different configurations for the surveys and/or descriptions of 21-cm signals. \cite{Xu:2017rfo} and our work reach a consensus that future HI IM surveys will be comparable to current CMB measurements in probing IDE. Moreover, it indicates the usefulness of HI IM in detecting/ruling out other non-standard cosmologies, especially those general extensions to our IDE model (e.g., a conformal/disformal coupling of dark sectors~\citep{vandeBruck:2017idm}), which we leave for future works.

\section*{Acknowledgements}

We thank Jiajun Zhang for productive discussions. This work was partially supported by the key project of NNSFC under contract No. 11835009. A.A.C. acknowledges financial support from the China Postdoctoral Science Foundation, grant number 2020M671611.

\section*{Data availability}
The data underlying this article will be shared on reasonable request to the corresponding author.



\bibliographystyle{mnras}
\bibliography{HiIM_IDE} 








\bsp	
\label{lastpage}
\end{document}